%% file: New_IEEEtran_how-to.tex
\documentclass[lettersize,journal]{IEEEtran}
\usepackage{amsmath,amsfonts}
\usepackage{pgf-pie}
\usepackage{algorithmic}
\usepackage{array}
\usepackage[caption=false,font=normalsize,labelfont=sf,textfont=sf]{subfig}
\usepackage{textcomp}
\usepackage{stfloats}
\usepackage{url}
\usepackage{verbatim}
\usepackage{graphicx}
\usepackage{amstext,amssymb,amsfonts,latexsym}
\usepackage{graphicx}
\usepackage{enumitem} 
\usepackage{array} 
\usepackage{caption} 
\usepackage{ccicons}
\usepackage{animate}
\usepackage{setspace}
\usepackage{longtable}
\usepackage{graphicx}
\usepackage{booktabs}
\usepackage{pifont}
\usepackage[utf8]{inputenc}
\usepackage{rotating}
\usepackage{tabularx}
\usepackage{multirow}
\usepackage{color}
\usepackage[normalem]{ulem}
\usepackage{tabularray}
\usepackage{amsmath}
\usepackage{float}
\usepackage{placeins}
\def\BibTeX{{\rm B\kern-.05em{\sc i\kern-.025em b}\kern-.08em
    T\kern-.1667em\lower.7ex\hbox{E}\kern-.125emX}}
\usepackage{balance}
\usepackage[colorlinks=true, linkcolor=black, citecolor=black, urlcolor=black]{hyperref}
\usepackage{cite}
\begin{document}
\title{From LLMs to LLM-based Agents for Software Engineering: A Survey of Current, Challenges and Future}
\author{Haolin Jin, Linghan Huang, Haipeng Cai, Jun Yan, Bo Li, Huaming Chen
\thanks{Haolin Jin, Linghan Huang and Huaming Chen are with the School of Electrical and Computer Engineering, The University of Sydney, Sydney, 2006, Australia. (email: huaming.chen@sydney.edu.au)}
\thanks{Haipeng Cai is with the Department of Computer Science and Engineering at University at Buffalo, SUNY, USA}
\thanks{Jun Yan is with the School of Computing and Information Technology at University of Wollongong, Australia}
\thanks{Bo Li is with the Computer Science Department at the University of Chicago, US}
}

\markboth{Journal of \LaTeX\ Class Files,~Vol.~18, No.~9, September~2020}%
{How to Use the IEEEtran \LaTeX \ Templates}

\maketitle

\begin{abstract}
With the rise of large language models (LLMs), researchers are increasingly exploring their applications in various vertical domains, such as software engineering. LLMs have achieved remarkable success in areas including code generation and vulnerability detection. However, they also exhibit numerous limitations and shortcomings. LLM-based agents, a novel technology with the potential for Artificial General Intelligence (AGI), combine LLMs as the core for decision-making and action-taking, addressing some of the inherent limitations of LLMs such as lack of autonomy and self-improvement. Despite numerous studies and surveys exploring the possibility of using LLMs in software engineering, it lacks a clear distinction between LLMs and LLM-based agents. It is still in its early stage for a unified standard and benchmarking to qualify an LLM solution as an LLM-based agent in its domain. In this survey, we broadly investigate the current practice and solutions for LLMs and LLM-based agents for software engineering. In particular we summarise six key topics: requirement engineering, code generation, autonomous decision-making, software design, test generation, and software maintenance. We review and differentiate the work of LLMs and LLM-based agents from these six topics, examining their differences and similarities in tasks, benchmarks, and evaluation metrics. Finally, we discuss the models and benchmarks used, providing a comprehensive analysis of their applications and effectiveness in software engineering. We anticipate this work will shed some lights on pushing the boundaries of LLM-based agents in software engineering for future research.
\end{abstract}

\begin{IEEEkeywords}
Large Language Models, LLM-based Agents, Software Engineering, Benchmark, Software Security, AI System Development
\end{IEEEkeywords}

\input{pages/intro.tex} 
\input{pages/preliminaries.tex}

\input{pages/requirement.tex}

\input{pages/code.tex}
\input{pages/autonomous.tex}
\input{pages/design.tex}
\input{pages/test.tex}

\input{pages/maintenance.tex}
\input{pages/discussion.tex}

\input{pages/conclusion.tex}
\bibliography{ref.bib}
\bibliographystyle{ieeetr}
\end{document}

%% file: pages/intro.tex
\section{Introduction}\label{cha:introduction}
\IEEEPARstart{S}{oftware} engineering (SE) has seen its booming research and development with the aid of artificial intelligence techniques. Traditional approaches leveraging neural networks and machine learning have facilitated various SE topics such as bug detection, code synthesis, and requirements analysis~\cite{wang2016bugram,vogelsang2019requirements}. However, they often present limitations, including the need for exclusive feature engineering, scalability issues, and the adaptability across diverse codebases. The rise
of Large Language Models (LLMs) has embarked on new solutions and findings in this landscape. LLMs, such as GPT~\cite{chatgpt2022} and Codex~\cite{chen2021evaluating}, have demonstrated remarkable capabilities in handling downstream tasks in SE, including code generation, debugging, and documentation. These models leverage vast amounts of training data to generate human-like text, offering unprecedented levels of fluency and coherence. Studies have shown that LLMs can enhance productivity in software projects by providing intelligent code suggestions, automating repetitive tasks, even generating entire code snippets from natural language descriptions~\cite{10.1145/3510003.3510203}.

Despite their potential, there are significant challenges in applying LLMs to SE. One major issue is their limited context length~\cite{li2024longcontextllmsstrugglelong}, which restricts the model's ability to comprehend and manage extensive codebases, making it challenging to maintain coherence over prolonged interactions. Hallucinations is another main concern, where the model generates code that appears plausible but is actually incorrect or nonsensical~\cite{10.1145/3649506}, potentially introducing bugs or vulnerabilities if not carefully reviewed by experienced developers. Additionally, the inability of LLMs to use external tools restricts their access to real-time data and prevent them from performing tasks outside their training scope. It diminishes their effectiveness in dynamic environments. These limitations significantly impact the application of LLMs in SE, and also highlight the need for expert developeers to critically refine and validate LLM-generated code for accuracy and security~\cite{10449667}. In complex projects, the static nature of LLMs can hinder their ability to adapt to changing requirements or efficiently incorporate new information. Moreover, LLMs typically cannot interact with external tools or databases, further limits their utility in dynamic and evolving SE contexts. 

To address these challenges, LLM-based agents have emerged \cite{10.1007/s11704,xi2023risepotentiallargelanguage}, combining the strengths of LLMs with external tools and resources to enable more dynamic and autonomous operations. These agents leverage recent advancements in AI, such as Retrieval-Augmented Generation (RAG) and tool utilization, to perform more complex and contextually aware tasks~\cite{NEURIPS2020_6b493230}. For instance, OpenAI's Codex has been integrated into GitHub Copilot~\cite{copilot}, enabling real-time code suggestions and completion within development environments. Unlike static LLMs, LLM-based agents can perform a wide range of tasks, such as autonomously debugging code by identifying and fixing errors, proactively refactoring code to enhance efficiency or readability, and generating adaptive test cases that evolve alongside the codebase. These features make LLM-based agents a powerful tool for SE, capable of handling more complex and dynamic workflows than traditional LLMs. 

Historically, AI agents focused on autonomous actions based on predefined rules or learning from interactions ~\cite{russell2016artificial,jennings2000survey}. The integration of LLMs has presented new opportunities in this area, providing the language understanding and generative capabilities needed for more sophisticated agent behaviors. \cite{xi2023risepotentiallargelanguage} shows that LLM-based agents are capable of autonomous reasoning and decision-making, achieving the third and fourth levels of WS (World Scope)~\cite{bisk2020experiencegroundslanguage}, which outlines the progression from natural language processing (NLP) to general AI. In software engineering, LLM-based agents show promise in areas such as autonomous debugging, code refactoring, and adaptive test generation, demonstrating capabilities that approach artificial general intelligence(AGI). 
\begin{figure}
    \centering
    \includegraphics[width=\linewidth]{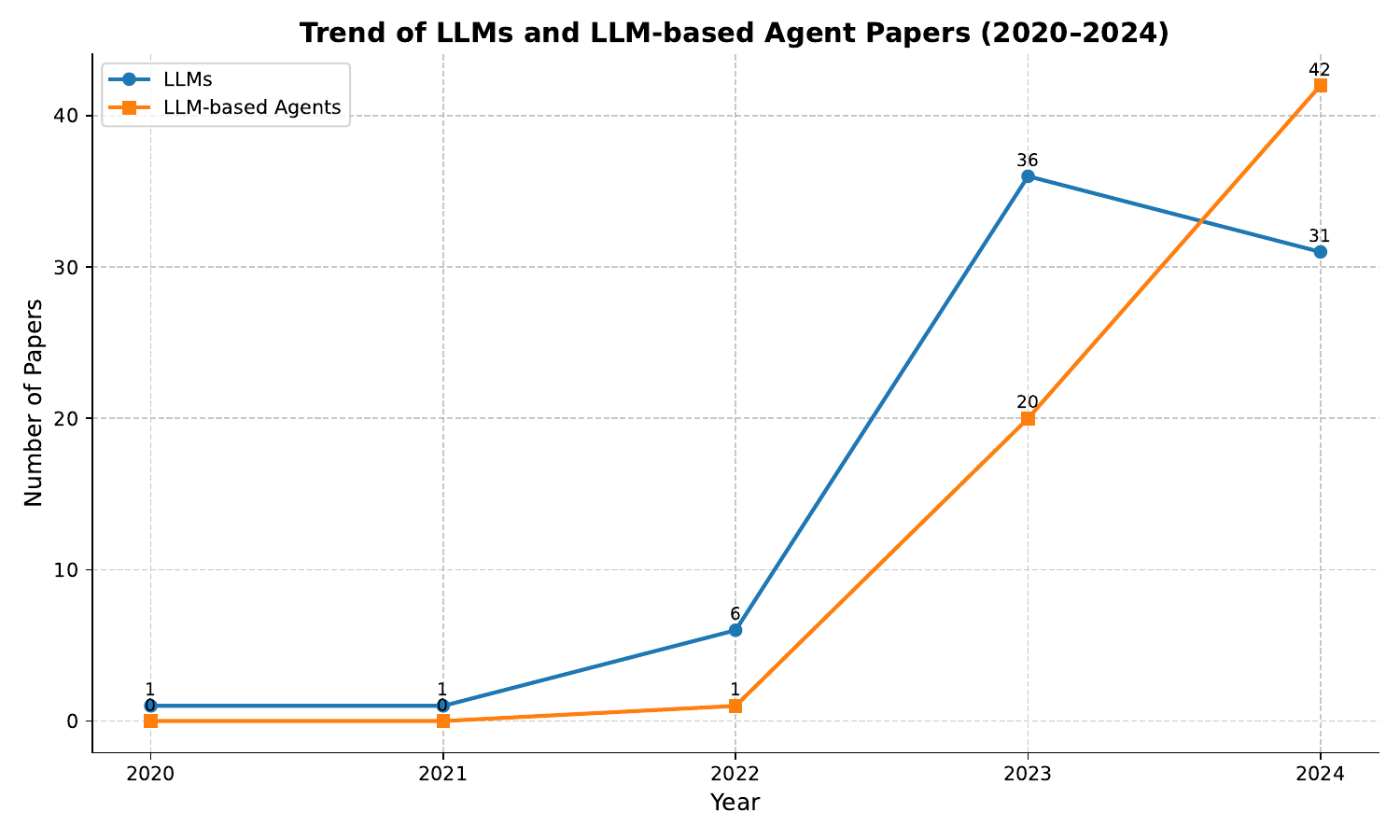}
    \caption{Number of Papers Collected on LLMs and LLM-based Agents from 2020 to 2024.}
    \label{fig:year}
\end{figure}

In this work, we present, to the best of our knowledge, a first survey outlining the integration and transformation of LLMs to LLM-based agents in the domain of SE. Our survey covers six key themes in SE:
\begin{enumerate}
    \item \textbf{Requirement Engineering and Documentation}: Capturing, analyzing, and documenting software requirements, as well as generating user manuals and technical documentation.
    \item \textbf{Code Generation and Software Development}: Automating code generation, assisting in the development lifecycle, refactoring code, and providing intelligent code recommendations.
    \item \textbf{Autonomous Learning and Decision Making}: Highlighting the capabilities of LLM-based agents in autonomous learning, decision-making, and adaptive planning within SE contexts.
    \item \textbf{Software Design and Evaluation}: Contributing to design processes, architecture validation, performance evaluation, and code quality assessment.
    \item \textbf{Software Test Generation}: Generating, optimizing, and maintaining software tests, including unit tests, integration tests, and system tests.
    \item \textbf{Software Security \& Maintenance}: Enhancing security protocols, facilitating maintenance tasks, and aiding in vulnerability detection and patching.
\end{enumerate}

In detail, we aim to address following research questions:
\begin{itemize}
    \item RQ1: What are the state-of-the-art techniques and practices in LLMs and LLM-based agents for SE? (Section~\ref{cha:requirement}-~\ref{cha:softwareSecurity})
    \item RQ2: What are the key differences in task performance between LLMs and LLM-based agents in SE applications? (Sections~\ref{cha:requirement}-~\ref{cha:softwareSecurity})
    \item RQ3: Which benchmark datasets and evaluation metrics are most commonly used for assessing the performance of LLMs and LLM-based agents in SE tasks? (Section~\ref{cha:requirement}-~\ref{cha:softwareSecurity} and Section~\ref{cha:results})
    \item RQ4: What are the predominant experimental models employed when utilizing LLMs in SE? (Section~\ref{cha:results})
\end{itemize}
\section{EXISTING Surveys AND THE SURVEY STRUCTURE}
\subsection{Existing Works} 
In recent years, large language models have been primarily applied to help programmers generate code and fix bugs. These models understand and complete code or text based on the user's input, leveraging their training data and reasoning capabilities. In previous survey papers, such as Angela Fan's research~\cite{10449667}, there has not been much elaboration on requirement engineering. As mentioned in the paper, software engineers are generally reluctant to rely on LLMs for higher-level design goals. However, with LLMs achieving remarkable improvements in contextual analysis and reasoning abilities through various methods like prompt engineering and Chain-of-Thought (COT)~\cite{wei2022chain}, their applications in requirement engineering are gradually increasing. Table~\ref{tab:se_tasks} summarizes and categorizes the tasks in software engineering. Since the collection primarily focuses on the latter half of 2023 and before December 2024, and some papers address multiple tasks, the table reflects the number of tasks after accounting for overlaps.
\begin{table*}
\centering
\caption{Distribution of SE tasks}
\label{tab:se_tasks}
\scriptsize 
\begin{tabular}{|c|l|l|c|} 
\hline
\multicolumn{1}{|c|}{Category}& \multicolumn{1}{c|}{LLMs}              & \multicolumn{1}{c|}{LLM-based agents}                      & Total  \\ 
\hline
\begin{tabular}[c]{@{}l@{}}Requirement \\Engineering~and~\\Documentation\end{tabular}  & \begin{tabular}[c]{@{}l@{}}\\\\Requirement Classification~and Extraction (4)~ \\Requirement Generation and Description (6)\\Requirements Satisfaction Assessment (1)\\Specification Generation (5)~ ~\\Quality Evaluation (3)\\Ambiguity Detection (2)~ ~\\~~\\\end{tabular}                                                                                 & \begin{tabular}[c]{@{}l@{}}\\\\Generation of Semi-structured Documents (1)\\Generate safety requirements (1)\\Automatically generating use cases based on \\high-level requirements (1)\\ Requirements Satisfaction Assessment (1)\\Automated User Story Quality Enhancement (3)\\\\\end{tabular} & 28     
\\ 
\hline
\begin{tabular}[c]{@{}l@{}}Code Generation \\and \\software \\development\end{tabular} & \begin{tabular}[c]{@{}l@{}}\\\\Code Generation Debugging (3)~ ~\\Code Evaluation (2)~~\\Implement HTTP server (1) \\Enhancing Code Generation Capabilities (5)\\Specialized Code Generation (3) \\Human Feedback Preference Simulation (1)\\\\\end{tabular}   & \begin{tabular}[c]{@{}l@{}}\\\\Automating the Software Development Process (5)~\\Large-Scale Code and Document Generation (2)\\Tool and External API Usage (4)\\Multi-Agent Collaboration and Code Refine (6)\\Improving Code Generation Quality (3)\\\\\end{tabular}  & 35     
\\ 
\hline
\begin{tabular}[c]{@{}l@{}}Autonomous\\Learning \\and Decision \\Making\end{tabular}   & \begin{tabular}[c]{@{}l@{}}\\\\Multi-LLMs Decision-Making (1)~ ~ ~ \\Creativity Evaluation (1)\\Self-Identify and Correct Code (1)~ ~ \\Judge Chatbot Response (1)\\Mimics Human Scientific Debugging (1)~\\Deliberate Problem Solving(1)~ ~ ~ ~ ~ ~ ~ ~~\\\\\end{tabular}                                                                                   & \begin{tabular}[c]{@{}l@{}}\\\\Collaborative Decision-Making and Multi-Agent \\Systems (6)\\Autonomous Reasoning and Decision-Making (12)\\Learning and Adaptation Through Feedback (4)\\Simulation and Evaluation of Human-like \\Behaviors (2)\\\\\end{tabular}                                                & 30     \\ 
\hline
\begin{tabular}[c]{@{}l@{}}Software Design \\and Evaluation\end{tabular}               & \begin{tabular}[c]{@{}l@{}}\\\\Creative Capabilities Evaluation (1)~ ~ ~\\Performance in SE Tasks (1)\\Educational Utility and Assessment (1) \\Efficiency Optimization (2)\\\\\end{tabular}                                                                                                                                                                 & \begin{tabular}[c]{@{}l@{}}\\\\Automation of Software Engineering Processes (3)\\Enhancing Problem Solving and Reasoning (4)\\Integration and Management of AI Models and \\Tools (3)\\Optimization and Efficiency Improvement (2)\\Performance Assessment in Dynamic Environments \\(2)\\\\\end{tabular}       & 19     \\ 
\hline
\begin{tabular}[c]{@{}l@{}}Software Test \\Generation\end{tabular}                     & \begin{tabular}[c]{@{}l@{}}\\\\Bug Reproduction and Debugging (2) \\Security Test (2)\\Test Coverage (3)\\Test-Informed Code Generation (1) \\Universal Fuzzing (1)\\\\\end{tabular}                                                                                                                                                                                                  & \begin{tabular}[c]{@{}l@{}}\\\\Multi-agent Collaborative Test Generation (3)\\Autonomous Testing and Conversational Interfaces \\(3)\\\\\end{tabular}                                                                                                                                                           & 15     \\ 
\hline
\begin{tabular}[c]{@{}l@{}}Software Security \\\&\\Maintainance\end{tabular}           & \begin{tabular}[c]{@{}l@{}}\\\\Vulnerability Detection (7) \\Vulnerability Repair (2)\\Program Repair (5)~ ~ ~ ~ ~ ~ ~\\Robustness Testing (1)\\Requirements Analysis (1)~ \\Fuzzing (1)\\Duplicate Entry (1)~\\Code Generation and Debugging (4)\\Penetration Testing and Security Assessment (2)\\Program Analysis and Debugging (1)~ ~~~\\\\\end{tabular} & \begin{tabular}[c]{@{}l@{}}\\\\Autonomous Software Development and \\Maintenance (6)\\Debugging and Fault Localization (4)\\Vulnerability Detection and Penetration \\Testing (3)\\Smart Contract Auditing and Repair (2)\\Safety and Risk Analysis (2)\\Adaptive and Communicative Agents (1)\\\\\end{tabular} & 43     \\
\hline
\end{tabular}
\end{table*}

\begin{figure}
    \centering
    \includegraphics[width=1\linewidth]{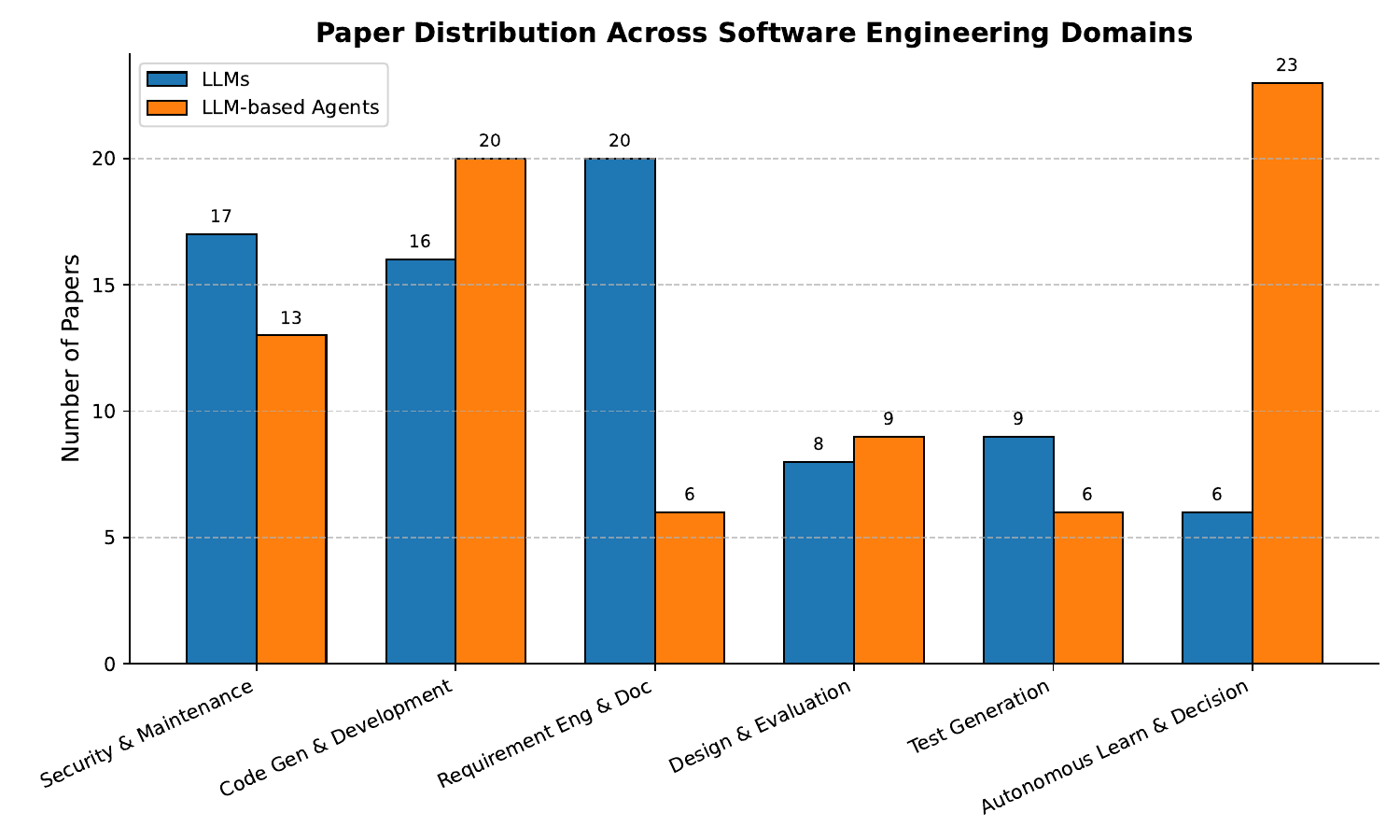}
    \caption{Paper Distribution.}
    \label{fig:paperDistribution}
\end{figure}
While other works have surveyed LLMs applications in some SE tasks~\cite{hou2024largelanguagemodelssoftware,10449667,zheng2023understandinglargelanguagemodels}, they lack a wider coverage of the general SE area to incorporate recent research developments. More importantly, a focus of LLMs is the main contributions of these works, but there is no distinguish the capabilities between LLMs and LLM-based agents. We summarize the difference between our work and others in Table~\ref{tab:paper_comparison}, this survey addresses these limitations by distinctly analyzing LLMs and LLM-based agents applications across six SE domains, providing a thorough and up-to-date review. 
From previous research, it is evident that the performance of LLMs in various applications and tasks heavily depends on the model's inherent capabilities~\cite{xi2023risepotentiallargelanguage}. More importantly, earlier surveys often present findings from papers spanning in a wide range of publication dates, leading to significant content disparities for LLMs in different SE tasks. For instance, research in requirement engineering was relatively nascent, resulting in sparse content in this area in previous surveys. The recent rise of LLM-based agents, with their enhanced capabilities and autonomy, fills these gaps. By focusing on the latest researches and clearly differentiating between LLMs and LLM-based agents, our survey provides a thorough and in-depth overview of how these technologies are applied and new opportunities they bring to SE. 

In summary, we have collected a total of 139 papers directly relevant to this topic, covering the six SE domains mentioned earlier as shown in Figure.\ref{fig:year}. Our analysis distinguishes between LLM and LLM-based agent contributions, offering a comparative overview and addressing the limitations of previous surveys. Considering the novelty nature of the LLM-based agents field and the lack of standardized benchmarks, this work seeks to offer a detailed review that can guide future research and provide a clearer view of the potential of these technologies in SE.
\begin{table}
\centering
\caption{Comparison between Our Work and the Existing Work for LLM in SE.}
\label{tab:paper_comparison}
\tiny
\renewcommand{\arraystretch}{1.6} 
\setlength{\tabcolsep}{5pt} 
\begin{tabular}{|c|c|c|c|c|c|c|} 
\hline
Paper & Year & Domain                             & Benchmarks & Metrics & \begin{tabular}[c]{@{}c@{}}Agent \\in SE\end{tabular} & \begin{tabular}[c]{@{}c@{}}Agent\\Distinction\end{tabular}  \\ 
\hline
{~\cite{nguyenduc2023generativeartificialintelligencesoftware}}    
& 2023 & GenAI in SE  & \ding{51}  & \ding{51} & \ding{51} & \ding{55} 
\\
{~\cite{10449667}}  
& 2023 & LLM in SE  & \ding{51}  & \ding{51} & \ding{51} & \ding{55}  
\\
{~\cite{zheng2023understandinglargelanguagemodels}}     
& 2023 & Generation task by LLM in SE  & \ding{51} & \ding{51} & \ding{55}  & \ding{55}                                                          
\\
{~\cite{ma2024lmsunderstandingcodesyntax}}     
& 2023 & LLM in syntax comprehension & \ding{51} & \ding{51} & \ding{55}    & \ding{55}                                                          
\\
{~\cite{yang2024robustnesssecurityprivacyexplainability}}     
& 2024 & LLM4Code in SE & \ding{51} & \ding{51} & \ding{55} & \ding{55}     \\
{~\cite{hou2024largelanguagemodelssoftware}}     
& 2024 & LLM for process optimization in SE & \ding{51} & \ding{51} 
& \ding{55} & \ding{55}                                                     \\
{~\cite{huang2024generativesoftwareengineering}}     
& 2024 & Generation task by LLM in SE & \ding{51} & \ding{51} & \ding{55}   & \ding{55}                                                          
\\ 
\hline
Ours  & 2024 & LLM \& LLM-based Agent in SE  & \ding{51} & \ding{51}      & \ding{51}  & \ding{51} \\
\hline
\end{tabular}
\end{table}

\subsection{Methodology}
By following previous research efforts~\cite{yang2024robustnesssecurityprivacyexplainability,10449667,hou2024largelanguagemodelssoftware}, we collected relevant papers from DBLP and arXiv, focusing on publications from the latter half of 2023 to December 2024. DBLP offers comprehensive, curated coverage of computer science venues, particularly in software engineering. We chose arXiv for its open-access nature, which makes the latest research results (including preprints) rapidly available~\cite{DBLP2024,arXiv2024}. This ensures we capture the most up-to-date studies, often before they appear in broader databases such as Scopus or Web of Science, thus enabling a timely and inclusive survey of emerging trends in LLM-based research.

\subsubsection{Database Selection and Rationale}
Most other survey papers collect papers from common databases such as Scopus and Web of Science. In contrast, our survey specifically selects DBLP and arXiv due to their distinct advantages:

\begin{itemize}
    \item \textbf{DBLP}: offers comprehensive, curated coverage of computer science venues, particularly in software engineering, thus ensuring relevance and high quality of retrieved papers.
    \item \textbf{arXiv}: provides open-access preprints, which enables rapid access to the most recent research findings in the field of computer science.
    \item \textbf{IEEE Xplore and ACM}: Additional databases such as IEEE Xplore and ACM Digital Library could further complement our search by including more peer-reviewed papers from recognized conferences and journals.
\end{itemize}

\subsubsection{Paper Retrieval Process}
We applied a systematic filtering process to ensure the relevance and quality of the papers included in this study. Specifically, we conducted a keyword-based filtering process to identify SE-related works. We utilized the keywords listed in Table~\ref{tab:keywords}, combining them with Boolean operators to perform precise and comprehensive searches:

\begin{itemize}
    \item Within each topic, keywords were joined using the OR operator, e.g., ("Software security" OR "Vulnerability detection").
    \item To ensure relevance, we combined topic keywords with LLM-related terms using the AND operator, e.g., ("Code generation" AND "Large Language Models").
\end{itemize}

\begin{table}
\scriptsize
\centering
\caption{Inclusion and Exclusion Criteria for Paper Selection.}
\label{tab:criteria}
\begin{tabular}{l} 
\hline
Inclusion criteria                                                                                                                                                                                                                                                                                                                                                              \\ 
\hline
\begin{tabular}[c]{@{}l@{}}1) The paper explicitly involves the application or study of LLMs.\\2) The paper demonstrates relevance to Software Engineering tasks.\\3) The paper provides sufficient experimental results and evaluations.\end{tabular}                                                                                                                          \\ 
\hline
Exclusion criteria                                                                                                                                                                                                                                                                                                                                                              \\ 
\hline
\begin{tabular}[c]{@{}l@{}}1) Papers with fewer than 7 pages.\\2) Papers that focus on general Artificial Intelligence not relate to LLMs.\\3) Studies that do not address Software Engineering applications or workflows.\\4) Grey literature, such as blog posts, white papers.\\5) Duplicate publications.\\6) Papers written in languages other than English.\end{tabular}  \\
\hline
\end{tabular}
\end{table}
\subsubsection{Selection Criteria}
Following the initial keyword-based retrieval, we applied additional inclusion and exclusion criteria to refine the selected papers further. We manually reviewed titles, abstracts, and, where necessary, the full texts to confirm that each paper specifically addressed the intersection of software engineering and large language models. We clearly established inclusion and exclusion criteria to filter papers, as detailed in Table~\ref{tab:criteria}. Papers primarily focused on unrelated domains or those mentioning LLMs without substantial relevance to software engineering tasks were excluded. Furthermore, we ensured that selected publications originated from reputable venues, including peer-reviewed journals, conferences, and workshops recognized by the software engineering and artificial intelligence research communities. This rigorous filtering process ensured the final set of papers provided a representative and high-quality overview of recent advancements in applying large language models to software engineering. 

Overall, we included papers that explicitly focused on topics involving LLMs, contained keywords from Table~\ref{tab:keywords} related to Software Engineering applications, and were at least seven pages long to ensure substantial contributions. We excluded papers unrelated to SE or LLMs, shorter than seven pages, or lacking relevant keywords.
\subsubsection{Manual Screening and Paper Distribution}
After the initial keyword-based retrieval and selection criteria, we also manually reviewed titles, abstracts, and, where necessary, the full texts to confirm that each paper specifically addressed the intersection of software engineering and large language models. Papers primarily focused on unrelated domains or those mentioning LLMs without substantial relevance to software engineering tasks were excluded. Furthermore, we ensured that selected publications originated from reputable venues, including peer-reviewed journals, conferences, and workshops recognized by the software engineering and artificial intelligence research communities. Additionally, we employed a snowballing search technique, systematically reviewing references cited within relevant papers to identify other significant studies that might have been overlooked initially \cite{hou2024largelanguagemodelssoftware}. This filtering process ensured the final set of papers provided a representative and high-quality overview of recent advancements in applying large language models to software engineering.

Overall, we identified 139 relevant papers. Figure~\ref{fig:paperDistribution} presents the distribution of these papers across the six SE domains and the proportion of LLM-based agent studies. However, some papers can be counted as multi-class fields, so the literature review in the figure includes more than 139 total paper instances.

\begin{figure}
    \centering
    \includegraphics[width=1\linewidth]{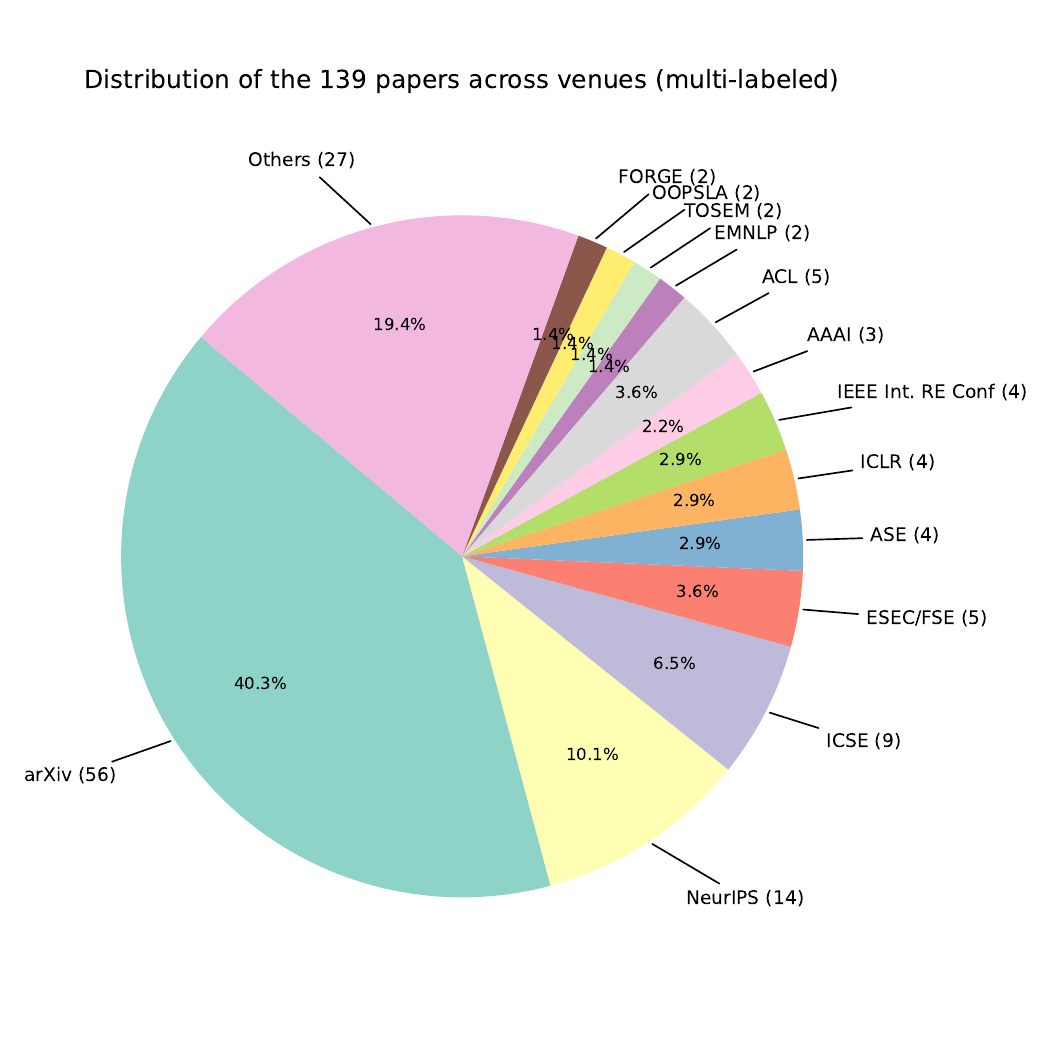}
    \caption{Venue Distribution.}
    \label{fig:venue_distribution}
\end{figure}

The venue distribution revealed in Figure~\ref{fig:venue_distribution} demonstrates both the interdisciplinary nature and rigorous academic foundation of this research domain. The prominence of arXiv (40.3\%) reflects the field's rapid evolution, where researchers pr ioritize immediate community dissemination of LLM advancements in software engineering. Notably, 59.7\% of papers originate from established peer-reviewed venues, with strong representation in software engineering flagships (ICSE: 6.5\%, ESEC/FSE: 3.6\% combined) and AI/ML top conferences (NeurIPS: 10.1\%, ICLR: 2.9\%). This dual concentration underscores how LLM-based SE research bridges fundamental AI innovation with practical engineering applications.

The distribution reveals three notable patterns: First, the presence of requirements engineering venues (IEEE RE Conference: 2.9\%) alongside code-centric conferences (OOPSLA: 1.4\%) illustrates LLMs' expanding role across the SE lifecycle. Second, NLP venues (ACL/EMNLP: 5.0\%) contribute specialized language modeling insights critical for code-text cross-modal tasks. Third, the inclusion of security (ACSAC) \cite{thapa2022transformer}, HCI (IUI) \cite{desmond2024evalullm}, and software practice (XP) \cite{rasheed2023autonomous} venues demonstrates the field's maturation into specialized subdomains. This balanced distribution ensures methodological rigor while capturing real-time innovation, validating our survey's comprehensive view of how LLM advancements redefine software engineering paradigms through contextual adaptation, open-ended reasoning, and human-AI collaboration.

\subsection{Overall Structure of the Work}
The remainder of this paper is organized as follows Section 2 Introduces the architectures and background of LLMs and LLM-based agents, including an overview of RAG, tool utilization, and their implications for SE. Section \ref{cha:Preliminaries}-\ref{cha:softwareSecurity} is the comparative analysis which Summarizes and compares the datasets, tasks, benchmarks, and metrics used in LLM and LLM-based agents studies across the six SE domains. Section \ref{cha:results} is the general discussion, Section \ref{cha:conclusion} is the final conclusion.

\begin{table*}[h!]
    \centering
    \caption{Keywords for Software Engineering Topics.}
    \scriptsize
    \label{tab:keywords}
    \begin{tabularx}{\textwidth}{X X}
         \toprule
         \textbf{Topic} & \textbf{Keywords} \\
         \midrule
         \textbf{Software Security \& Maintenance} & Software security, Vulnerability detection, Automated Program repair, Self-debugging, Vulnerability reproduction \\
         \midrule
         \textbf{Code Generation and Software Development} & Code generation, Automatic code synthesis, Code refactoring, Programming language translation, Software development automation, Code completion, AI-assisted coding, Development lifecycle automation \\
         \midrule
         \textbf{Requirement Engineering and Documentation} & Requirement engineering, Software requirements analysis, Automated requirement documentation, Technical documentation generation, User manual generation, Documentation maintenance, Requirements modeling, Requirements elicitation \\
         \midrule
         \textbf{Software Design and Evaluation} & Software design automation, Architectural validation, Design optimization, Performance evaluation, Code quality assessment,  Software metrics, Design pattern recognition, Architectural analysis, Code structure analysis \\
         \midrule
         \textbf{Software Test Generation} & Test case generation, Automated testing, Unit test generation, Integration test generation, System test generation, Test suite optimization, Fault localization, Test maintenance, Regression testing, Adaptive testing \\
         \midrule
         \textbf{Autonomous Learning and Decision Making} & Autonomous learning systems, Decision making, Adaptive planning, Project management automation, Self-improving software, Autonomous software agents \\
         \bottomrule
    \end{tabularx}
\end{table*}

%% file: pages/preliminaries.tex
\section{Preliminaries}\label{cha:Preliminaries}
In this section, we introduce the foundational concepts of large language models, including the evolution of their frameworks and an overview of their architectures. Subsequent to this, we will discuss LLM-based agents, exploring both single-agent and multi-agent systems. We will also covers the background of these systems and their applications and distinctions in the field of software engineering.
\subsection{Large Language Model}
There is an inherent connection between large language models and natural language processing (NLP), with the historical development of natural language technologies tracing back to the 1950s. The earliest attempts to generate language dialogues through machines using specific rules can be traced to the period between 1950 and 1970. The advent of machine learning technologies in the 1980s and the groundbreaking introduction of neural networks in the 1990s indicated a new era for NLP ~\cite{manning1999foundations}. These advancements facilitated significant progress in the NLP field, especially in the development of technologies for text translation and generation. The development of Long Short-Term Memory (LSTM) and Recurrent Neural Networks (RNN) during this period enabled more effective handling of the sequential nature of language data~\cite{6795963,hochreiter1997long}. These models addressed challenges associated with the lack of dependency in context, thereby enhancing the application of NLP in various domains.

In 2017 the new framework called "Transformer" introduced by Google’s research team~\cite{vaswani2017attention}. The transformer model based on the self-attention mechanism which significantly improved the effectiveness of language models. The inclusion of positional encoding not only solved the long-sequence dependency issue but also enabled parallel computation, which was a considerable improvement over previous models. In 2018, OpenAI developed the Generative Pre-trained Transformer (GPT)~\cite{chatgpt2022}, a model based on the transformer architecture. The core idea behind GPT-1 was to utilize a large corpus of unlabelled text for pre-training to learn the patterns and structures of language, followed by fine-tuning for specific tasks. Over the next two years, OpenAI released GPT-2 and GPT-3 which increased the parameter count to 175 billions and also demonstrated strong capabilities in context understanding and text generation~\cite{floridi2020gpt}. GPT-4 launched by OpenAI in 2023, represents a milestone following GPT-3.5. Although GPT-4 maintains a similar parameter count of approximately 175 billion, its performance and diversity have seen considerable improvements. Through more refined training techniques and algorithm optimizations, GPT-4 enhanced the capability of language understanding and generation, particularly outperformed in handling complex texts and special contexts. Compared to other contemporary models like Google's PaLM or Meta's OPT, GPT-4 continues to stand out in multi-task learning and logical consistency in the text generation. While Google's PaLM model boasts up to 54 billion parameters, GPT-4 shows superior generalization abilities across a broader range of NLP tasks~\cite{chowdhery2023palm}. On the open-source large models, Meta’s OPT model with a parameter size similar to the GPT-4 offers direct competition. Despite OPT’s advantages in openness and accessibility, GPT-4 still maintains a lead in specific application areas such as creative writing and complex problem solving~\cite{zhang2022optopenpretrainedtransformer}.

\subsection{Model Architecture}
There are three common LLM architectures, the \textit{Encoder-Decoder} architecture, exemplified by the traditional transformer model. This architecture comprises six encoders and six decoders, data input into the system will first passes through the encoder, where it undergoes sequential feature extraction via the model's self-attention mechanism. Subsequently, the decoders utilize the word vectors produced by the encoders to generate outputs, this technique is common to see in machine translation tasks, where the encoder processes word vectors from one language through several attention layers and feed-forward networks, thereby creating representations of the context. The decoder then uses this information to incrementally construct the correct translated text. A recent example of this architecture is the CodeT5+ model, launched by Salesforce AI Research in 2023~\cite{wang2023codet5+}. This model is an enhancement of the original T5 architecture, which designed to improve performance in code understanding and generation tasks. It incorporates a flexible architecture and diversified pre-training objectives to optimize its effectiveness in these specialized areas. This development highlights the competency of Encoder-Decoder architectures in tackling increasingly complex NLP challenges.

The \textit{Encoder-only} architecture, as the name suggests it eliminates the decoder from the entire structure making the data more compact. Unlike RNNs, this architecture is stateless and uses a masking mechanism that allows input processing without relying on hidden states, and also accelerating parallel processing speeds and providing excellent contextual awareness. BERT (Bidirectional Encoder Representations from Transformers) is a representative model of this architecture, this model is a large language model built solely on the encoder architecture. BERT leverages the encoder's powerful feature extraction capabilities and pre-training techniques to learn bidirectional representations of text, achieving outstanding results in sentiment analysis and contextual analysis ~\cite{devlin2018bert}.

The \textit{Decoder-only} archiecture, in the transformer framework primarily involves the decoder receiving processed word vectors and generating output. Utilizing the decoder to directly generate text accelerates tasks such as text generation and sequence prediction. This characteristic with high scalability is known as auto-regressiveness, which is why popular models like GPT use this architecture. In 2020, the exceptional performance of GPT-3 and its remarkable few-shot learning capabilities demonstrated the vast potential of the decoder-only architecture~\cite{brown2020language}. Given the enormous computational cost and time required to train a model from scratch, and the exponential increase in the number of parameters, many researchers now prefer to leverage pre-trained models for further research. The most popular open-source pre-trained language model LLaMA, developed by Meta AI also employs the decoder-only architecture~\cite{touvron2023llama}, as mentioned earlier, the autoregressiveness and simplicity of this structure make the model easier to train and fine-tune.
\subsection{Large Language Model Based Agent}
The concept of agents even trace back to the 19th century and is often referred to as intelligent agents, envisioned to possess intelligence comparable to humans. Over the past few decades, as AI technology has evolved, the capabilities of AI agents have significantly advanced, particularly with the reinforcement learning. This development has enabled AI agents to autonomously handle tasks and learn and improve based on specified reward/punishment rules. Notable milestones include AlphaGo~\cite{chen2016evolution}, which leveraged reinforcement learning to defeat the world champion in Go competition. 

The success of GPT has further propelled the field, with researchers exploring the use of large language models as the "brain" of AI agents, thanks to GPT's powerful text understanding and reasoning capabilities. In 2023, a research team from Fudan University~\cite{xi2023risepotentiallargelanguage} conducted a comprehensive survey on LLM-based agents, examining their perception, behavior, and cognition. Traditional LLMs typically generate responses based solely on given natural language descriptions, lacking the ability for independent thinking and judgment. LLM-based agents able to employ multiple rounds of interaction and customized prompts to gather more information, which enable the model to think and make decisions autonomously. In 2023, Andrew Zhao proposed the ExpeL framework~\cite{zhao2024expel}, which utilizes ReAct as the planning framework combined with an experience pool~\cite{yao2022react}. This allows the LLM to extract insights from past records to aid in subsequent related queries, by letting the LLM analyze why previous answers were incorrect, it learns from experience to identify the problems.

At the same time, the application of LLM-based embodied agents has also become a hot research area in recent years. LLM-based Embodied Agents are intelligent systems that integrate LLMs with embodied agents~\cite{huang2022language}. These systems can not only process natural language but also complete tasks through perception and actions in physical or virtual environments. By combining language understanding with actual actions, these agents can perform tasks in more complex environments. This integration often involves using visual domain technologies to process and understand visual data and reinforcement learning algorithms to train agents to take optimal actions in the environment. These algorithms guide the agent through reward mechanisms to learn how to make optimal decisions in different situations, while the LLM acts as the brain to understand user instructions and generate appropriate feedback. In 2023, Guanzhi Wang introduced VOYAGER, an open-ended embodied agent with large language models~\cite{wang2023voyager}. It uses GPT-4 combined with input prompts, an iterative prompting mechanism, and a skill library enabling the LLM-based agents to autonomously learn and play the game Minecraft, becoming the first lifelong learning agent in the game.

\begin{figure}[htbp]
    \centering
    \includegraphics[width=1\linewidth]{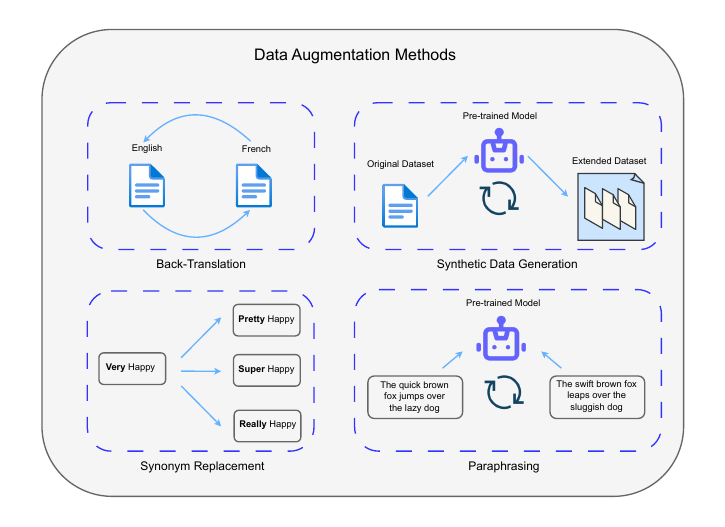}
    \caption{Illustration of Common Data Augmentation Methods.}
    \label{fig:data_aug}
\end{figure}
Nowadays, various agent systems are emerging and they rely on large language models to make judgments, combined with techniques such as few-shot learning and multi-turn dialogue for model fine-tuning. However, due to the lack of datasets many researchers employ different methods for data augmentation. Common approaches include synonym replacement, where words in the text are replaced with synonyms from the same domain to increase textual diversity; back-translation, where the text is translated into another language and then back to the original language to generate new texts with slightly different grammatical structures and word choices; Paraphrasing refers to the new dialogue which similar in context but slightly different in expression created through manual or automated means; Synthetic Data Generation refers to use pretrained model to make a synthetic data generation as shown in Figure.\ref{fig:data_aug}. In 2023, Chenxi Whitehouse explored using LLMs for data augmentation to enhance the performance of multilingual commonsense reasoning datasets, especially under conditions of extremely limited training data~\cite{whitehouse2023llmpowereddataaugmentationenhanced}. The study employed various LLMs (such as Dolly-v2, StableVicuna, ChatGPT, and GPT-4) to generate new data. These models were prompted to create new examples similar to the original data, thereby increasing the diversity and quantity of training data. Prompt engineering was mentioned as an essential skill for effectively interacting with LLMs. By applying these prompt patterns, users can efficiently customize their dialogues with LLMs ensuring the generation of high-quality outputs and achieving complex automated tasks. In 2023, Jules White introduced a set of methods and patterns to enhance prompt engineering, optimizing interactions with LLMs such as ChatGPT~\cite{white2023prompt}. This study categorizes prompt engineering into five main areas: Input Semantics, Output Customization, Error Identification, Prompt Improvement, and Interaction, to address a wide range of problems and adapt to different fields. 

One notable technique is Retrieval-Augmented Generation (RAG), the input question undergoes similarity matching with documents in an index library, attempting to find relevant results. If similar documents are retrieved, they are organized in conjunction with the input question to generate a new prompt, which is then fed into the large language model. Currently, large language models possess long-text memory capabilities, numerous studies have tested Gemini v1.5 against the Needle In A Haystack (NIAH) evaluation to explore whether RAG has become obsolete~\cite{reid2024gemini}. However, from various perspectives such as cost and expense, RAG still holds significant advantages (RAG could be 99 percent cheaper than utilizing all tokens). Additionally, long texts can negatively impact response performance, causing LLM to respond more slowly when the input text is too long. Therefore, the advancements in context length for LLM will not fully replace the role of RAG but treated as complements between each other. 
\subsection{Single and Multi-Agents}
Single agent systems leverage the capabilities of a LLM to perform various tasks, these agents typically use a single LLM to understand and respond to user queries, generate content, or execute automated tasks based on predefined instructions. Single agents are commonly used in scenarios where tasks accept a general answer and do not require complex decision-making. Examples include customer service chatbots, virtual assistants for scheduling, and automated content generation tools. However, single agents may struggle with dealing long context inputs, leading to inconsistent or irrelevant responses. The scalability of these systems is also limited when dealing with tasks that require extensive knowledge or context, this issue is often exacerbated by long texts as large language models cannot fully comprehend and analyze overly lengthy information in one turn. One of the primary issues with large language models is hallucination~\cite{10.1145/3649506}. Hallucination refers to the generation of fabricated information or definitions by LLMs, presented in seemingly logical and reasonable language to the user. Most research papers on LLMs have stated this problem, while prompt engineering or tool interventions can mitigate the affect caused by hallucination, but it cannot be entirely eliminated. In 2023, Ziwei Ji conducted an in-depth study on hallucination in natural language generation~\cite{ji2023survey}. This survey reviewed the progress and challenges in addressing hallucinations in NLG, providing a comprehensive analysis of hallucination phenomena across different tasks, including their definitions and classifications, causes, evaluation metrics, and mitigation methods.

Multi-agent systems involve the collaboration of multiple LLMs or agents to tackle complex tasks effectively. These systems fully utilize the advantages of multiple models, with each model specializing in specific aspects of the task to reduce overhead caused by multi-processes in single agents, the collaboration among agents allows for more sophisticated and robust problem-solving capabilities. Due to their exceptional capabilities, more researchers are beginning to explore the field of Multi-LLM based agents and start applying into software engineering domains. In 2024, a lot of researchers adopt the multi-agent system into the practical experiments~\cite{an2024nissist,li2024agentsneed}.

Multi-agent systems address the limitations of single-agent systems in the following ways:
\begin{itemize}
    \item \textbf{Enhanced Context Management}: Multiple agents can maintain and share context, generating more coherent and relevant responses over long interactions.
    \item \textbf{Specialization and Division of Labor}: Different agents can focus on specific tasks or aspects of a problem, improving efficiency and effectiveness.
    \item \textbf{Robustness and Error Correction}: Collaborative agents can cross-check and validate each other's outputs, reducing the likelihood of errors and improving overall reliability.
    \item \textbf{Contextual Consistency}: Multi-agent systems can better manage context over long dialogues. The collaboration of multiple agents improves the efficiency of incident mitigation.
    \item \textbf{Scalability and Flexibility}: These systems can integrate specialized agents to scale and handle more complex tasks. Through the division of labor among multiple agents, the quality of code generation is improved.
    \item \textbf{Dynamic Problem Solving}: By integrating agents with different expertise, multi-agent systems can adapt to a wider range of problems and provide more accurate solutions.
\end{itemize}

\subsection{LLM in Software Engineering}
Recently, there has been a shift towards applying general AI models to specific vertical domains such as medical and finance. In software engineering, new AI agents are emerging that are more flexible and intelligent compared to previous applications of LLMs, although they utilize different data and experiments. This continuous innovation underscores the transformative potential of AI agents across various fields, these models excel in text understanding and generation, promoting innovative applications in software development and maintenance.

LLMs profoundly impact software engineering by facilitating tasks such as code generation, defect prediction, and automated documentation. Integrating these models into development workflows not only simplifies the coding process but also reduces human errors. LLM-based agents enhance basic LLM capabilities by integrating decision-making and interactive problem-solving functions. These agents can understand and generate result by interacting with other software tools which optimize workflows, and make autonomous decisions to improve software development practices. In 2023, Yann Dubois introduced the AlpacaFarm framework~\cite{dubois2024alpacafarm}, where LLMs are used to simulate the behavior of software agents in complex environments. Moreover, significant research has been conducted in the field of automated program repair (APR). In 2024, Islem Bouzenia introduced RepairAgent~\cite{bouzenia2024repairagent}, another LLM-based tool designed for automatic software repair, this tool reduced the time developers spent on fixing issues. Additionally, in 2023, Emanuele Musumeci demonstrated a multi-agent LLM system~\cite{musumeci2024llm}, which involved a multi-agent architecture where each agent had a specific role in the generation of documents. This system significantly improved handling complex document structures without extensive human supervision. Besides these, LLMs have made outstanding contributions in software testing, software design, and emerging fields such as software security and maintenance. 

Currently, there is no comprehensive and widely accepted definition specifying the exact capabilities an LLM must exhibit to be considered an LLM-based agent, particularly within the context of software engineering. Given the extensive application of LLMs in software engineering and the varying levels of autonomy and intelligence exhibited by current frameworks, it is critical to establish a clear distinction between general LLM frameworks and those that can be defined as agents. This study synthesizes criteria from mainstream definitions and literature from the first half of 2024, reflecting the consensus found across multiple prominent research contributions.

Specifically, our criteria are derived from recent works emphasizing autonomous reasoning, decision-making, and adaptive planning capabilities in LLM-based systems, such as those outlined in frameworks including SWE-agent \cite{yang2024sweagentagentcomputerinterfacesenable}, CODEAGENT \cite{zhang2024codeagentenhancingcodegeneration}, AutoCodeRover \cite{AutoCodeRover}, and GOEX \cite{patil2024goex}. Further foundational concepts from broader agent literature, such as MetaGPT \cite{hong2023metagptmetaprogrammingmultiagent}, AgentVerse \cite{chen2023agentverse}, Reflexion \cite{shinn2024reflexion}, and ExpeL \cite{zhao2024expel}, support these definitions by consistently demonstrating advanced agent capabilities.

Based on this extensive review, we define an LLM architecture as an agent if it meets most of the criteria listed in Table~\ref{tab:criterias}. Criteria 1 to 4 (LLM as central reasoning core, decision-making and planning, autonomous tool usage, and evaluation capabilities) are considered fundamental for a basic level of agency. Criteria 5 and 6 (contextual multi-turn interaction and autonomous learning capability) indicate more advanced agentic behaviors, beneficial but not strictly necessary for an LLM to be initially classified as an agent. This distinction provides flexibility while preserving rigor, allowing diverse frameworks to be effectively and consistently evaluated.

\begin{table*}[h!]
\centering
\caption{Criteria for LLM-based agent}
\small
\label{tab:criterias}
\begin{tabular}{p{0.95\textwidth}}
\hline
\textbf{Criteria} \\ \hline
\begin{enumerate}[label=\arabic*)~, leftmargin=0.5cm, labelwidth=0.5cm, labelsep=0.2cm, itemsep=0.3cm, align=left]
    \item The LLM serves as the brain (the center of information processing and generation of thought).
    \item The framework not only relies on the language understanding and generation capabilities of LLMs but also possesses decision-making and planning abilities.
    \item If tools are available, the model can autonomously decide when and which tools to use and integrate the results into its predictions to enhance task completion efficiency and accuracy.
    \item The model can select the optimal solution from multiple homogeneous results (the ability to evaluate and choose among various possible solutions).
    \item The model can handle multiple interactions and maintain contextual understanding.
    \item The model has autonomous learning capabilities and adaptability.
\end{enumerate} \\ \hline
\end{tabular}
\end{table*}

%% file: pages/requirement.tex
\section{Requirement Engineering and and Documentation \label{cha:requirement}}
Requirement Engineering is a critical field within software engineering and plays an essential role in the software development process, its primary task is to ensure that the software system meets the needs of all relevant stakeholders. Typically, requirement engineering in project development involves many steps, where developers need to fully understand the users' needs and expectations to ensure that the development direction of the software system aligns with actual requirements. The collected requirements are then organized and evaluated by the development group. Requirements Specification is the process of formally documenting the analyzed requirements, the specification must be accurate and concise, and the requirement verification must be conducted to ensure that developers are building what users need and that it aligns with the specifications. Requirement engineering also includes requirement management, a task that spans the entire software development life-cycle, developers need to continuously track, control, and respond to any changes occurring during development, ensuring that these changes do not negatively impact the project's progress and overall quality.
\subsection{LLMs Tasks}
\textbf{Requirement Classification and Extraction.} In the field of requirement engineering, LLMs have demonstrated significant potential in automating and enhancing tasks such as requirement elicitation, classification, generation, specification generation, and quality assessment. Requirement classification and extraction is a crucial task in requirement engineering during the development process. It is common to encounter situations where clients present multiple requirements at once, necessitating manual classification by developers. By categorizing requirements into functional and non-functional requirements, developer can better understand and manage them, thanks to the strong performance of LLMs in classification tasks, many relevant frameworks have been developed. The PRCBERT framework, utilizing the BERT pre-trained language model, transforms classification problems into a series of binary classification tasks through flexible prompt templates, significantly improving classification performance~\cite{luo2022prcbert}. Studies have shown that the PRCBERT achieved an F1 score of 96.13\% on the PROMISE dataset which outperform the previous state-of-arts NoRBERT~\cite{9218141} and BERT-MLM models~\cite{devlin2018bert}. Additionally, the application of ChatGPT in requirement information retrieval has shown promising results, by classifying and extracting information from requirement documents, ChatGPT achieved comparable or even better $F\beta$ scores under zero-shot settings, particularly in feature extraction tasks, where its performance surpassed baseline models~\cite{zhang2023evaluation}. A comparative study further evaluated ChatGPT (GPT-3.5 and GPT-4) against traditional models such as SVM and LSTM across five datasets—PROMISE, Dronology, ReqView, Leeds Library, and WASP. Using a class-specific $F\beta$ scoring method to emphasize recall, the results demonstrated that GPT-3.5 Few-Shot prompting achieved the best overall classification performance, while GPT-4 Zero-Shot showed notable strength in specialized categories like OnlyFunctional requirements~\cite{ChatGPTvsSVMRE}. As seen in Table~\ref{tab:se_tasks}, there is also substantial literature and research on using LLMs to automatically generate requirements and descriptions in requirement engineering. 

\textbf{Requirement Elicitation and Specification Generation.  }By automating the generation and description of requirements, the efficiency and accuracy of requirement elicitation can be improved. Research indicates that LLMs hold significant potential in requirements generation task. For example, using ChatGPT to generate and gather user requirements, studies found that participants with professional knowledge could use ChatGPT more effectively, indicating the influence of domain expertise on the effectiveness of LLM-assisted requirement elicitation~\cite{arora2024advancing}. The study employed qualitative assessments of the LLMs' output against predefined criteria for requirements matches, including full matches, partial matches, and the relevancy of the elicited requirements, although their success varied depending on the complexity of the task and the experience of the users, the result showing that LLMs could effectively assist in eliciting requirements, and its particularly useful in identifying, and suggesting requirements based on the large corpus of training data they provided. The GeneUS pipeline~\cite{Genus} pushes this boundary by unifying user story and test case generation through a three-stage prompting process. Evaluated using the RUST framework (readability, understandability, specifiability, technical aspects), GPT-4-generated artifacts received high ratings, particularly in understandability and specifiability, showing LLMs' potential in supporting agile requirement processes. 

The SRS (Software Requirement Specification) generation is an important task which the developer normally spent a lot of time to refine and verified. In~\cite{krishna2024usingllmssoftwarerequirements}, researchers use both iterative prompting and a single comprehensive prompt to assess the performance of LLMs to generate SRS. The experiment conducted on GPT-4 and CodeLlama-34b one close-source LLM and one open-source LLM for comprehensive evaluation, the generated SRS will compare with human-crafted SRS and finally scored by the likert scale. The result indicate that, the human-generated SRS was overall superior, but CodeLlama often came close, sometimes outperforming in specific categories. The CodeLlama scored higher in completeness and internal consistency than GPT-4 but less concise, so this stuy demonstrated the potential of using fine-tuned LLMs to generate SRS and increase the overall project productivity. Another paper also explores using LLMs for generating specifications. In~\cite{ma2024specgenautomatedgenerationformal}, the authors introduce a framework called SpecGen for generating program specifications. The framework primarily uses GPT-3.5-turbo as the base model and employs prompt engineering combined with multi-turn dialogues to generate the specifications. SpecGen applies four mutation operators to modify these specifications and finally uses a heuristic selection strategy to choose the optimal variant. The results show that SpecGen can generate 70\% of the program specifications, outperforming traditional tools like Houdini~\cite{10.1007/3-540-45251-6_29} and Daikon\footnote{\href{https://github.com/codespecs/daikon}{https://github.com/codespecs/daikon}}. A significant advancement in formal modeling is shown in PathOCL~\cite{PathOCL}, which enhances OCL constraint generation from UML diagrams by injecting contextual navigation paths into GPT-4 prompts. Evaluated on 15 UML models and 168 English specifications, PathOCL significantly outperformed UML-Augmented prompting on Validity@K and Correctness@K metrics, while reducing hallucinations and improving alignment with formal intent.

\textbf{Prompt Engineering and Requirement Completeness. }Furthermore, designing prompt patterns can significantly enhance LLMs’ capabilities in tasks such as requirement elicitation and system design. The paper provides a catalog of 13 prompt patterns, each aimed at addressing specific challenges in software development~\cite{White2024}. The experiments test the efficacy of these patterns in real world scenarios to validate their usefulness. By applying different prompt patterns, the study found that these patterns could help generate more structured and modular results and reduce common errors. Automated requirement completeness enhancement is another important benefit brought by the LLMs in requirement generation. The study~\cite{luitel2024improving} use BERT's Masked Language Model (MLM) can detect and fill in missing parts in natural language requirements, significantly improving the completeness of requirements. BERT's MLM achieved a precision of 82\%, indicating that 82\% of the predicted missing terms were correct. 

\textbf{Ambiguity Detection in Requirement Documents.} There is also the application of LLMs in ambiguity detection tasks, aimed at detecting ambiguities in natural language requirement documents to improve clarity and reduce misunderstandings. This study primarily aims to address the issue of detecting term ambiguities within the same application domain (where the same term has different meanings in different domains). Although current models generally possess excellent contextual understanding capabilities, this was a common problem in machine learning at that time. This study provides an excellent paradigm for the subsequent application of LLMs in requirements engineering, study demonstrated the transformer-based machine learning models can effectively detect and identify ambiguities in requirement documents, thereby enhancing document clarity and consistency. The framework utilizes BERT and K-means clustering to identify terms used in different contexts within the same application domain or interdisciplinary project requirements documents~\cite{10.1145/3528588.3528651}. 

\textbf{Requirement Quality Assessment and Structured Evaluation.} In recent two years, more and more researchers use LLMs to help them to evaluate the requirement documentations, quality assessment tasks ensure that the generated requirements and code meet expected quality standards. The application of ChatGPT in user story quality evaluation has shown potential in identifying quality issues, but it requires further optimization and improvement~\cite{10.1007/978-3-031-48550-3_17}.
A similar study use LLM to automatically process the requirement satisfaction assessment, and evaluate whether design elements fully covered by the given requirements, but the the researcher indicated the necessity of further verification and optimization in practical applications~\cite{poudel2023leveragingtransformerbasedlanguagemodels}. Further extending this direction, Ronanki et al.~\cite{10371698} systematically compared ChatGPT-generated requirements with those written by RE professionals across seven quality attributes. ChatGPT excelled in understandability and correctness, but scored lower in feasibility and unambiguity, especially when domain-specific nuances were required. Complementing this, Yeow et al.~\cite{10459458} examined GPT-3.5’s ability to generate elicitation questions across multiple domains—including banking, healthcare, education, and retail. Through a combination of readability metrics and expert scoring, they found the questions to be generally clear, relevant, and complete. The model maintained consistent quality across domains, demonstrating potential to assist early-stage RE, though issues like specificity and depth still require refinement. Building on this progression, Wei~\cite{10628487} proposed a "progressive prompting" strategy that aligns with the waterfall model. Their system guided GPT-3.5 through structured stages—from requirements to use cases, designs, and code—using iterative prompts and knowledge bases. Applied to a case study, the approach improved traceability and explainability, positioning LLMs as semi-agentic partners in structured RE and software design workflows.

\subsection{LLM-based Agents Tasks}
\textbf{Multi-Agent Collaboration and Autonomous Iteration.} Currently the application of LLM-based agents in the requirement engineering is till quite nascent, but there are some useful researches to help us to see the potential possibility. LLM-based agents bring both efficiency and accuracy for tasks like requirement elicitation, classification, generation, and verification. Compared to traditional LLMs, these systems exhibit higher levels of automation and precision through task division and collaboration. The application of multi-agent systems in semi-structured document generation has shown significant effectiveness. In~\cite{10.1007/978-3-031-60615-1_7}, a multi-agent framework is introduced that combines semantic recognition, information retrieval, and content generation tasks to streamline the creation and management of semi-structured documents in the public administration domain. The proposed framework involves three main types of agents: Semantics Identification Agent, Information Retrieval Agent, and Content Generation Agent. By avoiding the overhead of a single model, each agent is assigned a specific task with minimal user intervention, following the designed framework and workflow.

Additionally, the AI-assisted software development framework (AISD) also showcases the autonomy brought by the LLM-based agents in requirement engineering.~\cite{zhang2024experimentingnewprogrammingpractice} proposes the AISD framework, which continuously improves and optimizes generated use cases and code through ongoing user feedback and interaction. In the process of the experiment, humans need to first give a fuzzy requirement definition, and then LLM-based agent will improve the requirement case according to this information, and then design the model and generate the system according to the case, and then the generated results will let humans judge whether the requirements are met or not. The study results indicate that AISD significantly increased use case pass rates to 75.2\%, compared to only 24.1\% without human involvement. AISD demonstrates the agents' autonomous learning ability by allowing LLMs to generate all code files in a single session, continually refining and modifying based on user feedback. This also ensures code dependency and consistency, further proving the importance of human involvement in the requirement analysis and system testing stages.

\textbf{Safety Requirements and Agile User Stories.} Furthermore, in generating safety requirements for autonomous driving, LLM-based agents have shown unique advantages by introducing multimodal capabilities. The system employs LLMs as automated agents to generate and refine safety requirements with minimal human intervention until the verification stage, which is unattainable with only LLMs.~\cite{nouri2024engineeringsafetyrequirementsautonomous} describes an LLM prototype integrated into the existing Hazard Analysis and Risk Assessment (HARA) process, significantly enhancing efficiency by automatically generating specific safety-related requirements. The study through three design iterations progressively improved the LLM prototype's efficiency by completing within a day compared to months manually. In agile software development, the quality of user stories directly impacts the development cycle and the realization of customer expectations.~\cite{10.1007/978-3-031-61154-4_8} demonstrates the successful application of the ALAS system in six agile teams at the Austrian Post Group IT.customized bjectives of user stories through automated analysis and enhancement. The entire agent framework allows the model to perform specific roles in the Agile development process, the study results indicated that the ALAS-improved user stories received high satisfaction ratings from team members. 

\textbf{Requirement Compliance and Workflow Automation.} Regulatory compliance challenges are addressed through advanced agent architectures. Masoudifard et al. \cite{masoudifard2024leveraging} integrate Graph-RAG \cite{edge2024local} with reasoning techniques (Chain-of-Thought, Tree-of-Thought) to validate requirements against regulations. Tested in financial and aerospace systems (e.g., NASA’s X-38), their framework achieves 87.93\% F1-score in accuracy, demonstrating scalability for mission-critical contexts. A comprehensive approach is exemplified by Sami et al. \cite{sami2024aibasedmultiagentapproach}, who simulate end-to-end requirement workflows using role-specific agents (Product Owner, QA, Developer, Manager). This multi-agent system automates user story generation, quality assessment, feasibility review, and prioritization, achieving coherence and efficiency comparable to human teams. Evaluations highlight GPT-3.5’s balance of performance and Mixtral-8B’s speed, while practitioners emphasize its potential to enhance human-AI collaboration. These advancements underscore LLM-based agents’ versatility across requirement engineering—from granular tasks like compliance checking to holistic pipeline automation—while emphasizing the enduring role of human expertise in validation and decision-making.

\subsection{Analysis}
The application of LLM-based agents in requirement engineering has demonstrated significant efficiency improvements and quality assurance \cite{ma2024specgenautomatedgenerationformal,nouri2024engineeringsafetyrequirementsautonomous}. Through multi-agent collaboration and automated processing, these systems not only reduce manual intervention but also enhance the accuracy and consistency of requirement generation and verification \cite{10.1007/978-3-031-60615-1_7}. We can see that the tasks of LLM-based agents are no longer limited to simply generating requirements or filling in the gaps in descriptions. Instead, they involve the implementation of an automated process, with the generation of requirement documents being just one part of it, integrating LLM into agents enhances the overall system's natural language processing and reasoning capabilities \cite{tao2024magis}. In the real-world application, many tasks can no longer be accomplished by simple LLMs alone, especially for high-level software design. The emergence of LLM-based agents addresses this issue through a multi-agent collaborative system centered around LLMs, these agents continuously analyze and refine the deficiencies in the requirement documents, this is might be the main application trend of LLM-based agents in requirements engineering in the future.

\begin{figure}[htbp]
    \centering
    \includegraphics[width=1\linewidth]{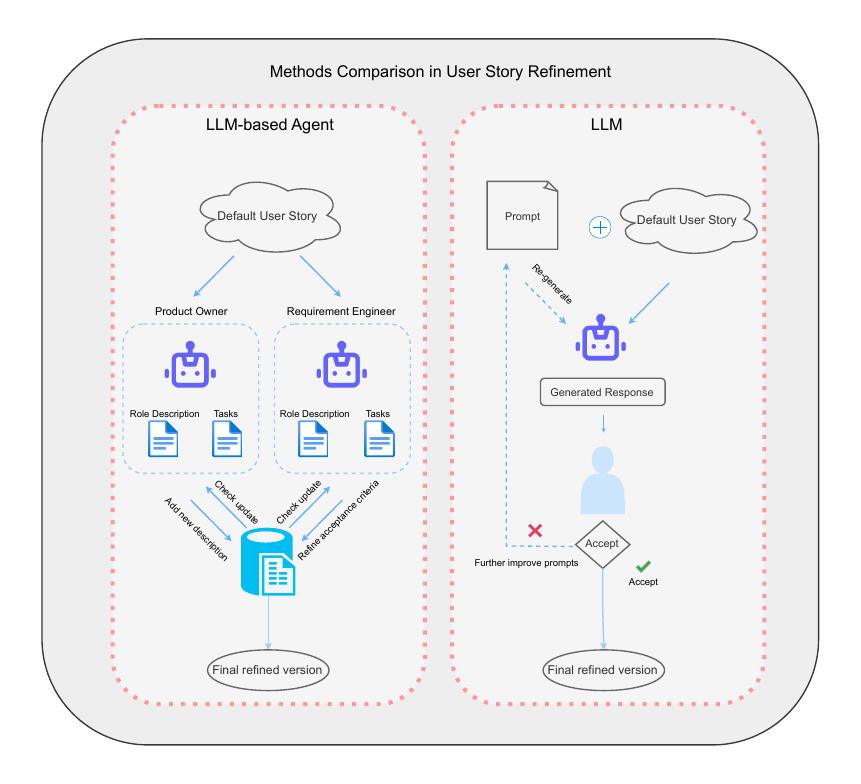}
    \caption{Illustration of Comparison Framework Between LLM-based Agent and LLM in User Story Refinement.}
    \label{fig:userStoryComp}
\end{figure}

The application of LLM-based agents in requirements engineering is still relatively limited, with most efforts focusing on leveraging the collaborative advantage of multi-agent systems to generate and refine requirements engineering documents. As illustrated in Figure.\ref{fig:userStoryComp}, which roughly simulates the architectures presented in \cite{10.1007/978-3-031-48550-3_17} and \cite{10.1007/978-3-031-61154-4_8}, both applied to the generation and refinement of user stories, we can clearly compare the differences between the two architectures. On the left is the architecture of the LLM-based agent, while on the right is the approach of using prompt engineering and LLMs alone to refine user stories. The figure omits more detailed and complex aspects of the architecture to highlight the core differences between the two approaches. LLM-based agents can continuously improve from different professional perspectives by utilizing a shared database. Although there are not many papers on LLM-based agents, we can observe the trend and benefits of transitioning from LLMs to LLM-based agents.

\subsection{Benchmarks}
Requirement engineering, unlike tasks such as bug fixing and code generation, does not have an abundance of public datasets available, such as HumanEval which commonly used for code generation assessment. Most training datasets for models in requirement engineering are self-collected by the authors and not all of them are open-sourced on Huggingface, resulting in a limited amount of dataset in requirement engineering. For instance, some papers do not mention a specific benchmark dataset but instead focus on practical examples and case studies to demonstrate the effectiveness of proposed prompt patterns~\cite{White2024}. Some researchers involve actual developers and requirements engineers to evaluate the accuracy, usability, and completeness of generated requirements and code artifacts through practical usage and feedback.

In~\cite{zhang2023evaluation}, four datasets are primarily used, characterized by average length, type-token ratio (TTR), and lexical density (LD). The NFR Multi-class Classification dataset includes 249 non-functional requirements (NFRs) across 15 projects from the PROMISE NFR dataset. The App Review NFR Multi-label Classification dataset comprises 1800 app reviews from Google Play and Apple App Store, labeled with various NFRs. The Term Extraction dataset contains 100 smart home user stories with 250 manually extracted domain terms. Lastly, the Feature Extraction dataset consists of 50 app descriptions across 10 application categories with manually identified feature phrases. In~\cite{luitel2024improving}, the PURE dataset consisting of 40 requirements specifications totaling over 23,000 sentences, is used to test BERT's ability to complete requirements. In~\cite{10371698}, the benchmark dataset comprised 36 responses to six questions: 6 responses generated by ChatGPT and 30 responses from five human RE experts (each expert provided 6 responses). These datasets serve as evaluation metrics for the models. Combining these papers, we can see that benchmark datasets for LLMs in requirement engineering mainly include various classifications of software requirements and functional and non-functional requirements to aid and assist models in learning this domain, the dataset utilization are quite flexible and diversifies.

In LLM-based agents' research in requirement engineering, the selection and construction of datasets are also important. In~\cite{musumeci2024llm}, the dataset mainly consists of semantic templates from the public administration domain. These templates cover various semi-structured forms of administrative documents, such as official certificates and public service forms. Although the detailed composition of the dataset is not specified, it can be inferred that these templates include a large number of practical cases and contextual information to ensure that the documents generated by the multi-agent system meet actual needs.

Additionally, in~\cite{zhang2024experimentingnewprogrammingpractice}, the CAASD (Capability Assessment of Automatic Software Development) dataset is introduced. This specially constructed benchmark dataset is used to evaluate the capabilities of AI-assisted software development systems. The CAASD dataset contains 72 tasks from various domains, such as small games and personal websites, each with reference use cases to define system requirements. The purpose of constructing this dataset is to provide a comprehensive evaluation benchmark that covers different types of development tasks, testing the performance of LLM-based agents in diverse tasks. In~\cite{nouri2024engineeringsafetyrequirementsautonomous}, the study mainly uses Design Science Methodology to design and evaluate the LLM prototype but does not mention a specific dataset, focusing on validating the model's effectiveness through practical application and case studies. Despite the lack of detailed dataset descriptions, this approach emphasizes iterative improvement and practical application to ensure that the safety requirements generated by LLM-based agents meet high safety standards. Finally, in~\cite{10.1007/978-3-031-61154-4_8}, 25 synthetic user stories are used, derived from a mobile delivery application project. The study evaluates the ALAS system's effectiveness by testing it in six agile teams at the Austrian Post Group IT. Although these user stories are synthetic data designed for the experiment, they realistically reflect the requirements in actual projects, providing a valuable testing benchmark.

From these papers, it can be seen that the selection and construction of datasets in LLM-based agents' research in requirement engineering often rely on practical projects and case studies, lacking standardization and large-scale datasets. Compared to LLM literature, the datasets used are broader and in a higher level like an actual system's files, not limited to the classification of non-functional requirements and pure software requirement specifications. Researchers focus more on validating the model's effectiveness through practical application and iterative improvement to enhance model performance. While this approach is flexible and targeted, it also highlights the field's shortcomings in dataset standardization and scaling. In the future, with more public datasets being constructed and shared, the application of LLM-based agents in requirement engineering is expected to achieve broader and deeper development.

A closer examination reveals a clear shift in benchmark design accompanying the transition from LLM-based RE methods to agent-based systems. Benchmarks in LLM studies tend to focus on isolated tasks, such as requirement classification, term extraction, or user story generation. These datasets are often small-scale, domain-specific, and emphasize text-level correctness or completeness—frequently evaluated through human ratings or token-level metrics. The scope is typically narrow, targeting single-turn input-output behavior of the model. In contrast, LLM-based agent research introduces a new benchmark paradigm: datasets are increasingly designed to simulate realistic RE workflows, involving multi-turn reasoning, multi-agent collaboration, and cross-document traceability. For instance, recent benchmarks such as SRS Broker and SRS Aero~\cite{masoudifard2024leveraging} are built around regulatory compliance, containing rich interdependencies among requirements and standards. Evaluation expands beyond accuracy to include reasoning traceability, compliance violation detection, and structured knowledge retrieval. Moreover, agent-based benchmarks often incorporate role-driven interaction settings, as seen in Sami et al.~\cite{sami2024aibasedmultiagentapproach}, where Product Owners, Developers, QA Engineers, and Project Managers coordinate through LLM agents. Evaluation thus moves toward semantic alignment, system response time, prioritization quality, and multi-perspective expert judgment, often grounded in international standards such as ISO/IEC 29148 or the INVEST framework. This evolution marks a significant step from task-oriented benchmarking to scenario- and collaboration-driven evaluation. This progression indicates that as LLMs evolve from passive tools to autonomous agents, benchmark datasets must likewise evolve—from measuring isolated task performance to evaluating contextual understanding, dynamic decision-making, and agentic collaboration in real-world RE environments.
\begin{table*}
\centering
\scriptsize
\caption{Evaluation Metrics in Requirement Engineering and Documentation}
\label{tab:re_metrics}
\begin{tblr}{
  row{1} = {c},
  cell{2}{4} = {c},
  cell{3}{4} = {c},
  cell{4}{4} = {c},
  cell{5}{4} = {c},
  cell{6}{4} = {c},
  cell{7}{4} = {c},
  cell{8}{4} = {c},
  cell{9}{4} = {c},
  cell{10}{4} = {c},
  cell{11}{4} = {c},
  cell{12}{4} = {c},
  cell{13}{4} = {c},
  cell{14}{4} = {c},
  cell{15}{4} = {c},
  cell{16}{4} = {c},
  cell{17}{4} = {c},
  cell{18}{4} = {c},
  cell{19}{4} = {c},
  hlines,
  vlines,
}
Reference Paper & Benchmarks                                                      & Evaluation Metrics                                                                                                                                                                        & Agent \\
{~\cite{arora2024advancing}}            & Evaluation on~ActApp                                            & {Precision and Recall for Elicitation\\Clarity, Consistency, and Compliance\\Completeness and accuracy of acceptance\\criteria}& No    \\
{~\cite{zhang2023evaluation}}            & NFR,~Smarthome user stories                                     & Precision, recall, and F\textbackslash{}beta (F1 or F2)& No    \\
{~\cite{luitel2024improving}}            & PURE                                                            & Precision, F1 Score, Recall& No    \\
{~\cite{krishna2024usingllmssoftwarerequirements}}            & Not Specified                                                       & Likert Scale & No    \\
{~\cite{White2024}}            & Case studies                                                    & {Accuracy in identifying missing requirements.\\Quality and Modularity of generated code.\\Correctness of refactoring suggestions.\\Efficiency in automating software engineering\\tasks}& No    \\
{~\cite{10371698}}            & 36 responses to the six questions                               & {Abstraction, Atomicity, Consistency, Correctness\\Unambiguity, Understandability, Feasibility}& No    \\
{~\cite{luo2022prcbert}}            & PROMISE NFR-Review, NFR-SO                                      & F1 Score, Weighted F1 Score (w-F))& No    \\
{~\cite{ma2024specgenautomatedgenerationformal}}            & SV-COMP,~SpecGenBench                                           & {Number of Passes, Success Probability\\Number of Verifier Calls, User Rating}& No    \\
{~\cite{xie2023impactlargelanguagemodels}}            & {Jdoctor-data,~DocTer-data,~\\SpecGenBench,~SV-COMP}            & Accuracy, Precision, Recall, F1 Score& No    \\
{~\cite{10.1007/978-3-031-48550-3_17}}            & {Benchmark evaluations of user \\stories using the AQUSA tool}  & {Agreement Rate, Precision, Recall, Specificity,\\F1 Score}& No    \\
{~\cite{10.1145/3528588.3528651}}            & Crawled Documents from Wikipedia                                & Manual validation& No    \\
{~\cite{poudel2023leveragingtransformerbasedlanguagemodels}}             & {CM1,~CCHIT, Dronology,~PTC-A,~\\PTC-B}                         & $F\beta$, Mean Average Precision (MAP)& No    \\
{~\cite{9218141}}            & PROMISE NFR                                                     & {Precision (P), Recall (R), F1-score (F1),\\Weighted average F1-score (A)}& No    \\
{~\cite{MOHARIL2023102994}}            & CS-specific corpora, PURE                                       & Contextual Clarity, User Feedback& No    \\
{~\cite{PathOCL}} & 15 UML models, 168 English specs & Validity@K, Correctness@K, McNemar’s Test & No \\
{~\cite{Genus}} & 7 RE documents & RUST framework, Likert scale & No \\
{~\cite{ChatGPTvsSVMRE}} & PROMISE, Dronology, ReqView, Leeds Library, WASP & $F\beta$, Accuracy, Recall & No \\
{~\cite{10371698}} & 36 requirements (6 AI, 30 human) & 7 RE Quality Attributes & No \\
{~\cite{10459458}} & 4 domains survey and QA  & {Flesch Reading Ease, FK Grade Level, Clarity, \\Relevance, Completeness}& No \\
{~\cite{10628487}} & SuperFrog Scheduler case study & {Accuracy, Adherence to Requirements, \\Practical Usability}& No \\
{~\cite{10.1007/978-3-031-60615-1_7}}            & {Semantic Templates from Public\\Administration}                & {Accuracy, Prompt Conformity, User Intervention\\Frequency, Hallucination Rate}& Yes   \\
{~\cite{zhang2024experimentingnewprogrammingpractice}}            & CAASD                                                           & Pass Rate, Token Consumption& Yes   \\
{~\cite{nouri2024engineeringsafetyrequirementsautonomous}}            & AEB, CAEM                                                       & {Performance Accuracy and Relevance, Efficiency,\\Feedback from industry}& Yes   \\
{~\cite{10.1007/978-3-031-61154-4_8}}            & {25 Synthetic User Stories for a \\Mobile Delivery Application} & {Independence, Negotiability, Value, Estimability,\\Smallness, Testability. \\Survey among professionals~}& Yes   
\end{tblr}
\end{table*}
\subsection{Evaluation Metrics}
The evaluation of LLMs and LLM-based agents in requirement engineering spans from classical NLP performance metrics to RE-specific quality attributes and agent-oriented behavioral indicators. As summarized in Table~\ref{tab:re_metrics}, evaluation methods are closely tied to the task nature—whether static classification, document generation, or interactive system behavior.

For traditional LLM applications, especially in classification and extraction tasks (e.g., functional vs. non-functional requirements), common metrics include \textbf{precision}, \textbf{recall}, and \textbf{F1 score}, with several studies applying class-weighted \textbf{F$_\beta$} scores to emphasize recall in imbalanced datasets~\cite{ChatGPTvsSVMRE, zhang2023evaluation}. In generative scenarios, such as user story or specification generation, human-centric evaluation becomes prominent. For instance, the RUST framework~\cite{Genus} scores GPT-4’s outputs on \textit{Readability}, \textit{Understandability}, \textit{Specifiability}, and \textit{Technicality} using Likert scales. Similarly, SRS evaluations in~\cite{krishna2024usingllmssoftwarerequirements} employ 1–5 Likert ratings for \textit{Conciseness}, \textit{Consistency}, and \textit{Completeness}, while~\cite{10371698} compares ChatGPT-generated requirements with expert-authored ones using seven RE-specific quality attributes including \textit{Correctness}, \textit{Unambiguity}, and \textit{Feasibility}. Domain-specific ambiguity detection models evaluate semantic divergence using contextual clustering~\cite{10.1145/3528588.3528651}, and completeness assessments for masked terms use precision-oriented metrics~\cite{luitel2024improving}.

As models shift from passive language tools to autonomous agents, the scope of evaluation expands significantly. In collaborative and role-specific agent settings~\cite{sami2024aibasedmultiagentapproach}, evaluation considers \textbf{semantic similarity} (via sentence embeddings), \textbf{response time}, and alignment with standards like \textbf{ISO/IEC 29148} and the \textbf{INVEST} criteria for user stories. The assessment includes human feedback across four real-world roles: Product Owner, Developer, QA, and Project Manager, offering multi-perspective validation. In compliance-oriented tasks such as those tackled by Graph-RAG agents~\cite{masoudifard2024leveraging}, evaluation focuses on \textbf{compliance classification accuracy}, \textbf{F1 scores} for non-compliance detection, and \textbf{reasoning interpretability} under various prompting strategies (e.g., Chain-of-Thought, Tree-of-Thought). These methods evaluate not only output correctness but also the explainability and robustness of agent decision paths.

Agent-driven frameworks like AISD~\cite{zhang2024experimentingnewprogrammingpractice} adopt system-level metrics such as \textbf{use case pass rate}, \textbf{code consistency}, and \textbf{human-LLM iteration count}. These reflect the agents’ ability to refine and adapt outputs over multiple feedback loops. Similarly, in safety-critical environments~\cite{nouri2024engineeringsafetyrequirementsautonomous}, evaluation involves task duration reduction and expert validation of safety requirement adequacy across design iterations. This transition from LLM to agent evaluation represents a fundamental shift: from static, single-turn outputs measured by token-level correctness to dynamic, multi-role collaboration assessed by behavioral robustness, human satisfaction, and contextual traceability. While LLMs excel in task precision, agent-based systems are evaluated for their autonomy, adaptability, and collaborative fluency. This shift implies that future RE systems must embrace richer evaluation schemas, integrating quantitative performance with qualitative feedback and scenario-driven benchmarks, to capture the full spectrum of agentic capability in evolving software engineering workflows.

%% file: pages/code.tex
\section{Code Generation and Software Development \label{cha:code}}
Code generation and software development are core areas within software engineering which plays a crucial role in the software development process. The primary objective of using LLMs in code generation is to enhance development efficiency and code quality through automation processes, thereby meeting the needs of both developers and users.

In recent years, the application of LLMs in code generation and software development has made significant progress, this has changed the way developers work and revealed a shift in automated development processes. Compared to requirement engineering, research on the application of LLMs and LLM-based agents in code generation and software development is more extensive and in-depth. Using natural language processing and generation technologies, LLMs can understand and generate complex code snippets, assisting developers in automating various stages from code writing and debugging to software optimization. The decoder-based large language models such as GPT-4 have shown significant potential in code generation by providing accurate code suggestions and automated debugging, greatly improving development efficiency. Recently, the application of LLM-based agents in software development is also gaining attention, these intelligent agents can not only perform complex code generation tasks but also engage in autonomous learning and continuous refinement, thereby offering flexible assist in dynamic development environments. Tools like GitHub Copilot~\cite{copilot}, which integrate LLMs, have already demonstrated their advantages in enhancing programming efficiency and code quality.

\subsection{LLMs Tasks}
\textbf{Code Generation and Translation from Natural Language.} Large language models have optimized various tasks in code generation and software development through automation and reasoning, covering areas such as code generation, debugging, code comprehension, code completion, code transformation, and multi-turn interactive code generation. The primary method is generating executable code from natural language descriptions, where models utilize previously learnt code snippets or apply few-shot learning to better understand user requirements. Nowadays the AI tools integrates deeply with IDEs like Visual Studio Code\footnote{\href{https://code.visualstudio.com/}{https://code.visualstudio.com/}} and JetBrains\footnote{\href{https://www.jetbrains.com/}{https://www.jetbrains.com/}} to enhance code writing and translation tasks such as OpenAI's Codex model~\cite{chen2021evaluatinglargelanguagemodels}. Codex fine-tuned on public code from GitHub, demonstrate the capability to generate Python functions from doc-strings also outperformed other similar models on the HumanEval benchmark.

In~\cite{ni2023l2cevalevaluatinglanguagetocodegeneration}, researchers comprehensively evaluated the performance of multiple LLMs on L2C(language to code) tasks. The results showed that GPT-4 demonstrates strong capability in tasks such as semantic parsing, mathematical reasoning, and Python programming. With instruction tuning and support from large-scale training data, the model can understand and generate code that aligns with user intent, achieving high-precision code generation. Applying LLMs to text-to-database management and query optimization is also a novel research direction in natural language to code generation task. By converting natural language queries into SQL statements, LLMs help developers quickly generate efficient database query code. In~\cite{sun2023sql}, proposed the SQL-PaLM framework which significantly enhances the execution accuracy and exact match rate for text-to-SQL tasks through a few-shot prompt and instruction fine-tuning, providing an effective solution for complex cross-domain SQL generation tasks. The improvements in accuracy and exact match achieved in the SQL-PaLM model are considered state-of-the-art (SOTA) in tested benchmarks, the SQL-PaLM performed promise results comparing with existing methods such as T5-3B + PICARD, RASAT + PICARD, and even GPT-4, achieving the highest test accuracy of 77.3\% and an execution accuracy of 82.7\%. Multilingual code generation is another important application of LLMs, particularly suited to the transformer architecture. In~\cite{zheng2024codegeexpretrainedmodelcode}, researchers introduced the CodeGeeX model, which was pre-trained on multiple programming languages and performed well in multilingual code generation and translation tasks. Experimental results showed that CodeGeeX outperformed other multilingual models on the HumanEval-X benchmark.

Although current LLMs possess excellent code generation capabilities, with accuracy and compile rates reaching usable levels, the quality of generated code often depends on the user's prompts. If the prompts are too vague or general, the LLM typically struggles to understand the user's true requirements, making it difficult to generate the desired code in a single attempt. In~\cite{hu2024leveragingprintdebuggingimprove}, researchers introduced "print debugging" technique, using GPT-4 to track variable values and execution flows, which enhancing the efficiency and accuracy by using in-context learning techniques. This method is particularly suitable for medium-difficulty problems on Leetcode, compared to the rubber duck debugging method, print debugging improved performance by 1.5\% on simple Leetcode problems and by 17.9\% on medium-difficulty problems.

\textbf{Real-Time Code Completion and Editing Tools.}
Additionally, the application of LLMs in improving programming efficiency has garnered widespread attention, the tools like GitHub Copilot which integrating OpenAI’s Codex model, provide real-time code completion and suggestions during coding. According to~\cite{peng2023impactaideveloperproductivity}, researchers present a controlled experiment with the Github Copilot, the result demonstrated that with developers completing HTTP server tasks 55.8\% faster when using Copilot. Another similar study also using LLM to be the programmer tools, in~\cite{fried2023incodergenerativemodelcode}, researchers introduced the INCODER model which capable of both program synthesis and editing. By leveraging bidirectional context, the model performs well in both single-line and multi-line code filling tasks, providing developers with smarter code editing tools. This real-time code generation and completion functionality not only improves programming efficiency but also reduce the burden on developers, allowing them to focus on higher-level design which is a common problem in software development where substantial workforce and time are wasted on tedious coding tasks.

\textbf{Multi-Turn Program Synthesis with Feedback.}
The multi-turn program synthesis tasks represent a significant breakthrough for LLMs in handling complex programming tasks, in ~\cite{nijkamp2023codegenopenlargelanguage},  researchers introduced the CODEGEN model, which iteratively generates programs through multiple interactions, significantly improving program synthesis quality and making the development process more efficient and accurate. By gradually generating and continuously optimizing code at each interaction, LLMs can better understand user intent and generate more precise and optimized code. In the experiments, comparisons were made with the Codex model, which was considered state-of-the-art in code generation at the time. CODEGEN-MONO 2.7B outperformed the Codex model of equivalent outcome in pass@k metrics for both k=1 and k=10. Furthermore, CODEGEN-MONO 16.1B exhibited performance that was comparable to or better than the best Codex model on certain metrics, further demonstrating its SOTA performance in the code generation.
By iteratively generating and optimizing code, LLMs continuously improve their output quality. In~\cite{10.1145/3649825}, researchers proposed the Cycle framework, which enhances the self-improvement capability of code language models by learning from execution feedback, improving code generation performance by 63.5\% on multiple benchmark datasets. Although Cycle has a certain degree of autonomy, its decision-making and planning capabilities are mainly limited to code generation and improvement tasks without overall planning, and the execution sequence is completely followed a fixed pattern, so it's better to classified as an advanced LLM application. Another representative multi-turn code generation framework is TICODER~\cite{fakhoury2024llm}, which integrates test-driven interaction to disambiguate user intent and improve LLM code generation quality. Rather than directly surfacing candidate code completions, TICODER generates tests based on the user’s natural language intent and asks users to validate them. Based on user responses (PASS, FAIL, or OUTPUT), it incrementally prunes invalid code suggestions and surfaces a refined, ranked list of consistent completions. A user study with 15 programmers showed that TICODER significantly improved correctness (84\% vs. 40\% in baseline), reduced cognitive load (mean TLX score 28 vs. 45), and did not increase task time. Furthermore, in large-scale benchmark evaluations on MBPP and HumanEval, TICODER showed average absolute improvements of 45.73\% in pass@1 accuracy within five user interactions across diverse LLMs (including GPT-3.5, GPT-4, CodeGen, and Codex). These results confirm LLM framework's strength as a semi-structured, test-guided workflow that supports LLMs in interactive, high-precision code generation. Another framework that embraces multi-phase refinement is AlphaCodium~\cite{ridnik2024codegenerationalphacodiumprompt}, which proposes a test-driven, flow-based approach specifically optimized for solving complex competitive programming tasks. Unlike one-shot prompting or static few-shot examples, AlphaCodium introduces a structured workflow that includes (1) pre-processing through problem reflection and reasoning over public tests, and (2) iterative refinement via both public and AI-generated test cases. The process incorporates design patterns such as semantic bullet-point analysis, modular code generation, and structured YAML outputs. A key insight from the study is that generating diverse test cases is often easier and more reliable for LLMs than producing complete correct code solutions. AlphaCodium thus leverages this by iteratively running candidate programs against a curated test suite and refining the outputs until convergence. Another iterative refinement method is RepoCoder~\cite{zhang2023repocoderrag}, which addresses repository-level code completion through multi-round retrieval-augmented generation. RepoCoder retrieves relevant code snippets from the same repository based on unfinished code, generates candidate completions using an LLM (e.g., GPT-3.5), and reuses generated code to iteratively refine retrieval. Evaluated on the RepoEval benchmark with function-, API-, and line-level tasks, it achieved up to 10\% improvements in exact match and functional correctness over one-shot and in-file methods. The study demonstrates that iterative retrieval and synthesis significantly improve contextual grounding and code quality without requiring agentic coordination.

\textbf{LLM-Orchestrated Algorithm Synthesis with Verifiers.} While traditional LLM-based code generation excels at translating natural language into executable programs, it struggles with algorithm synthesis tasks that require not just implementation but also the inference of the correct algorithmic idea. To address this, ALGO~\cite{zhang2023algosynthesizingalgorithmicprograms} introduces a novel framework where LLMs are used not only as code generators but also as oracle verifiers. ALGO consists of two modules: a \textit{verifier} that generates a brute-force but correct oracle using LLMs, and a \textit{coder} that synthesizes efficient solutions guided by the verifier’s outputs. The oracle is created by prompting LLMs to exhaustively enumerate the search space and is used to validate candidate programs by comparing outputs on a batch of generated inputs. Experimental results show that ALGO significantly boosts synthesis performance across various LLM coders: achieving an 8× increase in one-shot pass rate over Codex and a 2.6× gain over CodeT.

While LLMs have demonstrated impressive performance across various static and feedback-driven code generation tasks, their ability to autonomously plan, reason across stages, or coordinate within larger development workflows remains limited. This gap has motivated recent research into LLM-based agents that embody higher-level autonomy and collaboration capabilities, as discussed in the following section.
\subsection{LLM-based Agents Tasks} \label{sec:code-agent}
\textbf{Multi-Agent Role Division for Software Development.} LLM-based agents have shown significant potential and advantages by substantially improving task efficiency and effectiveness through multi-agent collaboration. Unlike traditional LLMs, LLM-based agents adopt a division of labor approach, breaking down complex tasks into multiple subtasks handled by specialized agents, this method can enhance task efficiency and improves the quality and accuracy of generated code to mitigate the hallucination from the single LLM. 

In~\cite{dong2024selfcollaborationcodegenerationchatgpt}, researchers proposed a self-collaboration framework where multiple ChatGPT (GPT-3.5-turbo) agents act as different roles to collaboratively handle complex code generation tasks. Specifically, the introduction of Software Development Methodology (SDM) divides the development process into three stages: analysis, coding, and testing. Each stage is managed by specific roles, and after completing their tasks, each role provides feedback and collaborates with others to improve the quality of the generated code. Experiment shows that this self-collaboration framework significantly improves performance on both the HumanEval and MBPP benchmarks, with the highest improvement reaching 29.9\% in HummanEval compared to the SOTA model GPT-4. This result demonstrating the potential of collaborative teams in complex code generation tasks. Although it lacks external tool integration and dynamic adjustment capabilities, this framework exhibits common characteristics of LLM-based agents, such as role distribution, self-improvement ability, and excellent autonomous decision-making, these combined capabilities qualify it to be considered an LLM-based agent. Similarly, In~\cite{lin2024llm}, the LCG framework improved code generation quality also through multi-agent collaboration and chain-of-thought techniques, once again demonstrating the effectiveness of multi-agent collaboration in the software development process. Extending the multi-agent paradigm further, recent research has introduced the MapCoder framework~\cite{islam2024mapcoder}, specifically designed for competition-level code generation tasks. MapCoder employs four distinct agents—Retrieval, Planning, Coding, and Debugging—each performing specialized tasks reflective of human programming practices. The Retrieval agent autonomously identifies relevant prior solutions, while the Planning agent generates step-by-step solution strategies. Subsequently, the Coding agent translates these strategies into executable code, and the Debugging agent iteratively refines this code based solely on sample input-output pairs, eliminating dependency on externally generated test cases. This robust division of responsibilities among specialized agents highlights the strength of multi-agent frameworks in significantly enhancing both code quality and problem-solving efficiency in complex programming environments.


\textbf{Overcoming Context Limitations for Large Codebases.} The limitations of context windows were not discussed in previous studies, this has been thoroughly explored in 2024 by the University of Cambridge team. In~\cite{holt2024lmac}, researchers introduced the L2MAC framework, which dynamically manages memory and execution context through a multi-agent system to generate large codebases, and achieved SOTA performance in generating large codebases for system design tasks. The framework is primarily divided into the following components: the processor, which is responsible for the actual generation of task outputs; the Instruction Registry, which stores program prompts to solve user tasks; and the File Storage, which contains both final and intermediate outputs. The Control Unit periodically checks the outputs to ensure that the generated content is both syntactically and functionally correct. The researchers conducted multiple experiments and compared with many novel methods like GPT-4, Reflexion, and AutoGPT, achieving a Pass@1 score of 90.2\% on the HumanEval benchmark. Addressing similar scalability challenges, the Self-Organized multi-Agent framework (SoA)~\cite{ishibashi2024selforganizedagentsllmmultiagent} proposes a novel solution based on dynamically scalable agent collaboration. Unlike traditional single-agent models constrained by limited context length, SoA employs multiple self-organized agents, each independently responsible for generating and refining smaller code components. This design allows the framework to automatically scale the number of agents in response to increasing complexity, thus maintaining manageable workloads per agent regardless of total codebase size. Evaluated against strong single-agent baselines like Reflexion \cite{shinn2024reflexion}, SoA achieves a significant improvement in Pass@1 accuracy, surpassing Reflexion by 5\% on the HumanEval benchmark. The decentralized, scalable approach of SoA illustrates an effective strategy for overcoming single-agent context limitations, particularly beneficial in ultra-large-scale software development.

\textbf{Simulating the Software Engineering Lifecycle.} Recently, many studies have begun to use LLM-based agents to simulate real software development processes, the paper ~\cite{hong2023metagptmetaprogrammingmultiagent} introduced the MetaGPT framework, which enhanced problem-solving capabilities through standard operating procedures (SOPs) encoded in multi-agent collaboration. The entire process of the multi-collaboration framework simulates the waterfall life-cycle of software development, with each agent playing different roles and collaborating to achieve the goal of automating software development. LLM-based agents have also shown strong ability in automated software development,~\cite{rasheed2023autonomous} proposed a multi-GPT agent framework that automates tasks such as project planning, requirement engineering, software design, and debugging, illustrating the potential for automated software development. Similarly ~\cite{rasheed2024codepori} introduced the model called CodePori, which is a novel model designed to automate code generation for extensive and complex software projects based on natural language prompts. In~\cite{huang2023agentcoder} the AgentCoder framework collaborates with programmer agents, test design agents, and test execution agents to generate and optimize code, outperforming existing methods, achieved SOTA performance on the HumanEval-ET benchmark with pass@1 of 77.4\% compared to the previous state-of-the-art result of 69.5\%, this result showcasing the advantages of multi-agent systems in code generation and testing. Building upon these foundations, more recent work explores how LLM-based agents can go beyond fixed workflows to emulate agile, dynamic, and context-aware development practices. One such framework is AGILECoder~\cite{nguyen2024agilecoderdynamiccollaborativeagents}, a dynamic multi-agent architecture that explicitly incorporates Agile roles—PM, SM, Developer, Senior Developer, and Tester—and iteratively drives software development through structured sprints. It introduces a Dynamic Code Graph Generator (DCGG), which tracks cross-file dependencies to enhance context awareness for code generation, test case design, and debugging. Compared to waterfall-style pipelines like MetaGPT \cite{hong2023metagptmetaprogrammingmultiagent} or ChatDev \cite{qian2023communicative}, AGILECoder enables fine-grained planning, adaptive role-based review, and execution-based validation. Experiments on HumanEval, MBPP, and a new ProjectDev benchmark show that AGILECoder outperforms previous multi-agent systems in both pass@1 and executability, especially in large-scale, repository-level development tasks. This result demonstrates the importance of agile coordination and graph-based memory in driving scalable, real-world agent collaboration.Complementing this approach, a lightweight multi-agent framework proposed by~\cite{manish2024autonomous} simplifies agile software development by assigning agents to handle user stories, task planning, coding, and pull request submissions in a structured yet less resource-intensive manner. Although it handles lower-complexity tasks without external tools or complex memory management, it reinforces the practicality of Agile-based multi-agent coordination for reducing manual interventions and maintaining coherent software outputs. These studies illustrate the versatility and scalability potential of LLM-based agents when explicitly aligned with Agile development methodologies.

\textbf{Enhancing Feedback, Tool Use, and Open-Source Model Performance.} The purpose of integrating LLMs into agents from many framework is to enhance the self-feedback and reflection capabilities of the entire agent system. Because the current open-source LLMs generally have much lower capabilities in this aspect compared to proprietary models, the emergence of LLM-based agents can help bridge the gap between open-source models and the advanced capabilities of proprietary systems like GPT-4.~\cite{zheng2024opencodeinterpreter} introduced the OpenCodeInterpreter framework, which improved the accuracy of code generation models by integrating code generation, execution, and human feedback. Based on CodeLlama and DeepSeekCoder, this framework performed close to the GPT-4 Code Interpreter on the HumanEval and MBPP benchmarks. The abbility of using external tools or APIs is another significant advantage of LLM-based agents,~\cite{schick2024toolformer} proposed the Toolformer model, which significantly enhanced task performance by learning to call APIs through self-supervision. The framework Based on GPT-J (6.7B parameters) achieved significant performance improvements across multiple benchmark tasks, demonstrating the possibility of LLM-based agent brought by the external tool, the diverse choice of tools and architectures, allowing LLMs to continuously learn new things and improve themselves. 
Similarly,~\cite{qin2023toolllm} enhanced LLMs' interaction with external APIs through the ToolLLM framework, outperforming Text-Davinci-003 and Claude-2 on the ToolBench and APIBench benchmarks and excelling in multi-tool instruction processing. 

Extending this paradigm, CodeAgent \cite{zhang2024codeagentenhancingcodegeneration} addressed complex real-world, repository-level coding challenges through extensive integration of external programming tools, including retrieval systems, documentation parsing, and symbol navigation tools. By strategically orchestrating tool usage via multiple agent strategies (e.g., ReAct and Tool-Planning), CodeAgent significantly enhanced context-awareness and practical usability. Evaluations on the CODEAGENTBENCH benchmark demonstrated superior performance over commercial tools like GitHub Copilot, emphasizing the effectiveness of structured tool integration for tackling intricate software engineering tasks. A further advancement in this domain is provided by SWE-agent~\cite{yang2024sweagentagentcomputerinterfacesenable}, which introduces a dedicated agent-computer interface (ACI) designed explicitly for LM-agents interacting with software engineering tasks. Unlike traditional Linux shells or general-purpose interfaces, the ACI provides simplified, LM-friendly commands, immediate feedback, and built-in guardrails for common software tasks such as file navigation, code editing, and test execution. This carefully tailored interface significantly enhances LLM-agent efficiency, achieving state-of-the-art pass@1 rates on SWE-bench (12.5\%) and HumanEvalFix (87.7\%), clearly demonstrating that interface design is critical in enabling effective, autonomous software engineering by LLM-based agents. The CodeAct framework proposed in~\cite{wang2024executable} allows agents to generate Python code directly executable by a Python interpreter, enabling agents to dynamically revise actions and integrate existing Python libraries seamlessly. CodeAct's structured executable actions allow LLMs to effectively utilize control flow mechanisms such as loops and conditionals to handle complex multi-step tool invocations within a unified action framework. This framework underline the potential of executable code actions in bridging the gap between open-source and proprietary models by capitalizing on LLMs' inherent code comprehension skills, thus enhancing their practical usability in diverse, complex software engineering tasks.

\textbf{Interactive Requirement Clarification and Self-Planning.} ClarifyGPT~\cite{mu2024clarifygpt} is a novel LLM-based agent framework designed to address the ambiguity in user-provided natural language requirements during code generation. Unlike conventional prompting methods that directly generate code regardless of input clarity, ClarifyGPT first determines whether a requirement is ambiguous via a two-step code consistency check, which samples multiple code completions and evaluates their behavioral divergence on synthesized test inputs. If inconsistency is detected, the system engages in a reasoning-driven prompt strategy to generate targeted clarifying questions. Upon receiving responses (either from users or a high-fidelity simulator), the agent refines the original requirement and re-prompts the LLM to generate code. This pipeline, involving perception, planning, action, and reflection, exemplifies typical LLM-based agent characteristics. Evaluation on five benchmarks (MBPP, HumanEval, CoderEval, etc.) demonstrates that ClarifyGPT significantly improves the Pass@1 accuracy, achieving a relative gain of 11.66\% with GPT-4 and 15.00\% with ChatGPT. These improvements are particularly notable on ambiguous tasks, confirming the effectiveness of proactive clarification in agent-driven development workflows.

\subsection{Analysis} The evolution from traditional LLM-based approaches to more autonomous LLM-based agent frameworks in software development underscores a fundamental shift in how code generation is conceptualized, orchestrated, and evaluated. Traditional LLM methods typically center on improving a single model’s ability to generate, debug, or refine code. While powerful, these methods often face context window constraints, limited memory, and a linear development workflow that depends heavily on human-provided test cases or clarifications. In contrast, LLM-based agent systems distribute responsibilities across multiple specialized agents (or roles) and often integrate external tools to emulate more realistic and adaptive software engineering practices.

One key difference is in the degree of autonomy and collaboration. Traditional LLMs, even those enhanced with techniques like few-shot prompting \cite{brown2020language}, chain-of-thought \cite{wei2022chain}, or feedback-driven refinement \cite{10.1145/3672456}, tend to operate as a single model with minimal division of labor. They excel in generating snippets or providing incremental improvements, but often struggle in orchestrating an entire development process, especially when large codebases or complex cross-file dependencies are involved. By contrast, the newer LLM-based agent frameworks discussed in Section \ref{sec:code-agent} (e.g., SoA, L2MAC, AGILECoder, CodeAgent) simulate multi-step development pipelines through distributed roles or agents for planning, coding, testing, and debugging. This division of labor allows for context expansion beyond a single model’s window limit, as well as iterative feedback and refinement that better aligns with real-world team-based software development processes.

Another prominent distinction lies in tool integration. Although LLMs can incorporate limited forms of external knowledges, such as print debugging~\cite{hu2024leveragingprintdebuggingimprove}, test case execution~\cite{fakhoury2024llm, ridnik2024codegenerationalphacodiumprompt}, or retrieval mechanisms~\cite{zhang2023repocoderrag}, they rarely exhibit flexibility. In agent systems, each specialized role (e.g., a “retrieval agent” or a “debug agent”) can invoke dedicated libraries, file navigation interfaces, or testing harnesses without human intervention. This substantially boosts effectiveness in tasks like repository-level code completion, where scanning large amounts of context or retrieving relevant snippets from extensive codebases becomes essential. Moreover, it helps address the gap between open-source LLMs and advanced proprietary systems by enhancing self-reflection and externally verifiable feedback mechanisms.

Scalability and life-cycle simulation further differentiate the two paradigms. Traditional LLM-based methods do handle multi-turn queries, such as iterative code completion or user-guided debugging, but they do not generally organize these interactions under a coherent, larger software development methodology. In contrast, many agentic systems like MetaGPT \cite{li2023metaagents}, ChatDev \cite{qian2023communicative} and AGILECoder \cite{nguyen2024agilecoderdynamiccollaborativeagents} explicitly model entire development cycles, whether it is the classic waterfall model or agile “sprints”. By dynamically coordinating specialized agents, these frameworks emulate real-world team strategies: clarifying requirements, planning sprints, writing code, and designing or running tests. As a result, tasks that go beyond single-file code generation (like orchestrating multiple modules, dealing with cross-dependencies, or generating large-scale projects) become more tractable.

\begin{figure}[htbp]
\scriptsize
    \centering
    \includegraphics[width=1\linewidth]{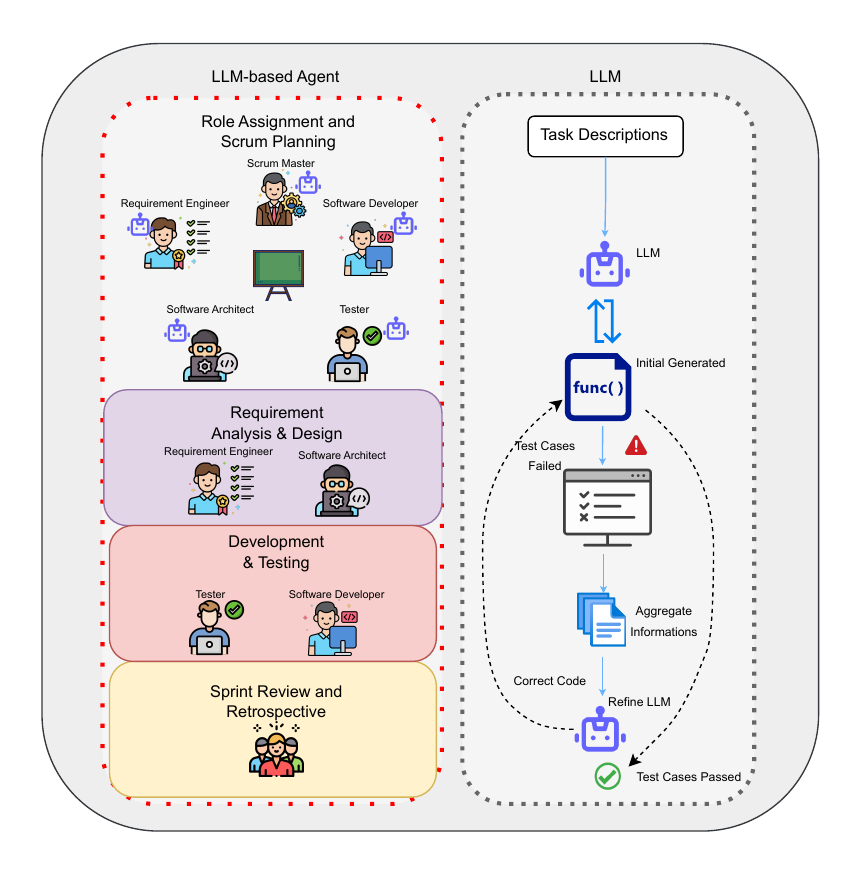}
    \caption{Illustration of Comparison Framework Between LLM-based Agent and LLM in Code Generation and Software Development.}
    \label{fig:codeGenGraph}
\end{figure}
As illustrated in Figure~\ref{fig:codeGenGraph} (adapted from \cite{lin2024llm} and \cite{10.1145/3649825}), the left-hand flow represents an LLM-based agent system that divides software tasks among specialized roles simulating an industry Scrum team. These roles collaboratively implement, test, and refine code through iterative coordination. In contrast, the right-hand flow depicts how traditional LLMs might process multiple prompts or code analyses in isolation, lacking the continuous autonomous coordination and tool-driven synergy inherent to multi-agent environments. While single-model approaches can achieve high pass@k performance on established benchmarks, the agent-based paradigm demonstrates broader coverage of large-scale, complex tasks, frequently surpassing state-of-the-art results.

This shift toward LLM-based agents reflects the growing demand for holistic, end-to-end solutions in software development. Traditional LLMs excel at isolated tasks but face limitations due to constrained context awareness, insufficient feedback mechanisms, and minimal autonomy. In contrast, LLM-based agents leverage role specialization, iterative refinement, tool integration, and dynamic context management. These capabilities address critical performance gaps in large-scale, multi-file coding challenges and enable autonomous workflows that increasingly mirror human-led engineering processes.

\subsection{Benchmarks}
In the field of code generation and software development, there are notable differences and commonalities in the dataset used for research on LLMs and LLM-based agents. These datasets provide important benchmarks for evaluating model performance, the HumanEval dataset, widely used for assessing code generation models, is handcrafted by OpenAI and contains 164 programming problems, each including a function signature, problem description, function body, and unit tests. This dataset is primarily used to evaluate a model's ability to generate correct code, particularly in tasks that involve converting natural language descriptions into executable code. Many studies have utilized HumanEval to test the performance of code generation models~\cite{dong2024selfcollaborationcodegenerationchatgpt}. The MBPP (Mostly Basic Python Programming) dataset is another common benchmark, comprising 427 Python programming problems that cover basic concepts and standard library functions, this dataset is used to evaluate model performance across various programming scenarios. In~\cite{huang2023agentcoder}, researchers used the MBPP dataset to test the performance of multi-agent systems in code generation and optimization, improving the accuracy and robustness of generated code through agent collaboration. The HumanEval-ET and MBPP-ET datasets are extensions of the original HumanEval and MBPP datasets, adding more test cases and more complex problems for a comprehensive evaluation of model performance~\cite{10.1145/3672456}. The Spider and BIRD datasets focus on converting natural language to SQL queries, evaluating the model's ability to handle complex query generation tasks. In~\cite{sun2023sql}, researchers used these datasets to test the SQL-PaLM framework, which evaluating the execution accuracy and exact match rate for SQL generation tasks through few-shot prompt and instruction fine-tuning. ToolBench and APIBench datasets are used to evaluate a model's capability in using tools and APIs, ToolBench contains 16,464 real-world RESTful API instructions, and APIBench normally tests a model's generalization ability to unseen API instructions~\cite{qin2023toolllm}. The CAASD (Capability Assessment of Automatic Software Development) dataset is a newly developed benchmark comprising 72 software development tasks from various domains, each with a set of reference use cases to evaluate AI-assisted software development systems~\cite{zhang2024experimentingnewprogrammingpractice}. 

There are some obvious commonalities in dataset selection for LLMs and LLM-based agents, the HumanEval and MBPP datasets are widely used to assess code generation capabilities, covering a variety of programming tasks and languages. Moreover, many studies have adopted multilingual and cross-domain datasets, such as HumanEval-X and CodeSearchNet, to evaluate model performance across different languages and tasks. For the differences, LLM-based agents tend to use multi-agent collaboration frameworks to handle complex tasks, thus favoring benchmark datasets that emphasize multi-turn interactions and iterative optimization, also focus on tool usage and API integration capabilities, the framework TOOLLLM used ToolBench and APIBench to assess its tool usage capabilities, while Toolformer demonstrated its ability to autonomously learn to use tools. These differences primarily from the different approaches to task handling between LLMs and LLM-based agents, LLMs typically optimize a single model's performance by fine-tuning on relevant datasets.

\subsection{Evaluation Metrics}
Various evaluation metrics are used to assess the performance of LLMs and LLM-based agents in code generation and software development. These metrics measures the models' performance in specific tasks and how they improve the code generation and software development process. Table~\ref{tab:cg_metrics} includes the distribution of evaluation metrics cited in this paper, encompassing both LLMs and LLM-based agents.

In research on LLMs and LLM-based agents, Pass@k is a common evaluation metric used to measure the proportion of generated code that passes all test cases within the first k attempts, this metric is widely applied across various datasets. In~\cite{10.1145/3672456}, Pass@k was used to evaluate the quality of code generation in multi-turn interactions, showing that the model's Pass@k significantly improved by introducing a planning phase. Besides Pass@k, BLEU score is another common evaluation metric, mainly used to measure the syntactic similarity and correctness between generated code and reference code. In~\cite{fried2023incodergenerativemodelcode}, BLEU score was used to evaluate the quality of generated code. Complete Time and Success Rate are other important evaluation metrics, particularly when assessing the productivity impact of AI-assisted development tools, these metrics are crucial as we expect LLMs to generate accurate code while maintaining expected speed. Confidence Calibration and Execution Rate are metrics used to evaluate the confidence level and execution success rate of the model when generating code. Researchers often use these metrics to assess various LLMs' performance in understanding user intent and generating correct code with high precision.

Compared to the evaluation metrics for LLMs in software development, LLM-based agents also use Pass@k but more diverse to reflect their multi-agent collaboration characteristics. Win Rate and Agreement Rate are important metrics for evaluating the effectiveness of multi-agent collaboration. Additionally, LLM-based agents often use metrics like Execution Effectiveness and Cost Efficiency to evaluate their performance in real-world applications. For instance, in MetaGPT~\cite{hong2023metagptmetaprogrammingmultiagent}, researchers evaluated not only the correctness of code generation but also analyzed the execution effectiveness, development costs, and productivity. Results indicated that MetaGPT significantly improved development efficiency and reduced development costs while generating high-quality code. Overall both are using traditional metrics such as Pass@k, Win Rate, and task completion time to evaluate their code generation capabilities, these metrics directly reflect the accuracy and efficiency of the model in generating code. But LLM-based agents normally requiring more comprehensive and diverse metrics for evaluation to help assess the performance of multiple agents and the whole development process, that's why we can see the human revision cost, qualitative feedback in the evaluation metrics. Researchers consider user or developer satisfaction metrics, as agent applications often involve extensive projects rather than isolated small-scale development, these metrics focus on the correctness of code generation and also resource utilization efficiency of the agent system. 
\begin{table*}
\centering
\scriptsize
\caption{Evaluation Metrics in Code Generation and Software Development}
\label{tab:cg_metrics}
\begin{tblr}{
  column{4} = {c},
  cell{1}{1} = {c},
  cell{1}{3} = {c},
  cell{1}{2} = {c},
  hlines,
  vlines,
}
Reference Paper & Benchmarks                                                                                                            & Evaluation Metrics                                                                    & Agent \\
{~\cite{hu2024leveragingprintdebuggingimprove}}            & Leetcode problem                                                                                                      & Accuracy                                                                              & No    \\
{~\cite{peng2023impactaideveloperproductivity}}            & {HTTP server in JavaScript by\\95 programmer}                                                                         & {Task Completion Time,~\\Task Success}                                                & No    \\
{~\cite{ni2023l2cevalevaluatinglanguagetocodegeneration}}            & {Spider,~WikiTQ,~GSM8k,~\\SVAMP,~MBPP,~MBPP,~DS-1000}                                                                 & {Execution Accuracy,\\Confidence Calibration\\Execution Rate}                         & No    \\
{~\cite{dubois2024alpacafarm}}            & Alpaca Data                                                                                                           & {Win-Rate,~Agreement \\Rate}                                                          & No    \\
{~\cite{10.1145/3672456}}            & {HumanEval/-X/-ET,~\\MBPP-sanitized/-ET}                                                                              & {Pass@k,~AvgPassRatio, \\CodeBLEU}                                                    & No    \\
{~\cite{fried2023incodergenerativemodelcode}}            & HumanEval,~CodeXGLUE                                                                                                  & {Pass rate, Exact Match\\BLEU Score}                                                  & No    \\
{~\cite{sun2023sql}}            & Spider                                                                                                        & {Accuracy,~\\Exact Match}                                                             & No    \\
{~\cite{nijkamp2023codegenopenlargelanguage}}            & HumanEval,~MTPB                                                                                                       & Pass@k,~Pass rate                                                                     & No    \\
{~\cite{wang2023codet5+}}            & {HumanEval,~MathQA-Python,\\GSM8K-Python,~CodeSearchNet,\\CosQA, AdvTest}                                             & {Pass@k,~BLEU-4, \\Exact Matcha\\Edit Similarity,\\Mean Reciprocal Rank (MRR)}        & No    \\
{~\cite{zheng2024codegeexpretrainedmodelcode}}            & HumanEval/-X                                                                                                          & Pass@k                                                                                & No    \\
{~\cite{chen2021evaluatinglargelanguagemodels}}            & HumanEval                                                                                                             & Pass@k,~BLEU Score                                                                    & No    \\
{~\cite{10.1145/3649825}}            & HumanEval,~MBPP-S, APPS                                                                                               & {Pass Rate,~Token Edit Distance,\\~Exact Copy Rate}                                   & No    \\
{~\cite{fakhoury2024llm}} & {MBPP, HumanEval} & {pass@k, pass@1@m} & No \\
{~\cite{ridnik2024codegenerationalphacodiumprompt}} & CodeContests & pass@k & No \\
{~\cite{zhang2023repocoderrag}} & None (real-world prompts + OSCAT lib) & {Manual simulation and correctness \\via OpenPLC} & No \\
{~\cite{zhang2023algosynthesizingalgorithmicprograms}} & CodeContests, LeetCode (recent) & n@k (1@k, 2@k, 10@k, 100@k) & No \\
{~\cite{dong2024selfcollaborationcodegenerationchatgpt}}            & MBPP/-ET,~HumanEval/-ET                                                                                               & Pass@k                                                                                & Yes   \\
{~\cite{holt2024lmac}}            & HumanEval                                                                                                             & Pass@1                                                                                & Yes   \\
{~\cite{zhang2024experimentingnewprogrammingpractice}}            & CAASD                                                                                                                 & Pass Rate,~Token Consumption                                                          & Yes   \\
{~\cite{schick2024toolformer}}            & {CCNet,~SQuAD, Google-RE,~T-REx,\\ASDiv, SVAMP, MAWPS, \\Web Questions, Natural Questions,\\TriviaQA, MLQA, TEMPLAMA} & {Zero-shot performance,~\\Perplexity,~Tool usage effectiveness}                       & Yes   \\
{~\cite{huang2023agentcoder}}            & MBPP/-ET, HumanEval/-ET                                                                                               & Pass@1                                                                                & Yes   \\
{~\cite{hong2023metagptmetaprogrammingmultiagent}}            & {HumanEval, HumanEval,\\SoftwareDev}                                                                                  & {Pass@k,~Executability,~Cost,~\\Code Statistics,~Productivity,~\\Human Revision Cost} & Yes   \\
{~\cite{rasheed2024codepori}}            & HumanEval, MBPP                                                                                                       & {Pass@k,~Practitioner-Based \\Assessment}                                             & Yes   \\
{~\cite{qin2023toolllm}}            & ToolBench, APIBench                                                                                                   & Pass Rate,~Win Rate                                                                   & Yes   \\
{~\cite{rasheed2023autonomous}}            & No Specificed                                                                                                         & Pass Rate,~Win Rate                                                                   & Yes   \\
{~\cite{lin2024llm}}            & MBPP/-ET, HumanEval/-ET                                                                                               & Pass@1                                                                                & Yes   \\
{~\cite{zheng2024opencodeinterpreter}}            & HumanEval,~MBPP,~EvalPlus                                                                                             & Pass@1                                                                                & Yes   \\
{~\cite{mu2024clarifygpt}} & {MBPP-sanitized, MBPP-ET, HumanEval/-ET, CoderEval} & Pass@1 & Yes \\
{~\cite{nguyen2024agilecoderdynamiccollaborativeagents}} & {HumanEval, MBPP, ProjectDev} & {Pass@1, Executability Rate, \#Errors, Runtime Stats} & Yes \\
{~\cite{manish2024autonomous}} & Real-world case studies & Human Review, Task Success Rate & Yes \\
{~\cite{ishibashi2024selforganizedagentsllmmultiagent}} & HumanEval & Pass@1 & Yes \\
{~\cite{yang2024sweagentagentcomputerinterfacesenable}} & {SWE-bench, SWE-bench Lite, HumanEvalFix} & {Pass@1, \%Resolved, pass@k, \$Avg. Cost} & Yes \\
{~\cite{wang2024executable}} & {API-Bank, M3ToolEval} & {Success Rate, Avg. Number of Turns, Tool Correctness} & Yes \\
{~\cite{islam2024mapcoder}} & {HumanEval/-ET, EvalPlus, MBPP/-ET, \\APPS, xCodeEval, CodeContests} & Pass@k & Yes \\
{~\cite{zhang2024codeagentenhancingcodegeneration}} & {CODEAGENTBENCH, HumanEval} & Pass@1 & Yes \\
{~\cite{murali2023codecompose}}            & {First-party data from Meta's \\code repositories and notebooks}                                                      & {Acceptance Rate, Percentage \\of Code Typed,~\\Qualitative Feedback}                 & Yes  
\end{tblr}
\end{table*}

%% file: pages/autonomous.tex
\section{Autonomous Learning and Decision Making \label{cha:autonomous}}
Autonomous Learning and Decision Making is a critical and evolving field in modern software engineering, especially under the influence of artificial intelligence and big data. The core task of autonomous learning and decision making is to achieve automated data analysis, model building, and decision optimization through machine learning algorithms and intelligent systems, thereby enhancing the autonomy and intelligence of systems.

In this process, LLMs and LLM-based agents bring numerous possibilities, following the development of NLP technology, a lot of achievements have been made in the application of LLMs in this field. These models can handle complex language tasks and also demonstrate powerful reasoning and decision-making abilities, the research on voting inference using multiple LLMs calls has revealed new methods for optimizing performance, with the frequently used method called majority vote~\cite{huang2022large}, this improves the accuracy of inference systems and ensures the selection of the optimal possibility. Additionally, the performance of LLMs in tasks such as automated debugging and self-correction has enhanced the system's autonomous learning capabilities, achieving efficient error identification and correction. At the same time, the application of LLM-based agents in autonomous learning and decision-making is also a novel but popular topic, these agents can perform complex reasoning and decision-making tasks with the help from the LLM, and also improve their adaptability in dynamic environments through continuous learning and optimization. In this context, we have collected 23 research papers on LLM-based agents in this field. This survey will provides a general review of these studies, analyzing the specific applications and technical implementations in autonomous learning and decision making.

\subsection{LLMs Tasks}
\textbf{API Utilization and Optimization in LLM Workflows.} The API call for the LLMs is one common applications, often requiring continuous calls to enable the model to make judgments and inferences, but does continuously increasing the number of calls always improve performance? In~\cite{chen2024more}, researchers explored the impact of increasing LLM calls on the performance of composite reasoning systems. Paper analyze the voting inference design systems, the result showed that there is a non-linear relationship between the number of LLM calls and system performance; performance improves initially with more calls but declines after reaching a certain threshold. This research provides a theoretical basis for optimizing LLM calls, helping to allocate resources reasonably in practical applications to achieve optimal performance. However, the performance of Voting Inference Systems shows a non-monotonic trend due to the diversity of query difficulties, and the continuously increasing cost also needs to be considered. 

\textbf{Autonomous learning Learning from Feedback.} Autonomous learning is also applied in bug fixing, where researchers hope LLMs can continuously learn to fix bugs and eventually identify human oversights or common errors. In~\cite{chen2023teaching}, the SELF-DEBUGGING method was proposed, enabling LLMs to debug code by analyzing execution results and natural language explanations. This method significantly improved the accuracy and sample efficiency of code generation tasks especially for complex problems. Experimental results on the Spider and TransCoder benchmarks showed that the SELF-DEBUGGING method increase the model's accuracy by 2-12\% which demonstrates the potential of LLMs in autonomous learning to debug and correct any erros. Another similar study introduced the AutoSD (Automated Scientific Debugging) technique~\cite{kang2023explainable}, which simulates the scientific debugging process through LLMs, generating explainable patched code. Researchers evaluated AutoSD's capabilities from six aspects: feasibility, debugger ablation, language model change, developer benefit, developer acceptance, and qualitative analysis. Result have shown that AutoSD can generate effective patches and also improve developers' accuracy in evaluating patched code by providing explanations, its explainability function makes it easier for developers to understand and accept automatically generated patches. Although the above two studies primarily focus on automated debugging techniques, the frameworks designed in these studies automatically determine the optimal repair solution based on the debugging results after collecting sufficient information, and provide specific code implementations, which demonstrated the capability of autonomous decision-making and learning. Another framework ALGO \cite{zhang2023algosynthesizingalgorithmicprograms}, which was previously discussed under the theme of code generation and software development due to its significant improvements in program synthesis. However, ALGO is also highly relevant to autonomous learning and decision-making. It embodies a verifier-guided learning paradigm in which the LLM autonomously generates brute-force oracles, constructs test cases, and iteratively compares outputs to verify correctness and guide efficient code synthesis. This self-supervised refinement loop allows the model to make informed decisions about code quality and correctness without external supervision, exemplifying autonomous reasoning and learning in practice.

\textbf{LLM-Based Decision Making and Evaluation.} Since the rise of LLMs applied to various fields, one research direction has been the rational analysis of their creativity and the exploration of their potential for continuous learning, this creativity also highly determined by the decision making capability of the models.~\cite{franceschelli2023creativity} analyzed the outputs of LLMs from the perspective of creativity theory, exploring their ability to generate creative content, the study used metrics such as value, novelty, and surprise, finding that current LLMs have limitations in generating combinatorial, exploratory, and transformative creativity. Although LLMs can generate high-quality creative content, further research and improvement are needed to achieve true creative breakthroughs. Additionally, innovative responses generated by LLMs may come with the possibility of hallucination, a long-standing issue for large language models. Despite many techniques to mitigate its downsides, it still cannot be entirely prevented. There are many interesting experiments in decision making, such as having LLMs act as judges to determine whether a person has committed a crime~\cite{lai2023largelanguagemodelslaw}. A familiar attempt is to have a primary LLM interact with other LLMs.~\cite{zheng2024judging} explored the effectiveness of using LLMs as judges to evaluate other LLM-driven chat assistants. The study validated the consistency of LLM judgements with human preferences through the MT-Bench and Chatbot Arena benchmarks, with results showing that GPT-4's judgments were highly consistent with human judgments across various tasks. This research demonstrates the potential of LLMs in simulating human evaluation, providing new ideas for automated evaluation and optimization.

\subsection{LLM-based Agents Tasks}
\textbf{Multi-Agent Discussion and Reasoning Enhancement.} Multi-agent collaboration and dialogue frameworks also demonstrated strong capabilities in both decision making and autonomous learning.~\cite{wang2024rethinking} explores whether multi-agent discussions can enhance the reasoning abilities of LLMs. The proposed CMD framework simulates human group discussion processes, showing that multi-agent discussions can improve performance in commonsense knowledge and mathematical reasoning tasks without task-specific examples. Additionally, the study found that multi-agent discussions also correct common errors in single agents, such as judgment errors and the propagation of incorrect answers, thereby enhancing overall reasoning accuracy.
~\cite{chen2023towards} researchers explored the potential of multi-modal large language models (MLLMs) like GPT4-Vision in enhancing agents' autonomous decision-making processes. The paper introduce the PCA-EVAL benchmark, and evaluated multi-modal decision-making capabilities in areas such as autonomous driving, home assistants, and gaming. The results showed that GPT4-Vision exhibiting outstanding performance across the dimensions of perception, cognition, and action.

\textbf{Self-Reflection and Experience-Based Learning.}~\cite{shinn2024reflexion} proposes the Reflexion framework, a novel approach that strengthens learning through language feedback rather than traditional weight updates to avoid expensive re-train costs. The framework uses self-reflection and language feedback to help language agents learn from mistakes, significantly improving performance in decision-making, reasoning, and programming tasks. The Reflexion's first-pass success rate on the HumanEval Python programming task increased from 80.1\% to 91.0\%, success rates in the ALFWorld decision-making task improved by 22\%, and performance in the HotPotQA reasoning task increased by 14\%. These results indicate that the Reflexion framework demonstrate the state-of-art performance in various tasks through self-reflection and language feedback.

Another agent framework~\cite{zhao2024expel} introduces the ExpeL agent framework which enhances decision-making capabilities by autonomously collecting experiences and extracting knowledge from a series of training tasks using natural language, this experience collection process is similar to how humans gain insights through practice and apply them in exams. By accessing internal databases, ExpeL also reduces hallucinations, employing the RAG technique discussed in \ref{cha:Preliminaries}. The ExpeL framework doesn't require parameter updates it enhances decision-making capabilities by recalling past successes and failures which fully leveraging the advantages of ReAct framework~\cite{yao2022react}. Experimental showed that ExpeL can continuous improvement across tasks in multiple domains and exhibited cross-task transfer learning capabilities. The combination of ExpeL and Reflexion even further enhance the performance in iterative task attempts, highlighting the importance of autonomous learning and experiential accumulation in developing intelligent agents. The ExpeL framework demonstrates its potential as a state-of-the-art (SOTA) LLM-based agents in several aspects, particularly in cross-task learning, self-improvement, and memory mechanisms. By comparing ExpeL with existing SOTA agents like Reflexion~\cite{shinn2024reflexion}, ExpeL outperforms baseline methods in various task environments. These studies collectively indicate the importance of autonomous learning and improvement in LLM-based agents, agent systems continuously optimize and improve decision-making processes through self-feedback, self-reflection, and experiential accumulation which shows higher autonomy and flexibility in handling dynamic and complex tasks compared to traditional LLMs. Unlike traditional LLMs, which mainly rely on pre-training data and parameter updates, LLM-based agents adapt and improve their performance in real-time through continuous self-learning and feedback mechanisms, thus demonstrating outstanding performance in various tasks.

\textbf{Multi-Agent Collaboration and Emergent Behavior.}~\cite{chen2023agentverse} proposes the AGENTVERSE multi-agent framework, designed to improve task completion efficiency and effectiveness through collaboration. The framework draws on human group dynamics by designing a collaborative system of expert agents that exhibit outstanding performance in tasks such as text understanding, reasoning, coding, and tool usage. Experiments showed that the AGENTVERSE framework performed well not only in independent task completion but also significantly improved performance through group collaboration especially in coding tasks where the framework use GPT-4 to be the brain of the agent groups. The framework also observed emergent behaviors in agents during collaboration, such as voluntary actions, conformity, and destructive behaviors, providing valuable insights for understanding and optimizing multi-agent systems.

Another multi-agents study ~\cite{li2023camel} Introducing the CAMEL framework, this is a well known agent framework, which explores building scalable techniques to facilitate autonomous collaborative agent frameworks. The study proposes a role-playing collaborative agent framework that guides dialogue agents to complete tasks through embedded prompts while maintaining alignment with human intentions. The CAMEL framework generates dialogue data to study behaviors and capabilities within the agent society, the study further enhanced agent performance by fine-tuning the LLaMA-7B model, validating the effectiveness of generated datasets in enhancing LLM capabilities.
~\cite{liu2023bolaa} investigates the comprehensive comparison of LLM-augmented autonomous agents and proposes a new multi-agent coordination strategy for solving complex tasks through efficient communication and coordination called BOLAA. The experiment showed that the BOLAA outperforms other agent architectures in the WebShop environment especially in high-performance LLMs The above three studies focus on achieving a multi-agent collaboration architecture by increasing the number of agents. This trend indicates that more frameworks are beginning to explore the potential of multi-agent systems.~\cite{li2024agentsneed} explores methods to enhance LLMs performance by increasing the number of agents. Using sampling and voting methods, the study showed that as the number of agents increased, LLM performance in arithmetic reasoning, general reasoning, and code generation tasks improved significantly. This method proves the effectiveness of multi-agent collaboration in enhancing model performance. These studies collectively indicate the importance of multi-agent collaboration and dialogue frameworks in autonomous learning and decision-making tasks. Compared to traditional LLMs, these multi-agent frameworks enhance reasoning accuracy under zero-shot learning and demonstrate higher autonomy and flexibility which reduce the burden on developers.

LLM-based agents not only perform complex data analysis tasks but also demonstrate potential in simulating and understanding human trust behaviors.~\cite{lu2024selfselfevolutionlanguagefeedback} introduces a framework named SELF, designed to achieve self-evolution of LLMs through language feedback, use RLHF to train agent behavior to meet the human alignment. The framework enhances model capabilities through iterative processes of self-feedback and self-improvement without human intervention. In experiments, the test accuracy on GSM8K and SVAMP datasets increasing by 6.82\% and 4.9\%, respectively and the overall task win rates on the Vicuna test set and Evol-Instruct test set also increased by 10\% and 6.9\%.
Another similar study exploring the potential of LLM-based agents to simulate the human trust behaviors.
~\cite{xie2024can} also examines whether LLM-based agents can simulate human trust behaviors. The study aims to determine if LLM-based agents exhibit trust behaviors similar to humans and explore whether these behaviors can align with human trust. Through a series of trust game variants such as initial fund allocation and return trust games, the research analyzes LLM-based agents' trust decisions and behaviors in different contexts. Results show that particularly for GPT-4, LLM-based agents exhibit trust behaviors consistent with human expectations in these trust games, validating the potential of LLM-based agents in simulating human trust behaviors. The efficient and accurate handling of diverse datasets highlights the broad application prospects in fields such as software engineering. In terms of simulating trust behaviors, LLM-based agents demonstrate human-like behavior patterns through complex trust decisions and behavior analysis providing an important theoretical foundation for future human-machine collaboration and human behavior simulation. 

\textbf{Tooling and Engineering Frameworks for Agents.} Integrating LLMs into agents allows for more complex task processing.~\cite{liu2024agentlite} proposes a lightweight user-friendly library named AgentLite whic designed to simplify the development, prototyping and evaluation of task-oriented LLM-based agents systems. The main goal of the study is to enhance the capabilities and flexibility of LLM-based agents in various applications by introducing a flexible framework. This framework enhances task-solving capabilities through task decomposition and multi-agent coordination, using a hierarchical multi-agent coordination approach where a managing agent supervises the task execution of each agent.~\cite{zhuge2024language} introduces a framework, GPTSwarm, that represents LLM-based agents as computational graphs to unify existing prompt engineering techniques and introduces methods for optimizing these graphs to enhance agent performance. The study verifies the effectiveness of the framework through various benchmarks such as MMLU, Mini Crosswords, and HumanEval. The framework demonstrated significant performance improvements on the GAIA benchmark with an improvement margin of up to 90.2\% compared to the best existing methods. ~\cite{bassner2024iris} presents Iris, an LLM-based virtual tutor embedded within the Artemis learning platform, designed to support students in programming exercises. The system integrates context-awareness by automatically gathering code, problem descriptions, and test feedback, enabling it to provide relevant and personalized assistance without requiring manual input from students, empirical results show that Iris effectively enhances learning engagement and comprehension, while students maintain confidence to work independently in exams. This study represents a practical instantiation of LLM-based tool frameworks in real-world software education environments. 
~\cite{patil2024goex} presents \textbf{GoEx}, an open-source runtime specifically designed for executing actions generated by LLM agents in a secure, reversible, and isolated manner. GoEx targets the infrastructure-level trust and reliability challenges of autonomous LLM deployment in real-world systems. The core concept proposed is \textit{post-facto validation}, allowing agents to act first while enabling undo and damage-confinement mechanisms for safety. GoEx supports actions across RESTful APIs, databases, and file systems, providing containerized execution environments, credential vaults, reversion sets, and blast-radius control. Additionally, agents have shown strong capabilities in autonomous learning and decision-making in software engineering and security, which will be introduced in the subsequent Software Security section \ref{cha:autonomous} ~\cite{feldt2023towards,happe2023getting,ma2024combining,fang2024llm}. 

\textbf{Bridging Code Generation and Autonomous Learning.}
While the majority of works in this section emphasize decision-making, feedback, or learning strategies, several frameworks also demonstrate strong capabilities in real-world code generation tasks, thus forming natural intersections between \textit{Autonomous Learning and Decision Making} and \textit{Code Generation and Software Development}. For example, \textbf{CodeAct}~\cite{wang2024executable}, this approach goes beyond traditional code generation by enabling LLMs to reason through control flows and conditionals, showing strong decision-making ability. Similarly, \textbf{CODEAGENT}~\cite{zhang2024codeagentenhancingcodegeneration} and the multi-agent Agile framework~\cite{manish2024autonomous} both automates not only code creation but also iterative task decomposition and coordination, blending collaborative decision-making with procedural learning. Another representative cross-topic work is \textbf{ToolCoder}~\cite{ding2025toolcoder}, which reformulates tool learning as a code generation problem and systematically integrates software engineering principles into the agent planning and execution pipeline. Rather than relying on natural language chains-of-thought, ToolCoder converts user tasks into structured Python function scaffolds, enabling modular task decomposition, tool selection, and precise action planning. Its modular code-centric design demonstrates strong capabilities in both software-oriented code execution and adaptive decision making under uncertainty, bridging the gap between traditional tool usage and agent-style autonomous behavior.

\subsection{Analysis}
Overall, LLMs and LLM-based agents exhibit strong capability on the autonomous learning and decision making but slightly different view. These differences are reflected in the focus of task execution and also in autonomy, interactivity, learning and adaptation mechanisms, and the integration with other systems and modalities. From the perspective of task execution focus, LLMs primarily concentrate on enhancing specific functions in software engineering, such as debugging, problem-solving and automated reasoning. The tasks they perform are usually static and well-defined, such as automatic debugging, enhancing debugging capabilities to autonomously identify and correct errors, evaluating creativity and judging responses from other chatbots. In contrast, LLM-based agents not only focus on specific tasks but also manage multiple tasks simultaneously, often involving dynamic decision-making and interaction with other agents or systems. Examples of these agents' tasks include enhancing reasoning through multi-agent discussions, continuous learning from experiences, requiring real-time dynamic decision-making, and also LLM-based agents can get in touch with the multimodal task in the visual environment.

We can conclude that, the application of LLM-based agents in the topic of autonomous learning and decision-making primarily involves exploring their performance in specific tasks through various framework designs. These studies evaluate the agents' autonomy and decision-making capabilities to determine whether they align with human behavior and decision-making processes. If we dive into the specific task deigns, in terms of autonomy and interactivity LLMs are usually designed to perform highly specific tasks without needing to adapt to external input or environmental changes, they mainly operate as single models focusing on processing and responding within predefined boundaries, this also applied to all LLM applications. On the other hand, LLM-based agents exhibit higher autonomy which are typically designed to interact with or adapt to the environment in real-time, they are often part of multi-agent systems where collaboration and communication are key components, for example use extra model or tools to further help with the planning phases. 
In terms of integration with other systems and modalities, LLMs typically operate in text input-output scenarios and even in multi-modal settings, their role is usually limited to processing and generating text-based content. Also, LLM-based agents are more likely to integrate with other systems and modalities such as visual input or real-world perception data, enabling them to perform more complex and context-based decision-making tasks. 

\begin{figure}[htbp]
\scriptsize
    \centering
    \includegraphics[width=1\linewidth]{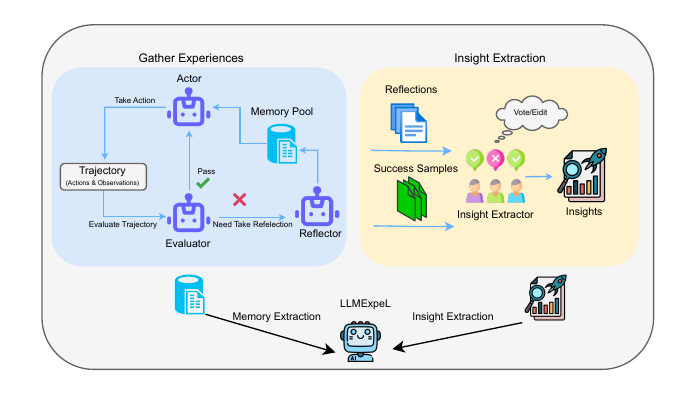}
    \caption{Expel\cite{zhao2024expel} Framework with Reflexion\cite{shinn2024reflexion} in Experience Gathering.}
    \label{fig:Expel}
\end{figure}
Regarding learning and adaptation mechanisms, LLMs' adaptation and learning are usually confined to the model's training data and parameter range, although they can adapt through new data updates, they lack the ability to continuously learn from real-time feedback, they are more focused on using existing knowledge to solve problems and generate responses. In contrast, LLM-based agents are often equipped with experiential learning and real-time feedback adaptation mechanisms, allowing them to optimize strategies and responses based on continuous interactions. One good example of LLm-based agents framework is Expel~\cite{zhao2024expel}, which utilize the previous researches ReAct~\cite{yao2022react} and Reflexion~\cite{shinn2024reflexion} as shown in Figure.~\ref{fig:Expel}. This framework utilizes a memory pool and insights pool to enable the LLM to learn from past knowledge, thereby aiding subsequent decision-making. This autonomous decision-making capability is something that traditional LLM frameworks cannot achieve.
\subsection{Benchmarks}
In the field of autonomous learning and decision-making, the benchmark datasets used by LLMs and LLM-based agents are quite similar in task handling and application requirements. We can gain a deeper understanding of the strengths and weaknesses of both approaches in different tasks and their application contexts. The specific dataset references, please see Table~\ref{tab:al_metrics}.

In the research on LLMs, the main datasets include Defects4J, MMLU, TransCoder, and MBPP. These datasets are primarily used to evaluate model performance in specific domains and tasks. Defects4J is a widely used in the software engineering, this software defect dataset containing 525 real defects from 17 Java projects. It's designed to test the effectiveness of automated program repair and defect detection tools by providing a standardized benchmark that allows researchers to compare the performance of different methods. MMLU (Massive Multitask Language Understanding) is a large-scale benchmark dataset covering 57 subjects, testing models on a broad spectrum of knowledge and reasoning abilities in multitask language understanding. It includes questions ranging from elementary education to professional level such as College Mathematics, Business Ethics, and College Chemistry, challenging the models' diverse knowledge base and reasoning capabilities. The TransCoder dataset focuses on code translation across programming languages which evaluate the model's ability to automatically translate code from one programming language to another. This is crucial for multilingual software development and maintenance, as it can greatly enhance development efficiency. MBPP (Mostly Basic Python Programming) has been introduced in previous section, it's a dataset containing 427 Python programming problems, covering basic concepts and standard library functions, it's widely used to test the model's performance in different programming scenarios, evaluating its ability to generate correct and efficient code.

In contrast, LLM-based agents use datasets that emphasize multitasking and decision-making capabilities in complex scenarios. The main datasets include HotpotQA, ALFWorld, FEVER, WebShop, and MGSM. HotpotQA is a multi-hop question-answering dataset that requires models to reference content from multiple documents when answering questions, evaluating their information synthesis and reasoning abilities, this dataset challenges the model's performance in complex reasoning tasks. ALFWorld is a text-based environment simulation dataset requiring multi-step decision-making where the model completes tasks in a virtual home environment. The dataset combines natural language processing and decision-making, evaluating the model's performance in dynamic and interactive tasks. The FEVER (Fact Extraction and VERification) dataset is used for fact verification tasks, where the model needs to verify the truthfulness of given statements and provide evidence, it assesses the model's capabilities in information retrieval and logical reasoning. WebShop is an online shopping environment simulation dataset containing 1.18 million real-world products and human instructions, it used to test the model's performance in complex decision-making tasks such as completing shopping tasks and attribute matching. MGSM (Multimodal Generalized Sequence Modeling) is a multimodal dataset containing tasks related to dialogue, creative writing, mathematical reasoning, and logical reasoning, evaluating the model's comprehensive abilities in multimodal tasks.

Comparatively, LLM datasets typically focus on single, static tasks such as code generation, mathematical reasoning and creative writing, which suitable for models working within predefined task scopes. Datasets like Defects4J, MMLU, and MBPP help evaluate model capabilities in specific domains. LLM-based agents are more suited for complex, multitasking, and dynamic environments where datasets require models to handle multimodal inputs and real-time decision-making, it can showcase their advantages in handling complex interactions and multitasking scenarios. Datasets like HotpotQA, ALFWorld, FEVER, and WebShop challenge the models' performance in information synthesis, dynamic decision-making/interaction and multimodal tasks. This difference arises from the distinct design goals of the two: LLMs aim to optimize performance on single tasks, while LLM-based agents are designed to handle complex or multi-modal task, this require higher autonomy and adaptability. It's also reflects modern applications' demand for highly interactive, adaptive, and multifunctional AI systems, driving the development from single LLM models to multi-agent systems. Through these analyses, we can identify the different application of LLMs and LLM-based agents in autonomous learning and decision-making, it's important to choose the appropriate framework to meet different task requirements in the real world applications.

\subsection{Evaluation Metrics}
various evaluation metrics are used in the research on LLMs and LLM-based agents, these metrics used to evaluate the models' performance in specific tasks and analyze their application effectiveness in this domain. Below, we discuss several representative studies analyzing the evaluation metrics they employed and exploring the differences between LLMs and LLM-based agents in this field.

In research on LLMs, evaluation metrics primarily focus on model accuracy and task completion. In~\cite{chen2024more}, researchers used the accuracy of a voting inference system which measured by the expected 0/1 loss (the proportion of correct responses) to assess model performance. This metric evaluates the accuracy of models through multiple calls, reflecting the ability of LLMs to improve result accuracy via iterative reasoning. Common evaluation metrics in the literature include accuracy and sample efficiency, accuracy refers to the proportion of correct predictions made by the model, while sample efficiency measures the number of samples required to achieve a certain accuracy level. These metrics assess both the predictive and decision making ability of the model and its data utilization efficiency during training. In~\cite{kang2023explainable}, evaluation metrics include possible patches, correct patches, precision, and developer accuracy. Possible patches refer to patches that pass all tests, while correct patches are semantically equivalent to the original developer patches. Precision measures the proportion of correct patches among the possible patches, and developer accuracy assesses the correctness of patches with and without explanations through human evaluation. These metrics emphasize the model's explanatory capability and practical effectiveness in automated code repair, increasing reliance on human evaluation. To assess model creativity, value, novelty and surprise are used as creativity dimensions. Quality, social acceptability, and similarity of generated works, as well as the ability to generate creative product, are also included in the evaluation.~\cite{yao2024tree} used the success rate in the Game of 24 and the coherence of generated paragraphs in creative writing as evaluation metrics. These metrics assess the model's performance in problem-solving and text generation, showcasing LLMs' potential in solving complex problems and generating coherent text. In~\cite{zheng2024judging}, consistency and success rate were used as evaluation metrics, the consistency calculates the probability of agreement between two judges on randomly selected questions which measures the alignment of LLM judges with human preferences. Success rate is used for specific tasks (such as the Game of 24) to measure the correct response rate.

In contrast, LLM-based agents use more diverse evaluation metrics to reflect their multi-agent collaboration characteristics. In~\cite{chen2023towards}, evaluation metrics include Perception Score (P-Score), Cognition Score (C-Score), and Action Score (A-Score). These metrics comprehensively assess the model's perception, cognition and action capabilities, demonstrating the comprehensive performance of LLM-based agents in handling multimodal tasks. In multimodal applications, success rate (SR) is often used as a primary metric, evaluated through tasks such as HotpotQA and FEVER to assess precise matching success. These metrics focus on task completion success and accuracy, showcasing the practical execution capabilities of LLM-based agents in different task environments. In~\cite{rasheed2024can}, evaluation metrics include practitioner feedback, efficiency, and accuracy. Practitioner feedback uses the Likert scale to collect satisfaction and performance feedback, the Likert scale is a commonly used psychometric tool designed to measure an individual's attitude or opinion toward a particular statement. The scale typically consists of the following five options: Strongly Disagree, Disagree, Neutral, Agree, Strongly Agree. While efficiency and accuracy are measured through the effectiveness of model-executed qualitative data analysis validated by practitioners. These metrics assess the agents' performance in qualitative data analysis, demonstrating their utility and accuracy in practical applications.

By comparing these metrics, we find that LLMs using traditional metrics such as accuracy and sample efficiency to assess their capabilities. In contrast, LLM-based agents handle more complex algorithm through multi-agents, which requires more comprehensive and diverse metrics to evaluate their performance from multiple directions. LLM-based agents in multimodal tasks and self-evolution tasks emphasize the integrated performance of perception, cognition, and action capabilities. This difference reflects LLMs' strengths in single-task optimization and LLM-based agents' potential in collaborative handling of complex tasks with higher capability of autonomous learning. Additionally, practical application evaluation metrics for LLM-based agents, such as practitioner feedback, efficiency, and accuracy, demonstrate their utility and user satisfaction in real-world scenarios. This evaluation approach assesses task completion but also consider a comprehensive evaluation of user experience, which can also evaluate the human alignment of their decision making capabilities. 

\begin{table*}
\centering
\caption{Evaluation Metrics in Autonomous Learning and Decision Making}
\scriptsize 
\label{tab:al_metrics}
\begin{tblr}{
  cell{1}{1} = {c},
  cell{1}{2} = {c},
  cell{1}{3} = {c},
  cell{2}{4} = {c},
  cell{3}{4} = {c},
  cell{4}{4} = {c},
  cell{5}{4} = {c},
  cell{6}{4} = {c},
  cell{7}{4} = {c},
  cell{8}{4} = {c},
  cell{9}{4} = {c},
  cell{10}{4} = {c},
  cell{11}{4} = {c},
  cell{12}{4} = {c},
  cell{13}{4} = {c},
  cell{14}{4} = {c},
  cell{15}{4} = {c},
  cell{16}{4} = {c},
  cell{17}{4} = {c},
  cell{18}{4} = {c},
  cell{19}{4} = {c},
  cell{20}{4} = {c},
  cell{21}{4} = {c},
  cell{22}{4} = {c},
  cell{23}{4} = {c},
  cell{24}{4} = {c},
  cell{25}{4} = {c},
  hlines,
  vlines,
}
Reference Paper                                                                                                                    & Benchmarks                                                                                             & Evaluation Metrics                                          & Agent \\
{~\cite{chen2024more}}                                                                                                & MMLU                                                                                                   & Accuracy                                                    & No    \\
{~\cite{chen2023teaching}}                                                                                 & Spider, TransCoder, MBPP                                                                               & Accuracy,~Sample Efficiency                                 & No    \\
{~\cite{kang2023explainable}}                                       & {Defects4J v1.2, Defects4J v2.0,\\Almost-Right HumanEval}                                              & {Plausible Patches,\\Correct Patches,\\Precision, Accuracy} & No    \\
{~\cite{franceschelli2023creativity}}                                                                                     & No Specific                                                                                            & Quality, Acceptance Rate                                    & No    \\
{~\cite{yao2024tree}}                                                      & {Game of 24, Creative Writing,\\5x5 Crosswords}                                                        & Success rate,~Coherency                                     & No    \\
{~\cite{zheng2024judging}}                                                                       & MT-Bench, Chatbot Arena                                                                                & {Agreement Rate,~Success Rate\\Human Judgement}             & No    \\
{~\cite{wang2024rethinking}}                                                 & ECQA, GSM8K, FOLIO-wiki                                                                                & Accuracy                                                    & Yes   \\
{~\cite{chen2023towards}} & PCA-EVAL                                                                                               & Accuracy,~P/C/A-Score                                       & Yes   \\
{~\cite{zhao2024expel}}                                                                                    & HotpotQA, ALFWorld, WebShop, FEVER                                                                     & Success Rate                                                & Yes   \\
{~\cite{li2024agentsneed}}                                                                                                        & GSM8K, MATH, MMLU, Chess, HumanEval                                                                    & Accuracy                                                    & Yes   \\
{~\cite{happe2023getting}}                                                            & MITRE ATTCK framework                                                                                  & Ability Identify Vulnerabilities                            & Yes   \\
{~\cite{lu2024selfselfevolutionlanguagefeedback}}                                                                                    & {GSM8K, SVAMP, Vicuna testset,\\Evol-Instruct testset}                                                 & Accuracy, Feedback Accuracy                                 & Yes   \\
{~\cite{shinn2024reflexion}}                                                                  & {HotPotQA, ALFWorld, HumanEval,MBPP, \\LeetcodeHardGym}                                                & Pass@1, Success Rate                                        & Yes   \\
{~\cite{rasheed2024can}}             & {Github Developer Discussions, BBC News, \\Social Media Conversations, \\In-depth Interviews}            & {Practitioner Feedback,\\Efficiency and Accuracy}           & Yes   \\
{~\cite{li2023camel}}                                           & {AI Society, Code, Math,Science, \\Misalignment}                                                       & {Human Evaluation,\\GPT-4 Evaluation}                       & Yes   \\
{~\cite{chen2023agentverse}}                                          & {FED, Commongen Challenge,\\MGSM, Logic Grid Puzzles,\\HumanEval}                                      & Pass@1, Task completion rate                                & Yes   \\
{~\cite{yao2022react}}                                                                     & HotpotQA, FEVER, ALFWorld, WebShop                                                                     & {Exact Match, Accuracy, \\Success rate, Average Score}      & Yes   \\
{~\cite{xie2024can}}                                                              & {Trust Game, Dictator Game,\\MAP Trust Game,\\Risky Dictator Game,\\Lottery Game, Repeated Trust Game} & {Valid Response Rate,\\Alignment}                           & Yes   \\
{~\cite{liu2024agentlite}}                                  & HotPotQA, WebShop                                                                                      & F1-Score, Average Reward                                    & Yes   \\
{~\cite{ma2024combining}}                         & 263 real smart contract vulnerabilities                                                                & {F1 Score, Accuracy\\Precision, Recall\\Consistency Rate.}  & Yes   \\
{~\cite{fang2024llm}}                                                                    & {15 real-world one-day vulnerabilities\\from CVE database}                                             & Success Rate, Cost                                          & Yes   \\
{~\cite{liu2023bolaa}}                                                        & WebShop, HotPotQA with Wikipedia AP                                                                    & Reward Score, Recall                                        & Yes   \\
{~\cite{zhuge2024language}}                                                                                              & {MMLU, Mini Crosswords, HumanEval, \\GAIA}                                                             & Accuracy,~Pass@1                                            & Yes   \\
{~\cite{patil2024goex}} & None & N/A (Design Perspective and Architectural Analysis) & Yes \\
{~\cite{bassner2024iris}} & Artemis Programming Exercises Dataset & {Correctness, Engagement, 
\\ Learning Confidence} & Yes \\
{~\cite{ding2025toolcoder}} & {RestBench (TMDB, Spotify), API-Bank, ToolAlpaca} & {Success\%, Path\%, Accuracy\%,\\Correctness\%, Procedure, Response} & Yes \\
\end{tblr}
\end{table*}

%% file: pages/design.tex
\section{Software Design and Evaluation \label{cha:softwareDesign}}
The application of LLMs to software design and evaluation has very similar overlaps with previous topics, software design is an early phase of software development, and the quality of the design directly impacts the quality of furture development. Modern software engineering methodologies emphasize the integration of design and development to ensure that decisions made during the design phase seamlessly translate into high-quality code. Consequently, the research on software design often explores aspects related to code generation and development by utilizing LLMs for software development with a certain framework and special architecture design. Software design frameworks often involve multiple stages of continuous refinement to achieve optimal results, which can be considered part of LLM applications in software development ~\cite{zheng2024opencodeinterpreter}. Similarly,~\cite{qin2023toolllm} and~\cite{schick2024toolformer} highlight the frequent use of tools or API interfaces when using LLMs to assist in development and design, demonstrating an overlap with the topic of code generation and software development.

LLMs in software design and evaluation also intersect extensively with autonomous learning and decision making, these two topics are interrelated fields. Software design needs to consider system adaptability and learning capabilities to handle dynamic environments, therefore design evaluations involving autonomous learning and decision making naturally become a focal point of intersection for these two topics. Many LLM techniques and methods find similar applications in both fields, for example LLMs based on reinforcement learning can be used for automated design decisions and evaluations, as well as for self-learning and optimization. Common applications of LLMs in software engineering involve fine-tuning models with prompt engineering techniques to continuously enhance performance particularly in software design and evaluation, more sample learning is often required to ensure that the model outputs align with user expectations~\cite{franceschelli2023creativity,lu2024selfselfevolutionlanguagefeedback,li2024agentsneed,rasheed2024can,zhuge2024language,wang2024rethinking}. Additionally, requirement elicitation and specification in requirement engineering can also be considered part of software design and evaluation~\cite{arora2024advancing,ataei2024elicitron}. This section reviews the main research achievements of LLMs in software design and evaluation in recent years, discussing their application scenarios and practical effects.

\subsection{LLMs Tasks}
\textbf{Software Task Evaluation and Assessment.} In recent years, there has been extensive research on the use of LLMs in tasks such as automation, optimization, and code understanding. ChatGPT has been widely utilized for various software engineering tasks and demonstrated excellent performance in tasks like log summarization, pronoun resolution, and code summarization, achieving a 100\% success rate in both log summarization and pronoun resolution tasks~\cite{sridhara2023chatgpt}. However, its performance on tasks such as code review and vulnerability detection is relatively poor, which shows that it needs further improvement for more complex tasks. 
Another framework EvaluLLM addresses the limitations of traditional reference-based evaluation metrics (such as BLEU and ROUGE) by using LLMs to assess the quality of natural language generation (NLG) outputs~\cite{desmond2024evalullm}. The EvaluLLM introduces a new evaluation method that compares generative outputs in pairs and uses win rate metrics to measure model performance, this approach can simplifies the evaluation process also ensures consistency with human assessments, showcasing the broad application prospects of LLMs in generative tasks. 
Similarly, in the LLMs evaluation domain, LLM-based NLG Evaluation provides a review and classification of current LLMs used for NLG evaluation, the paper summarizes four main evaluation methods: LLM-derived metrics, prompt-based LLMs, fine-tuned LLMs, and human-LLM collaborative evaluations~\cite{gao2024llm}. These methods demonstrate the potential of LLMs in evaluating generative outputs which also mention challenges such as the need for improved evaluation metrics and further exploration of human-LLM collaboration. 

\textbf{Software and Hardware Design Automation.} There are also many novel application design with the LLMs which applied in the engineering design, one study explores strategies for software/hardware co-design to optimize LLMs and applies these strategies to design verification~\cite{wan2024software}. Through quantization, pruning, and operation-level optimization, this research demonstrates applications in high-level synthesis (HLS) design functionality verification, GPT-4 was used to generate high-level synthesis (HLS) designs containing predefined errors to create a dataset called Chrysalis, this dataset provides a valuable resource for evaluating and optimizing LLM-based HLS debugging assistants. The optimized LLM significantly improves inference performance, providing new possibilities for error detection and correction in the electronic design automation (EDA) field. 
In~\cite{kolthoff2023data}, the researchers introduces RaWi, a data-driven GUI prototyping approach. The framework allows users to retrieve GUIs from this repository, edit them, and create new high-fidelity prototypes quickly. The experiment conducted by comparing RaWi with a traditional GUI prototyping tool (Mockplus) to measure how quickly and effectively users can create prototypes. The result demonstrated that RaWi outperformed on multiple benchmarks, with 40\% improvement on precision@k metric. This study proves the possibility of LLMs to improve the efficiency during prototyping phase of software design, which allows designers to quickly iterate on GUI designs, facilitating early detection of design flaws.

\textbf{Software Engineering Education.} With the new possibility brought by the LLMs, there has been much discussion in the education field, with researchers exploring the implications of the prevalence of large language models for education~\cite{kirova2024software}. Study indicates that ChatGPT shows significant potential but some limitations in answering questions from software testing courses~\cite{jalil2023chatgpt}. ChatGPT was able to answer about 77.5\% of the questions and provided correct or partially correct answers 55.6\% of the time. However, the correctness of its explanations was only 53.0\%, indicating the need for further improvement in educational applications. 

\subsection{LLM-based Agents Tasks}
\textbf{Assessment and Limitation Studies in LLM-Based Agents.} The application of LLM-based agents in software design and evaluation enhance the development efficiency and code quality, as well as showcase the broad applicability and immense potential of LLM-based agents in practical software engineering tasks.~\cite{suri2023software} explores the current capabilities, challenges, and opportunities of autonomous agents in software engineering. Study evaluate Auto-GPT's performance across different stages of the software development lifecycle (SDLC), including software design, testing, and integration with GitHub, the paper finds that detailed contextual prompts significantly enhance agent performance in complex software engineering tasks which mentions the importance of context-rich prompts in reducing errors and improving efficiency, underscoring the potential of LLM-based agents to automate and optimize various SDLC tasks, thereby enhancing development efficiency. This paper also evaluate the limitation of the Auto-GPT, 
includes task or goal skipping, generating unnecessary code or files (hallucinations), repetitive or looping responses, lack of task completion verification mechanisms. These limitations can lead to incomplete workflows, inaccurate outputs, and unstable performance in practical applications.

\textbf{Multi-Agent Collaboration for Software Engineering.} ~\cite{qian2023communicative} introduces ChatDev, the first virtual chat-driven software development company, a concept of using LLMs not just for specific tasks but as central coordinators in a chat-based, multi-agent framework. this approach allows for more structured, efficient, and collaborative software development processes, exploring how chat-driven multi-agent systems can achieve efficient software design and evaluation, reduce code vulnerabilities, and enhance development efficiency and quality. Experiments show that ChatDev can design and generate software in an average of 409.84 seconds at a cost of only \$0.2967 while significantly reducing code vulnerabilities. This indicates that chat-based multi-agent frameworks capable to improve software development efficiency and quality. ~\cite{chen2024llmarena} proposes the LLMARENA benchmark framework to evaluate LLMs' capabilities in dynamic multi-agent environments, the idea is similar to the ChatDev but innovates by shifting the focus from single-agent static tasks to dynamic and interactive multi-agent environments, providing a more realistic and challenging setting to assess the practical utility of LLMs, this approach mirrors real-world conditions where multiple agents (either AI or human) interact and collaborate. Experiments show that this framework can test LLMs' spatial designing, strategic planning, and teamwork abilities in gaming environments, offering new possibilities and tools for designing and evaluating LLMs in multi-agent systems. Similarly,~\cite{vallecillos2024agent} explores the application of LLM-based agents in software maintenance tasks, improving code quality and reliability through a collaborative framework. This study should origin be categorized under the software maintenance domain but exhibit the iterative manner of the design structure. The framework utilize the task decomposition and multi-agent strategies to tackle complex engineering tasks that traditional one-shot methods cannot handle effectively, multiple agents can learn from each other, leading to improved software maintenance outcomes. Experiments show that multi-agent systems outperform single-agent systems in complex debugging tasks, indicating that this new framework can be applied in software design to provide safer architectures.

\textbf{Agent-Orchestrated Software Design and Development.} ~\cite{josifoski2023flows} introduces the "Flows" conceptual framework for structuring interactions between AI models and humans to improve reasoning and collaboration capabilities. The study present the idea of conceptualizing processes as independent, goal-driven entities that interact through standardized message-based interfaces, enabling a modular and extensible design. This approach is inherently concurrency-friendly and supports the development of complex nested AI interactions without having to manage complex dependencies. Experiments in competitive coding tasks show that the "Flows" framework increases the AI model's problem-solving rate by 21 percentage points and the human-AI collaboration rate by 54 percentage points. This demonstrates how modular design can enhance AI and human collaboration, thereby improving the software design and evaluation process. Another similar collaboration framework introduced by Microsoft research team, ~\cite{shen2024hugginggpt} demonstrates the effectiveness of using LLMs, particularly ChatGPT as agent's controllers to manage and execute various AI tasks. The HuggingGPT system that uses ChatGPT to orchestrate the execution of tasks by various AI models available in Hugging Face, the purpose is to test how effectively the system can handle complex AI tasks, including language, vision, and speech tasks, by executing appropriate models based on user requests. The innovation lies in using LLMs not just as tools for direct task execution but as central orchestrators that leverage existing AI models to fulfill complex tasks, This approach expands the practical applicability of LLMs beyond typical language tasks. ~\cite{weber2024large} presents a new taxonomy to structurally understand and analyze LLM-integrated applications, providing new theories and methods for software design and evaluation. This taxonomy helps in understanding the integration of LLM components in software systems, laying a theoretical foundation for developing more effective and efficient LLM-integrated applications. 

\subsection{Analysis}
Overall, LLM applications in software design and evaluation typically focus on the automation of specific tasks, such as code generation and log summarization, with a tendency towards evaluation the capability rather than implementation during the design phases. The process of software design is largely intertwined with software development and requirements engineering. As previously mentioned, the use of LLMs to assist in software development often includes aspects of the software design process, particularly in generating related design documentation. Therefore, there is relatively limited research focused on using LLMs for higher-level software design tasks.

LLM-based agents expand the capabilities of LLMs by handling more complex workflows through intelligent decision-making and task execution, these agents can collaborate, dynamically adjust tasks and gather and utilize external information. In software design and evaluation, a single model often cannot comprehensively consider both design and evaluation aspects, which is why more software developers are reluctant to entrust high-level tasks to AI. LLM-based agents, through collaborative work and more refined role division, can efficiently complete design tasks and adapt to various application scenarios. However, the application of LLM-based agents in software design is commonly included in the software development, like previously discussed, the self-reflection and reasoning before action occurs during the software design phases. The Chatdev\cite{qian2023communicative} framework uses role distribution to create a separate software design phase which significantly increases the flexibility and accuracy in the later development phases. In terms of efficiency and cost, LLMs are still slightly superior to LLM-based agents in text generation and vulnerability detection. However, handling tasks similar to software maintenance and root cause analysis requires more complex architectures, such as multi-turn dialogues, knowledge graphs, and RAG techniques, which can further benefit the design and evaluation phases.

\subsection{Benchmarks}
The benchmarks include public datasets and datasets self-crafted by the authors themselves, and the application scenarios are also quite differently as shown in the Table~\ref{tab:sd_metrics}. BigCloneBench is a benchmark dataset for code clone detection, containing a large number of Java function pairs. These pairs are classified as clones and non-clones, used for training and evaluating clone detection models, with the main evaluation metric being the correct identification rate. The Chrysalis dataset created by~\cite{wan2024software}, it contains over 1000 function-level designs from 11 open-source synthesizable HLS datasets, primarily used to evaluate the effectiveness of LLM debugging tools in detecting and correcting injected errors in HLS designs, with the main evaluation metric being the effectiveness of error detection and correction. The CodexGLUE dataset is a comprehensive benchmark dataset covering various code generation and understanding tasks such as code completion, code repair, and code translation, used to evaluate the performance of code generation models in practical programming tasks. In addition to these public datasets, some artificially simulated datasets are used, such as a simulated job fair environment dataset. This dataset simulates a virtual job fair environment containing multiple task scenarios such as interviews, recruitment, and team project coordination. The dataset used to evaluate the coordination capabilities of generative agents in complex social tasks, with the main evaluation metrics being task coordination success rate and role matching accuracy.

Comparatively, LLMs research tends to use specific and publicly available datasets, such as BigCloneBench. These datasets provide standardized evaluation benchmarks, aiding in the reproducibility and comparability of results. Researches on LLM-based agents tends to use customized experimental settings or unspecified datasets, such as requirement documentations, without specifying particular datasets but emphasizing that the experiments involve 70 user requirements. This choice is usually because the research needs to evaluate the performance from multiple angles, and it is difficult to perfectly adapt to the vertical application scenarios if some general datasets are used. Both LLM and LLM-based agents use a variety of datasets to evaluate the performance of the model, these datasets cover tasks ranging from code generation, code understanding, to natural language generation and task management, due to the topic of software design and evaluation is relatively inter-related with others. However, because the LLM-based agents can be expanded to application scenarios such as videos and pictures, the agents like Auto-GPT and HuggingGPT also use multimodal datasets. These datasets not only contain code and text, but also involve multiple data types such as images and speech. Moreover, compared with a single LLM framework, LLM-based agents need to evaluate more areas, so benchmarks also need to be considered separately. For example, LLMARENA is specially designed to test the performance of LLM in dynamic, multi-agent environments, covering complex tasks such as spatial reasoning, strategic planning, and risk assessment.

\subsection{Evaluation Metrics}
In Software Design and Evaluation, various studies have employed different evaluation metrics to measure the performance of LLMs and LLM-based agents across a range of tasks. Both LLM and LLM-based agent research use more than one metrics to comprehensively assess model performance, LLMs research tends to focus on traditional metrics such as accuracy, win rate, and consistency, while LLM-based agent research still consider those fundamental metrics but further introduces complex evaluation methods, such as task coordination success rate and role matching accuracy. However, it cannot be definitively stated that future LLM-based agent research will always use more flexible evaluation metrics considering multiple dimensions, but more dependent on the specific task and dataset being used. The reason for this phenomenon, as observed in this survey, is primarily that tasks in LLMs research are relatively single-tasked, mainly focusing on static tasks such as log summarization with traditional evaluation methods. On the other hand, LLM-based agent research involves more general multi-agent tasks, and its evaluation methods emphasize interactivity and dynamics. LLM-based agent research focuses more on the model's collaboration and decision-making capabilities by using multi-dimensional evaluation metrics to comprehensively assess their potential in practical applications consider not only the accuracy. This explains why, despite the similarity in evaluation metrics such as accuracy and completion time, LLM-based agents use flexible evaluation metrics, including metrics like mutual exclusiveness and appropriateness.
\begin{table*}
\centering
\caption{Evaluation Metrics in Software Design and Evaluation}
\scriptsize 
\label{tab:sd_metrics}
\begin{tabular}{|l|l|l|c|} 
\hline
\multicolumn{1}{|c|}{Reference Paper}                                                                                                                                             & \multicolumn{1}{c|}{Benchmarks}                                                                                                                      & \multicolumn{1}{c|}{Evaluation Metrics}                                                                                                                             & \multicolumn{1}{l|}{Agent}  \\ 
\hline
\begin{tabular}[c]{@{}l@{}}{~\cite{sridhara2023chatgpt}}\end{tabular}                                                             & \begin{tabular}[c]{@{}l@{}}BigCloneBench,~\\Python functions,\\Java methods,~\\Random logs,~\\Bug reports,~\\Requirement specifications\end{tabular} & Accuracy~ ~                                                                                                                                                         & No                          \\ 
\hline
\begin{tabular}[c]{@{}l@{}}{~\cite{desmond2024evalullm}}\end{tabular}                                                                                 & Not Specified                                                                                                                                        & Win rate, Agreement score                                                                                                                                           & No                          \\ 
\hline
\begin{tabular}[c]{@{}l@{}}{~\cite{gao2024llm}}\end{tabular}                                                                                 & Not Specified                                                                                                                                        & \begin{tabular}[c]{@{}l@{}}Embedding-based metrics, \\probability-based metrics, \\Comparison, Ranking~\end{tabular}                                                & No                          \\ 
\hline
\begin{tabular}[c]{@{}l@{}}{~\cite{wan2024software}}\end{tabular}                                                       & Chrysalis                                                                                                                                            & Effectiveness~ ~                                                                                                                                                    & No                          \\ 
\hline
\begin{tabular}[c]{@{}l@{}}{~\cite{cheng2023batch}}\end{tabular}                                                                     & \begin{tabular}[c]{@{}l@{}}CommonsenseQA,\\~StrategyQA, GSM8K\end{tabular}                                                                           & \begin{tabular}[c]{@{}l@{}}Accuracy, \\Token, Time costs\end{tabular}                                                                                               & No                          \\ 
\hline
\begin{tabular}[c]{@{}l@{}}{~\cite{jalil2023chatgpt}}\end{tabular}                                                                              & \begin{tabular}[c]{@{}l@{}}31 Questions from \\software testing textbook.\end{tabular}                                                               & Correctness,~Effectiveness~ ~                                                                                                                                       & No                          \\ 
\hline
\begin{tabular}[c]{@{}l@{}}{~\cite{shankar2024validates}}\end{tabular}                                  & \begin{tabular}[c]{@{}l@{}}Medical transcripts,\\Amazon Product \\Descriptions\end{tabular}                                                          & \begin{tabular}[c]{@{}l@{}}Coverage, \\False Failure Rate\\Alignment.\end{tabular}                                                                                  & No                          \\ 
\hline
\begin{tabular}[c]{@{}l@{}}{~\cite{kolthoff2023data}}\end{tabular}                                                                                            & Rico~                                                                                                                                       & \begin{tabular}[c]{@{}l@{}}Precision@k,\\NDCG@k, \\Mean Reciprocal Rank,\\Average Precision, HITS@k\end{tabular}                                                                     & No                          \\ 
\hline
\begin{tabular}[c]{@{}l@{}}{~\cite{suri2023software}}\end{tabular}                                                                         & Not Specified                                                                                                                                        & \begin{tabular}[c]{@{}l@{}}Accuracy, Success rate, \\Consistency, Effectiveness\end{tabular}                                                                        & Yes                         \\ 
\hline
\begin{tabular}[c]{@{}l@{}}{~\cite{shen2024hugginggpt}}\end{tabular}                                                             & \begin{tabular}[c]{@{}l@{}}Hugging Face's \\Model Repository.\end{tabular}                                                                           & \begin{tabular}[c]{@{}l@{}}Accuracy, \\Precision, \\Recall, \\F1-Score, \\Edit Distance, \\GPT-4 Score, \\Passing Rate, \\Rationality, \\Success Rate.\end{tabular} & Yes                         \\ 
\hline
\begin{tabular}[c]{@{}l@{}}{~\cite{josifoski2023flows}}\end{tabular}                                                                                & Codeforces, LeetCode~                                                                                                                                & Pass@1                                                                                                                                                              & Yes                         \\ 
\hline
\begin{tabular}[c]{@{}l@{}}{~\cite{chen2024llmarena}}\end{tabular}                                                                                           & Seven different game environments.                                                                                                                                & \begin{tabular}[c]{@{}l@{}}Win Rate, ErrorRate\end{tabular}                                                                                & Yes                         \\ 
\hline
\begin{tabular}[c]{@{}l@{}}{~\cite{qian2023communicative}}\end{tabular}                                                                                          & Codeforces~                                                                                                                                          & \begin{tabular}[c]{@{}l@{}}Comprehensiveness, \\Robustness, Conciseness, \\Mutual exclusiveness, \\Explanatory power, \\Extensibility.\end{tabular}                 & Yes                         \\ 
\hline
\begin{tabular}[c]{@{}l@{}}{~\cite{weber2024large}}\end{tabular}                                               & Sample Applications.                                                                                                                                 & BERTScore, BLEU                                                                                                                                                     & Yes                         \\ 
\hline
\begin{tabular}[c]{@{}l@{}}{\cite{vallecillos2024agent}}\end{tabular}                                                                                             & CodexGLUE~ ~                                                                                                                                         & \begin{tabular}[c]{@{}l@{}}BLEU, METEOR, \\ROUGE-L, BERTScore\end{tabular}                                                                                          & Yes                         \\ 
\hline
\begin{tabular}[c]{@{}l@{}}{~\cite{roy2024exploring}}\end{tabular}                                                                                      & Production Incidents                                                                                                                                 & \begin{tabular}[c]{@{}l@{}}Success rate,\\Accuracy, Alignment,\\Appropriateness\end{tabular}                                                                        & Yes                         \\ 
\hline
\begin{tabular}[c]{@{}l@{}}{~\cite{li2023metaagents}}\end{tabular} & \begin{tabular}[c]{@{}l@{}}Simulated Job Fair \\Environment\end{tabular}                                                                             & \begin{tabular}[c]{@{}l@{}}Completion time,\\Task Progress, \\Understanding Level\end{tabular}                                                                      & Yes                         \\
\hline
\end{tabular}
\end{table*}

%% file: pages/test.tex
\section{Software Test Generation \label{cha:softwareTest}}
In software development, a crucial component is software testing, which needs to be conducted continuously from the initial system development to the final deployment. In industry, agile development is commonly used which tests the system continuously at every stage to ensure the robustness of the entire system, whenever new code is committed to GitHub, tests are conducted to ensure the usability of the updated version. A common approach is to use Jenkins\footnote{\href{https://www.jenkins.io/}{https://www.jenkins.io/}} to achieve continuous integration and continuous deployment. Jenkins automatically hooks into the developer's action of pushing code to GitHub and runs a test suite against the new version. Although the entire process leans towards automated development, creating and refining test cases still requires a large human effort. 

Typical roles in development involve software testing, such as writing unit tests, integration tests, and fuzz tests. Researchers have been attempting to use AI to help generate test cases since before 2000. Initial implementations typically involved simpler forms of AI and machine learning to automate parts of the test case generation process. Over time, more sophisticated methods such as natural language processing and machine learning models have been applied to improve the precision and scope of test case generation. Online tools like Sofy\footnote{\href{https://sofy.ai/}{https://sofy.ai/}}, which use machine learning to generate context-based paths in applications, also exist to aid in generating test suites. Using large language models to generate test cases is a relatively new attempt but has been developed rapidly. In 2020, researchers utilized pre-trained language models fine-tuned on labeled data to generate test cases. They developed a sequence-to-sequence transformer-based model called "ATHENATEST" and compared its generated results with EvoSuite and GPT-3, demonstrating better test coverage~\cite{tufano2020unit}. More research and models are being dedicated to test suite generation experiments, for instance, the Codex model~\cite{chen2021evaluatinglargelanguagemodels}, mentioned earlier in the code generation section, combined with chain-of-thought prompting, achieved high-quality test suite generation with CodeCoT, even in zero-shot scenarios. The introduction of LLMs aims to automate and streamline the testing process, making it more rigorous and capable of addressing aspects that humans might easily overlook.

\subsection{LLMs Tasks}
\textbf{Security and Vulnerability Test Generation.} The application of LLMs in software test generation is extensive and encompasses more than just test suite generation. The reviewed paper included in this survey covers several aspects, including security test generation, bug reproduction, general bug reproduction, fuzz testing, and coverage-driven test generation. These tasks are achieved through various models and techniques, significantly improving software quality and reducing developers' workload. ~\cite{zhang2023well} aims to evaluate the effectiveness of using GPT-4 to generate security tests, demonstrating how to conduct supply chain attacks by exploiting dependency vulnerabilities. The study experimented with different prompt styles and templates to explore the effectiveness of varying information inputs on test generation quality, the results showed that tests generated by ChatGPT successfully discovered 24 proof-of-concept vulnerabilities in 55 applications, outperforming existing tools TRANSFER\cite{kang2022test} and SIEGE\footnote{\href{https://siegecyber.com.au/services/penetration-testing/}{https://siegecyber.com.au/services/penetration-testing/}}. This research introduces a new method for generating security tests using LLMs and provides empirical evidence of LLM's potential in the security testing domain, offering developers a novel approach to handling library vulnerabilities in applications.~\cite{fakhoury2024llm} introduces TICODER, an interactive framework that leverages test generation to clarify user intent and improve code accuracy, this work exemplifies the synergy between software testing and code synthesis, offering a novel path toward human-in-the-loop development with LLMs.


\textbf{Automated Bug Reproduction from Natural Language Reports.} Another application is bug reproduction, which allows testers to locate and fix bugs more quickly and efficiently.~\cite{feng2024prompting} addresses the limitations of current bug reproduction methods, which are constrained by the quality and clarity of handcrafted patterns and predefined vocabularies. The paper proposes and evaluates a new method framework called AdbGPT, which uses a large language model to automatically reproduce errors from Android bug reports. AdbGPT is described as outperforming current SOTA approaches in the context of automated bug replay for only Android system. The experimental results show that AdbGPT achieved accuracies of 90.4\% and 90.8\% in S2R entity extraction and a success rate of 81.3\% in error reproduction, significantly outperforming the baseline ReCDroid and ablation study versions. By introducing prompt engineering, few-shot learning, and chain-of-thought reasoning, AdbGPT demonstrates the powerful capabilities of LLMs in automated error reproduction. It also uses GUI encoding to convert the GUI view hierarchy into HTML-like syntax, providing LLMs with a clear understanding of the current GUI state. While AdbGPT is specialized for Android systems,~\cite{kang2023large} proposes the LIBRO framework, which uses LLMs to generate bug reproduction tests from bug reports. The experimental results show that LIBRO successfully reproduced 33.5\% of bugs in the Defects4J dataset and 32.2\% in the GHRB dataset. By combining advanced prompt engineering and post-processing techniques, LIBRO demonstrates the effectiveness and efficiency of LLMs in generating bug reproduction tests. Although LIBRO has a lower absolute effectiveness compared to AdbGPT, it was tested across a more diverse set of Java applications and not limited to Android. Therefore, while AdbGPT excels in specialized bug replay for Android, LIBRO provides a wider range of bug reproduction for Java applications. The extensive application of LLMs in test generation tasks such as security test generation, bug reproduction, fuzz testing, program repair, and coverage-driven test generation highlights their significant potential in improving software quality and reducing the burden on developers. Through various models and techniques, these tasks demonstrate how LLMs can automate and enhance the software testing process, addressing aspects that are often overlooked by humans.

\textbf{Universal and Language-Agnostic Fuzz Testing.} Similarly, in fuzz testing, LLMs have shown promise potential.~\cite{xia2024fuzz4all} developed a universal fuzzing tool, Fuzz4All, which uses LLMs to generate and mutate inputs for various software systems. This tool addresses the issues of traditional fuzzers being tightly coupled with specific languages or systems and lacking support for evolving language features. The study conducted various experiments to test the tool's capabilities, including coverage comparison, bug finding, and targeted fuzzing. The results showed that Fuzz4All achieved the highest code coverage in all tested languages, with an average increase of 36.8\%, and discovered 98 bugs across nine systems, which considered as state-of-art technique in universal fuzzing with LLMs at that time. Through self-prompting and LLM-driven fuzzing loops, Fuzz4All demonstrated the effectiveness of LLMs in fuzz testing and showcased their capability across multiple languages and systems under test (SUTs) through comprehensive evaluations.

\textbf{Coverage-Driven Test Generation.} ~\cite{ryan2024code} introduced SymPrompt, a new code-aware prompting strategy aimed at addressing the limitations of existing Search-Based Software Testing (SBST) methods and traditional LLM prompting strategies in generating high-coverage test cases. By decomposing the original test generation process into a multi-stage sequence aligned with the execution paths of the method under test, SymPrompt generated high-coverage test cases. Experimental results indicated that SymPrompt increased coverage on CodeGen2 and GPT-4 by 26\% and 105\% respectively. Through path constraint prompting and context construction techniques, SymPrompt demonstrated the potential of LLMs in generating high-coverage test cases.~\cite{pizzorno2024coverup} also focused on test suite coverage, this study introduced the COVERUP system which generates high-coverage Python regression tests through coverage analysis and interaction with LLMs. The experimental results showed that COVERUP increased code coverage from 62\% to 81\% and branch coverage from 35\% to 53\% through iterative prompting and coverage-driven methods.~\cite{liu2024llm} proposed the AID method, which combines LLMs with differential testing to improve fault detection in "plausibly correct" software. By comparing the effectiveness of AID in generating fault-revealing test inputs and oracles, the experiments showed that AID improved recall and precision by 1.80 times and 2.65 times respectively, and increased the F1 score by 1.66 times. By integrating LLMs with differential testing, AID showcased the powerful capability of LLMs in detecting complex bugs. Beyond proposing new prompting strategies, \cite{tang2024chatgpt} conducts a comprehensive comparison between ChatGPT and EvoSuite\footnote{\href{https://github.com/EvoSuite/evosuite}{https://github.com/EvoSuite/evosuite}}(a SOTA SBST tool) in generating unit test suites. The evaluation spans correctness, readability, code coverage, and bug detection. ChatGPT achieved 55.4\% average statement coverage across 207 Java classes, while EvoSuite reached 74.2\%, outperforming ChatGPT in most cases. Nonetheless, ChatGPT showed advantages in generating diverse test data, semantically rich call chains, and context-aware inputs, outperforming EvoSuite in 17.9\% of the classes. The study suggests that while LLMs may lack feedback mechanisms, they offer strong generalization and usability, making them suitable as lightweight, accessible test generation tools and potential complements to SBST.

\subsection{LLM-based Agents Tasks}
\textbf{Collaborative Multi-Agent Systems for Test Case Generation.} In the field of software test generation, the application of LLM-based agents demonstrates their potential in automated test generation. While relying on LLM-based agents for software test generation might seem excessive, more research is directed towards vulnerability detection and system maintenance. LLM-based agents can enhance test reliability and quality by distributing tasks such as test generation, execution, and optimization through a multi-agent collaborative system. These multi-agent systems offer obvious improvements in error detection and repair, and coverage testing. An example of such a system is AgentCoder's multi-agent framework, as discussed in the code generation and software development section~\cite{huang2023agentcoder}. The primary goal of this system is to leverage multiple specialized agents to iteratively optimize code generation, overcoming the limitations of a single agent model in generating effective code and test cases. The paper introduce the test design agent, which creates diverse and comprehensive test cases; and the test execution agent, which executes the tests and provides feedback, it reached an 89.9\% pass rate on the MBPP dataset. Similarly, the paper collected for the software test generation topic are mostly multiple agents based system. The study~\cite{li2024large} evaluates the effectiveness of LLMs in generating high-quality test cases and identifies their limitations. It proposes a novel multi-agent framework called TestChain. The paper evaluates StarChat, CodeLlama, GPT-3.5, and GPT-4 on the HumanEval and LeetCode-hard datasets. Experimental results show that the TestChain framework, using GPT-4, achieved 71.79\% accuracy on the LeetCode-hard dataset, an improvement of 13.84\% over baseline methods. On the HumanEval dataset, TestChain with GPT-4 achieved 90.24\% accuracy. The TestChain framework designs agents to generate diverse test inputs, maps inputs to outputs using ReAct format dialogue chains, and interacts with the Python interpreter to obtain accurate test outputs.

\textbf{Interactive and Conversational Test Generation.} The SocraTest framework falls under the Autonomous Learning and Decision Making topic~\cite{feldt2023towards}. This framework automates the testing process through conversational interactions, the paper presents detailed examples of generating and optimizing test cases using GPT-4, emphasizing how multi-step interactions enhance testing methods and generate test code. Experimental results show that through conversational LLMs, SocraTest can effectively generate and optimize test cases and utilize middleware to facilitate interactions between the LLM and various testing tools, achieving more advanced automated testing capabilities.

\textbf{Integrated Testing in Software Development Pipelines.} ~\cite{nguyen2024agilecoderdynamiccollaborativeagents} proposes AGILECODER, a multi-agent framework that integrates Agile methodology into LLM-based software development which introduced in section \ref{sec:code-agent}. The Tester agent generates test suites and structured testing plans based on code changes, enabling efficient bug detection and feedback. Evaluated on HumanEval, MBPP, and a new ProjectDev benchmark, AGILECODER outperforms ChatDev and MetaGPT in both pass@1 and executability showcasing the synergy between collaborative development and automated testing in LLM-based systems. LLM-based agents can also be applied in user acceptance testing (UAT),~\cite{wang2024xuat} aims to enhance the automation of the WeChat Pay UAT process by proposing a multi-agent collaborative system named XUAT-Copilot, which uses LLMs to automatically generate test scripts. The study evaluates XUAT-Copilot's performance on 450 test cases from the WeChat Pay UAT system, comparing it to a single-agent system and a variant without the reflection component. Experimental results show that XUAT-Copilot achieved a Pass@1 rate of 88.55\%, compared to 22.65\% for the single-agent system and 81.96\% for the variant without the reflection component, with a Complete@1 rate of 93.03\%. XUAT-Copilot employs a multi-agent collaborative framework, including action planning, state checking, and parameter selection agents, and uses advanced prompting techniques. XUAT-Copilot demonstrates the potential and feasibility of LLMs in automating UAT test script generation.

\subsection{Analysis}
\begin{figure}[htbp]
    \centering
     \includegraphics[width=1\linewidth]{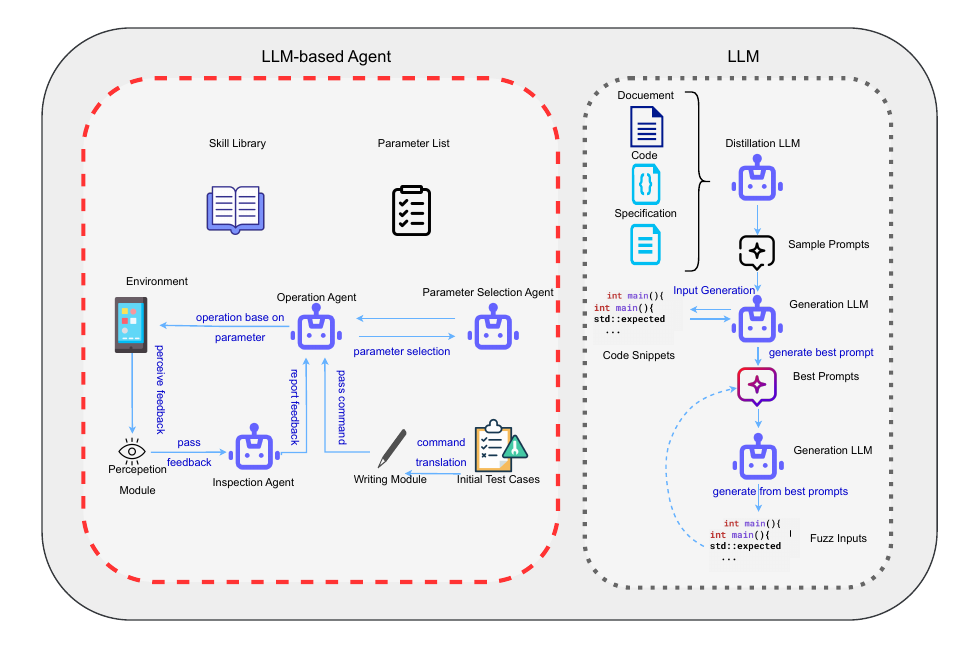}
    \caption{Illustration of Comparison Framework Between LLM-based Agent\cite{wang2024xuat} and LLM\cite{xia2024fuzz4all} in Software Test Generation.}
    \label{fig:testGen}
\end{figure}
In comparison, LLMs perform well in single-task implementations, generating high-quality test cases through techniques like prompt engineering and few-shot learning. The number of related studies is increasing as the capabilities of LLMs improve. On the other hand, LLM-Based Agents, through multi-agent collaborative systems, decompose tasks for specialized processing, significantly enhancing the effectiveness and efficiency of test generation and execution through iterative optimization and feedback. Considering the cost, using LLMs for test generation only is enough and more cost saving than using LLM-based agents. However, if a specific model performs poorly, it can affect the entire system's performance. 

A single LLM may struggle with complex, multi-step tasks. For example, in high-coverage test generation, LLMs may require more complex prompts and post-processing steps to achieve the desired results. Additionally, the quality of the generated results depends heavily on the prompt design and quality. For tasks requiring fine control and continuous optimization, a single LLM may find it challenging to deal with. As shown in Figure. \ref{fig:testGen}, the LLM framework uses \cite{xia2024fuzz4all} as an example to demonstrate the usage of LLMs in fuzz testing, the prompt will be optimized by given code snippets (fuzz inputs), and re-select by the LLM again to choose the best prompt for the future generation. The overall framework lacks autonomy, the LLM-based agent \cite{wang2024xuat} framework on the left fills this gap, as well as able to perceive the UI and interact with the skill library for the operations. The operation agent will receive any error reported by the inspection agent and do the self-reflection to refine the process autonomously. However, as previously discussed, build a LLM-based agents framework only for the software test generation task are "overkill", so the collected paper for LLM-based agents system generally focused on program repair by generated test cases or bug replay system, the LLM-based agent framework is actually used for automatically test the Wechat Pay system.

\subsection{Benchmarks}
In the tasks of LLMs in software test generation, the dataset Defects4J used to evaluate bug reproduction and program repair techniques. Other public datasets such as ReCDroid, ANDROR2+, and Themis are primarily used to evaluate mobile application bug reproduction and security test generation, particularly for the Android platform. GCC, Clang, Go toolchain, Java compiler (javac), and Qiskit involve fuzz testing datasets for various programming languages and toolchains, aimed at assessing the effectiveness of fuzz testing in multi-language environments. TrickyBugs and EvalPlus are datasets containing complex bug scenarios, used to evaluate the precision and recall of generated test cases, the benchmark applications evaluated by CODAMOSA are used to assess the effectiveness of coverage-based test generation tools. 

The datasets used in LLM-Based Agents research are also quite common, HumanEval, MBPP, and LeetCode-hard are mainly used to evaluate the accuracy and coverage of code generation and test generation, involving various programming problems and challenges which frequently appeared in previous sections. Datasets like Codeflaws, QuixBugs, and ConDefects are collected to familiarize LLMs with erroneous code and programs, containing multiple program errors and defects, and are used to evaluate the effectiveness of automated debugging and bug repair. A unique dataset is the WeChat Pay UAT system, which includes user acceptance test cases from actual applications and is used to evaluate the performance of multi-agent systems in user acceptance testing, focusing specifically on WeChat's security system.

Overall, the datasets used in LLM-based agents' research are broader covering a wide range of programming problems and challenges, LLM research is more focused on the actual generation tasks, such as bug reproduction on the Android platform and fuzz testing in multi-language environments. This because the LLM-Based agents not only focus on the quality of generated test cases and code but also evaluate the collaborative effects and iterative optimization capabilities of multi-agent systems, so the benchmarks also include the dataset used to evaluate performance of the framework. For instance, AgentCoder~\cite{huang2023agentcoder} improves the efficiency and accuracy of test generation and execution through multi-agent collaboration consider qualitative and quantitative evaluations and using MBPP,HummanEval to do the evaluations, researches on LLM-Based agents places more emphasis on verifying the effectiveness of the system through qualitative evaluation and user feedback.

\subsection{Evaluation Metrics}
As seen in Table~\ref{tab:st_metrics}, LLMs research predominantly utilizes traditional quantitative metrics such as bug reproduction rate, code coverage, precision, and recall, these metrics directly reflect the effectiveness and quality of test generation. In contrast, LLM-Based agents research not only focuses on quantitative metrics but also introduces qualitative evaluations, such as improvements through conversational interactions and the collaborative effects of multi-agent systems. This diversified evaluation approach provides a more comprehensive reflection of the system's practical application effects. From the task perspectives, LLMs are more inclined to single task processing, such as generating test sets and considering the coverage of generated test sets. However, because of the expansion of agents framework, LLM-based agents often tend to use the generated test sets to evaluate whether vulnerabilities can be found to achieve a more ideal practicality. From a design perspective, LLM systems are relying on prompt engineering and the generative capabilities of the models themselves, their evaluation metrics are also mainly focused on the quality and effectiveness of the model outputs, also their evaluation metrics include the collaborative effects and efficiency within the system, such as improving Pass@1 and Complete@1 rates through multi-agent collaboration. Overall, LLMs are more suited for rapid test generation and evaluation for specific tasks, with evaluation metrics directly reflecting the generation's effectiveness and quality. LLM-Based Agents excel in handling complex and diversified tasks, achieving higher system efficiency and effectiveness through multi-agent collaboration and iterative optimization.
\begin{table*}
\centering
\caption{Evaluation Metrics in Software Test Generation}
\scriptsize 
\label{tab:st_metrics}
\begin{tabular}{|l|l|l|c|} 
\hline
\multicolumn{1}{|c|}{Reference Paper}                                                                                                                      & \multicolumn{1}{c|}{Benchmarks}                                                                                      & \multicolumn{1}{c|}{Evaluation Metrics}                                                                                                                                   & \multicolumn{1}{l|}{Agent}  \\ 
\hline
\begin{tabular}[c]{@{}l@{}}{~\cite{zhang2023well}}\end{tabular}                                                                       & \begin{tabular}[c]{@{}l@{}}26 libraries and 55 \\applications with \\known vulnerabilities\end{tabular}              & \begin{tabular}[c]{@{}l@{}}Number of applications for \\which security tests were successfully \\generated.Number of tests that could\\demonstrate exploits.\end{tabular} & No                          \\ 
\hline
{~\cite{feng2024prompting}}                                                                                                                                  & \begin{tabular}[c]{@{}l@{}}ReCDroid,~ANDROR2+,\\Themis Empirical Study Dataset\end{tabular}                          & \begin{tabular}[c]{@{}l@{}}Accuracy of S2R Entity Extraction.\\Reproducibility of Bugs.\\Runtime Efficiency.\\User Satisfaction.\end{tabular}                             & No                          \\ 
\hline
\begin{tabular}[c]{@{}l@{}}{~\cite{kang2023large}}\end{tabular}                       & Defects4J,~GHRB                                                                                                      & \begin{tabular}[c]{@{}l@{}}Bug Reproduction Rate.\\Precision and Recall.\\Execution Time.\\Developer Effort.\end{tabular}                                                 & No                          \\ 
\hline
\begin{tabular}[c]{@{}l@{}}{~\cite{xia2024fuzz4all}}\end{tabular}                                                           & \begin{tabular}[c]{@{}l@{}}GCC and Clang.\\CVC5 and Z3.\\Go Toolchain.\\Java Compiler (javac).\\Qiskit.\end{tabular} & \begin{tabular}[c]{@{}l@{}}Code Coverage.\\Validity Rate.\\Hit Rate.\\Bugs Detected.\end{tabular}                                                                         & No                          \\ 
\hline
\begin{tabular}[c]{@{}l@{}}{~\cite{ryan2024code}}\end{tabular}               & \begin{tabular}[c]{@{}l@{}}897 focal methods from 26 \\widely used open-source \\Python projects.\end{tabular}       & \begin{tabular}[c]{@{}l@{}}Pass@1.\\FM Call@1.\\Correct@1.\\Line \& Branch Coverage.\end{tabular}                                                                         & No                          \\ 
\hline
\begin{tabular}[c]{@{}l@{}}{~\cite{liu2024llm}}\end{tabular}                                                       & \begin{tabular}[c]{@{}l@{}}TrickyBugs\\EvalPlus datasets.\end{tabular}                                               & \begin{tabular}[c]{@{}l@{}}Recall.\\Precision.\\F1 Score.\end{tabular}                                                                                                    & No                          \\ 
\hline
\begin{tabular}[c]{@{}l@{}}{~\cite{pizzorno2024coverup}}\end{tabular}                                                               & \begin{tabular}[c]{@{}l@{}}Benchmark applications originally\\~used to evaluate CODAMOSA.\end{tabular}               & \begin{tabular}[c]{@{}l@{}}Line Coverage.\\Branch Coverage.\\Line + Branch Coverage.\end{tabular}                                                                         & No                          \\ 
\hline
\begin{tabular}[c]{@{}l@{}}{~\cite{huang2023agentcoder}}\end{tabular}                                   & \begin{tabular}[c]{@{}l@{}}HumanEval.\\MBPP.\\HumanEval-ET.\\MBPP-ET.\end{tabular}                                   & Pass@1                                                                                                                                                                    & Yes                          \\ 
\hline
\begin{tabular}[c]{@{}l@{}}{~\cite{feldt2023towards}}\end{tabular}                                        & Not Specified                                                                                                        & \begin{tabular}[c]{@{}l@{}}Qualitative improvement through \\conversational interactions.\end{tabular}                                                                    & Yes                         \\ 
\hline
\begin{tabular}[c]{@{}l@{}}{~\cite{li2024large}}\end{tabular}                          & \begin{tabular}[c]{@{}l@{}}HumanEval.\\LeetCode-hard.\end{tabular}                                                   & \begin{tabular}[c]{@{}l@{}}Accuracy.\\Line Coverage (Line Cov).\\Code-with-Bugs (CwB).\end{tabular}                                                                       & Yes                         \\ 
\hline
\begin{tabular}[c]{@{}l@{}}{~\cite{lee2024unified}}\end{tabular}                                                   & \begin{tabular}[c]{@{}l@{}}Codeflaws.\\QuixBugs.\\ConDefects.\end{tabular}                                           & \begin{tabular}[c]{@{}l@{}}Number of Correct Patches.\\Number of Plausible Patches.\\Correctness Rate.\end{tabular}                                                       & Yes                         \\ 
\hline
\begin{tabular}[c]{@{}l@{}}{~\cite{tang2024chatgpt}}\end{tabular}                                                                & \begin{tabular}[c]{@{}l@{}}DynaMOSA benchmark\\Defects4J.\end{tabular}       & \begin{tabular}[c]{@{}l@{}}Correctness, Readability,\\Code Coverage, Bug Detection\\Vargha-Delaney A12.\end{tabular} & No                          \\ 
\hline
\begin{tabular}[c]{@{}l@{}}{~\cite{wang2024xuat}}\end{tabular} & \begin{tabular}[c]{@{}l@{}}450 test cases from the \\WeChat Pay UAT system\end{tabular}                              & \begin{tabular}[c]{@{}l@{}}Pass@1.\\Complete@1.\end{tabular}                                                                                                              & Yes                         \\
\hline
\end{tabular}
\end{table*}

%% file: pages/maintenance.tex
\section{Software Security and Maintenance \label{cha:softwareSecurity}}
In software engineering, software security and maintenance is a popular area for the application of LLMs, primarily aimed at enhancing the security and stability of software systems through existing technologies to meet the needs of users and developers. These models provide promising methods of vulnerability detection and repair, while also enabling automated security testing and innovative maintenance processes. The application of LLMs in software security and maintenance encompasses several aspects, including vulnerability detection, automatic repair, penetration testing, and system robustness evaluation. Compared to traditional methods, LLMs leverage natural language processing and generation technologies to understand and generate complex code and security policies, thereby automating detection and repair tasks. For example, LLMs can accurately identify potential vulnerabilities by analyzing code structures and contextual information and generate corresponding repair suggestions which improves the efficiency and accuracy of vulnerability recovery. 

Moreover, LLMs not only exhibit strong capabilities in vulnerability detection but also play a role in tasks like penetration testing and security evaluations. Automated penetration testing tools, such as PENTESTGPT~\cite{deng2023pentestgpt}. LLMs also demonstrate significant advantages in evaluating system robustness by simulating various attack scenarios to assess system performance under different conditions, helping developers better identify and address potential security issues. Research on LLM-based agents in software security and maintenance is also keep growing, these intelligent agents can execute complex code generation and vulnerability repair tasks and possess self-learning and optimization capabilities to handle issues encountered in dynamic development environments. Tools like RITFIS~\cite{xiao2024ritfis} and NAVRepair~\cite{wang2024navrepair} have shown potential in improving the precision and efficiency of program repairs by using LLM-based agents.

\subsection{LLMs Tasks}
In the field of software security and maintenance, research on LLMs can be categorized into three main areas: vulnerability detection, automatic repair, and penetration testing, along with some evaluation studies. The collected papers reviewed on LLMs in these domains illustrate their diverse applications and potential.

\textbf{Static Vulnerability Detection From Prompting to Structural Learning.} In the domain of vulnerability detection, researchers have fine-tuned LLMs to enhance the accuracy of source code vulnerability detection.~\cite{shestov2024finetuning} aims to investigate the potential of applying LLMs to the task of vulnerability detection in source code and to determine if the performance limits of CodeBERT-like models are due to their limited capacity and code understanding ability. The study fine-tuned the WizardCoder model (an improved version of StarCoder) and compared its performance with the ContraBERT model on balanced and unbalanced datasets. The experimental results showed that WizardCoder outperformed ContraBERT in both ROC AUC and F1 scores, significantly improving Java function vulnerability detection performance, which achieved the state-of-art performance at that time by improving ROC AUC from 0.66 in CodeBERT to 0.69. 

There are study mainly explored the applications of pure LLMs without any framework architecture in vulnerability detection, uncovering current challenges.~\cite{cheshkov2023evaluation} evaluated only the performance of ChatGPT and GPT-3 models in detecting vulnerabilities in Java code, the study compared text-davinci-003 (GPT-3) and gpt-3.5-turbo against a baseline virtual classifier in binary and multi-label classification tasks. The experimental results showed that while text-davinci-003 and gpt-3.5-turbo had high accuracy and recall rates in binary classification tasks, their AUC (Area Under Curve) scores were only 0.51, indicating performance equivalent to random guessing. In multi-label classification tasks, GPT-3.5-turbo and text-davinci-003 did not significantly outperform the baseline virtual classifier in overall accuracy and F1 scores. These findings indicate that the earlier model like GPT-3 has limited capabilities in practical vulnerability detection tasks, suggesting the need for further research and model optimization to improve their performance in real-world applications, fine-tuning and optimizing LLMs can significantly enhance their performance in source code vulnerability detection, However, these models still face many challenges in practical applications, requiring further research and technological improvements to enhance their real-world effectiveness and reliability. Recent work also emphasizes the importance of fair evaluation settings and advanced prompt engineering strategies in LLM-based program repair. ~\cite{zhang2024criticalreviewlargelanguage} constructs a new benchmark dataset, \textit{EvalGPTFix}, consisting of 151 real-world Java bug-fix pairs collected from 2023 AtCoder\footnote{\href{https://atcoder.jp/contests/ahc030}{https://atcoder.jp/contests/ahc030}} contests to eliminate data leakage risks inherent in previous benchmarks such as Defects4J. Through empirical studies, the paper demonstrates that ChatGPT, using only a basic prompt, can correctly repair 109 bugs (72.19\% recall), significantly outperforming state-of-the-art LLMs such as CodeT5 \cite{wang2021codet5} and PLBART by 27.5\% and 62.4\% respectively, This work highlights the critical role of prompt design and interaction in maximizing LLM capabilities for program repair.

In the later years,~\cite{lu2024grace} introduced a method to incorporate complex code structures directly into the model learning process, the GRACE framework combines graph structure information and in-context learning, using Code Property Graphs (CPGs) to represent code structure information. By integrating the semantic, syntactic, and lexical similarities of code, the framework GRACE addresses the limitations of text-based LLM analysis, improves the precision and recall rates of vulnerability detection tasks. The study utilized three vulnerability datasets, showing a 28.65\% improvement in F1 scores over baseline models, an important aspect of vulnerability detection is enhancing LLM performance in code security tasks.~\cite{li2023hitchhiker} fine-tuned LLMs for specific tasks and evaluated their performance against existing models such as ContraBERT. The researchers conducted numerous experiments to determine the optimal model architecture, training hyperparameters, and loss functions to optimize performance in vulnerability detection tasks. The study primarily focused on WizardCoder and ContraBERT, validating their performance through comparisons on balanced and unbalanced datasets and developing an efficient batch packing strategy that improved training speed. Results indicated that with appropriate fine-tuning and optimization, LLMs could surpass state-of-the-art models, contributing to more robust and secure software development practices.

Despite the development of numerous models, it is still necessary to investigate their practical effectiveness.~\cite{ding2024vulnerability} explored the effectiveness of code language models (code LMs) in detecting software vulnerabilities and identified significant flaws in existing vulnerability datasets and benchmarks. The researchers developed a new dataset called PRIMEVUL, and conducted experiments using it, they compared PRIMEVUL with existing benchmarks such as BigVul to evaluate several code LMs, including state-of-the-art base models like GPT-3.5 and GPT-4, using various training techniques and evaluation metrics. The results revealed that existing benchmarks significantly overestimated the performance of code LMs. For example, a state-of-the-art 7B model scored an F1 of 68.26\% on BigVul but only 3.09\% on PRIMEVUL, highlighting the gap between current code language models' performance and actual requirements for vulnerability detection. 

\textbf{Empirical Evaluation of Web Vulnerabilities in LLM-Generated Code.} While many vulnerability detection efforts focus on static code analysis or fine-tuning models, an emerging direction is the large-scale empirical evaluation of LLM-generated code for real-world deployment readiness and security risks. One recent study presents a comprehensive evaluation of 2,500 GPT-4-generated PHP web applications, analyzing their vulnerability using hybrid techniques including dynamic scanning, static analysis, and manual code audit~\cite{toth2024llms}. The authors constructed the \textit{ChatPHP-DB} dataset, which contains PHP code and corresponding SQL scripts automatically generated by LLMs. Vulnerability scanning was conducted on web services deployed in Docker environments, focusing on common web attack surfaces such as SQL Injection, Stored XSS, Reflected XSS, and Insecure File Upload. The authors highlighted that many vulnerabilities stemmed from overly simplistic coding patterns and missing sanitization, and that LLMs often fail to internalize secure coding practices, especially in zero-shot settings.

\textbf{Generalized and Specialized Program Repair.} In the domain of software security and maintenance, LLMs have not only been applied to vulnerability detection but also extensively used for automating program repair. One study proposed using Round-Trip Translation (RTT) for automated program repair, researchers translated defective code into another language and then back to the original language to generate potential patches. The study used various language models and benchmarks to evaluate RTT's performance in APR. The experiments explored how RTT performs when using programming languages as an intermediate representation, how RTT performs when using natural language (English) as an intermediate representation, and what qualitative trends can be observed in the patches generated by RTT. Three measurement standards and eight models were used in the experiments, the results showed that the RTT method achieved significant repair effects on multiple benchmarks, particularly excelling in terms of compilation and feasibility~\cite{ruiz2024novel}. Similarly, in automated program repair,~\cite{wang2024navrepair} introduced several innovative methods. For example, NAVRepair specifically targets C/C++ code vulnerabilities by combining node type information and error types. Due to the unique pointer operations and memory management issues in C/C++, this language poses complexities. The framework uses Abstract Syntax Trees (ASTs) to extract node type information and combines it with CWE-derived vulnerability templates to generate targeted repair suggestions, the study evaluated NAVRepair on several popular LLMs (ChatGPT, DeepSeek Coder, and Magicoder) to demonstrate its effectiveness in improving code vulnerability repair performance. The results showed that NAVRepair achieved state-of-art performance in C/C++ program repair task, which improved repair accuracy by 26\% compared to existing methods. 

In order to address the two main limitations of existing fine-tuning methods for LLM-based program repair—namely, the lack of reasoning about the logic behind code changes and the high computational costs associated with fine-tuning large datasets—~\cite{yang2024multi} introduced the MOREPAIR framework, this framework improve the performance of LLMs in automated program repair (APR) by simultaneously optimizing syntactic code transformations and the logical reasoning behind code changes, the study used techniques to enhance fine-tuning efficiency, such as QLoRA (Quantized Low-Rank Adaptation)~\cite{dettmers2023qloraefficientfinetuningquantized} to reduce memory requirements and NEFTune (Noisy Embedding Fine-Tuning)~\cite{jain2023neftunenoisyembeddingsimprove} to prevent overfitting during the fine-tuning process. The experiments evaluated MOREPAIR on four open-source LLMs of different sizes and architectures (CodeLlama-13B, CodeLlama-7B, StarChat-alpha, and Mistral-7B) using two benchmarks, evalrepair-C++ and EvalRepair-Java. The results indicated that CodeLlama improved by 11\% and 8\% on the first 10 repair suggestions for evalrepair-C++ and EvalRepair-Java respectively. Another study introduced the PyDex system, which automatically repairs syntax and semantic errors in introductory Python programming assignments using LLMs, the system combines multimodal prompts and iterative querying methods to generate repair candidates and uses few-shot learning to improve repair accuracy. PyDex was evaluated on 286 real student programs from an introductory Python programming course and compared against three baselines. The results showed that PyDex significantly improved the repair rate and effectiveness compared to existing baselines~\cite{zhang2024pydex}.

~\cite{joshi2023repair} introduced a new system named RING that leverages large language models (LLMCs) to perform multilingual program repair across six programming languages. RING employs a prompt strategy that minimizes customization efforts, including three stages: fault localization, code transformation, and candidate ranking. The results showed that RING was particularly effective in Python, successfully repairing 94\% of errors on the first attempt. The study also introduced a new PowerShell command repair dataset, providing valuable resources for the research community, this research demonstrated that AI-driven automation makes program repair more efficient and scalable. Another study,~\cite{xiang2024far} conducted a comprehensive investigation into function-level automated program repair, introducing a new LLM-based APR technique called SRepair. SRepair utilizes a dual-LLM framework to enhance repair performance, the SRepair framework combines a repair suggestion model and a patch generation model. It uses chain-of-thought to generate natural language repair suggestions based on auxiliary repair-related information and then utilizes these suggestions to generate the repaired function. The results showed that SRepair outperformed existing APR techniques on the Defects4J dataset, repairing 300 single-function errors, with an improvement of at least 85\% over previous techniques. This study demonstrated the effectiveness of the dual-LLM framework in function-level repair and, for the first time achieved multi-function error repair, highlighting the significant potential of LLMs in program repair. By extending the scope of APR, SRepair paves the way for applying LLMs in practical software development and evaluation.

\textbf{Penetration Testing.} LLMs can also be applied in the field of penetration testing, where they are used to enhance the efficiency and effectiveness of automated penetration testing. Although not as frequently studied as vulnerability detection and automated repair, this review includes two relevant papers.~\cite{deng2023pentestgpt} investigates the development and evaluation of an LLM-driven automatic penetration testing tool PENTESTGPT. The main purpose of this study is to evaluate the performance of LLMs in practical penetration testing tasks and address the issue of context loss during the penetration testing process, the paper introduces three self-interaction modules of PENTESTGPT (reasoning, generation, and parsing) and provides empirical research based on benchmarks involving 13 targets and 182 sub-tasks. It compares the penetration testing performance of GPT-3.5, GPT-4, and Bard. The experimental results show that PENTESTGPT's task completion rate is 228.6\% higher than GPT-3.5 and 58.6\% higher than GPT-4, this study demonstrates the potential of LLMs in automated penetration testing, helping to identify and resolve security vulnerabilities, thereby enhancing the security and robustness of software systems. 

A similar research paper explores the application of generative AI in penetration testing.~\cite{hilario2024generative} evaluates the effectiveness, challenges, and potential consequences of using generative AI tools (specifically ChatGPT 3.5) in penetration testing. Through practical application experiments. The research conducts a five-stage penetration test (reconnaissance, scanning, vulnerability assessment, exploitation, and reporting) on a vulnerable machine from VulnHub, integrating Shell\_GPT (sgpt) with ChatGPT's API to automate guidance in the penetration testing process. The experimental results demonstrate that generative AI tools can significantly speed up the penetration testing process and provide accurate and useful commands, enhancing testing efficiency and effectiveness. This study indicates that the need to consider potential risks and unintended consequences, emphasizing the importance of responsible use and human oversight. Assessing the robustness of systems is also a crucial part of development, LLMs are used to develop and evaluate new testing frameworks to detect and improve the robustness of intelligent software systems.~\cite{xiao2024ritfis} introduces a robust input testing framework named RITFIS, designed to evaluate the robustness of LLM-based intelligent software against natural language inputs. The study adapts 17 existing DNN testing methods to LLM scenarios and empirically validates them on multiple datasets to highlight the current robustness deficiencies and limitations of LLM software. The study indicate that RITFIS effectively assesses the robustness of LLM software and reveals its vulnerabilities in handling complex natural language inputs. This research underscores the importance of robustness testing for LLM-based intelligent software and provides directions for improving testing methods to enhance reliability and security in practical applications.

\subsection{LLM-based Agents Task}
\textbf{Autonomous Debugging and Program Repair Agents.} LLM-based Agents primarily appied in areas such as autonomous decision-making, task-specific optimization, and multi-agent collaboration, these frameworks showcasing their strong potential in proactive defense. 
~\cite{zhong2024ldb} aims to address the limitations of existing debugging methods that treat the generated program as an indivisible entity. By segmenting the program into basic blocks and verifying the correctness of each block based on task descriptions, the proposed method LDB (Large Language Model Debugger) provide a more detailed and effective debugging tool that closely reflects human debugging practices. The study's experiments covered testing LDB on several benchmarks and compared with baseline models without a debugger and those using traditional debugging methods (self-debugging with explanations and traces). LDB's accuracy increased from a baseline of 73.8\% to 82.9\% on the HumanEval benchmark, an improvement of 9.1\%. In the domain of vulnerability detection, researchers have enhanced detection accuracy by combining Role-Based Access Control (RBAC) practices with deep learning of complex code structures.
In ~\cite{lee2024unified}, developed and evaluated an automated debugging framework named FixAgent, which improves fault localization, repair generation, and error analysis through an LLM-based multi-agent system. Although this research primarily focuses on automated debugging, incorporating elements like fault localization and automated program repair (APR), it intersects with test generation, particularly in the validation phase for testing bug fixes. The study evaluates FixAgent's performance on the Codeflaws, QuixBugs, and ConDefects datasets, comparing it to 16 baseline methods, including state-of-the-art APR tools and LLMs. Experimental results show that FixAgent fixed 78 out of 79 bugs in the QuixBugs dataset, including 9 never-before-fixed bugs. In the Codeflaws dataset, FixAgent fixed 3982 out of 2780 defects, with a correctness rate of 96.5\%. The framework includes specialized agents responsible for localization, repair, and analysis tasks and uses the rubber duck debugging principle. FixAgent demonstrates the powerful capabilities of LLMs in automated debugging, improving the performance of existing APR tools and LLMs which can be considered as state-of-art framework in APR by LLM-based agent. 
In the field of software repair by multi-agents collaborations,~\cite{alhanahnah2024empirical} proposes a dual-agent framework that enhances the automation and accuracy of repairing declarative specifications through iterative prompt optimization and multi-agent collaboration. The researcher compare the effectiveness of the LLM-based repair pipeline with several state-of-the-art Alloy APR techniques (ARepair, ICEBAR, BeAFix, and ATR). In the result, framework repaired 231 defects in the Alloy4Fun benchmark which surpassing the 278 defects repaired by traditional tools.
\cite{bouzenia2024repairagent} introduces an automated program repair agent named RepairAgent, this agent can dynamically generates prompts and integrates tools to automatically fix software bugs. This researcher also address the limitations of current LLM-based repair techniques, which typically involve fixed prompts or feedback loops that do not allow the model to gather comprehensive information about the bug or code. RepairAgent is a LLM-based agent designed to alternately collect information about the bug, gather repair ingredients, and validate the repairs, similar to how human developers fix bugs. RepairAgent achieved impressive result by overall repaired 186 bugs in the Defects4J benchmark, with 164 being correctly repaired outperforming existing repair techniques achieved the state-of-art performances. Another notable work in autonomous program improvement is AutoCodeRover~\cite{zhang2024autocoderover}, which introduces a fully automated LLM-based agent framework for resolving GitHub issues through structured program understanding and modification. Different from previous agent frameworks that treat source code as flat files, AutoCodeRover leverages program representations such as Abstract Syntax Trees (AST) and introduces a two-stage architecture: a context retrieval agent and a patch generation agent. The context retrieval agent interacts with a set of designed APIs to iteratively collect relevant code context, while the patch generation agent synthesizes and validates patches based on the gathered information. This work marks a significant step toward scalable, low-cost, and semantically-aware autonomous software engineering. MAGIS~\cite{tao2024magisllmbasedmultiagentframework} addresses repository-level issue resolution, which introduces a novel LLM-based multi-agent framework to tackle the complex task of resolving GitHub issues. MAGIS simulates a realistic team collaboration setting by deploying four specialized agents—Manager, Repository Custodian, Developer, and Quality Assurance (QA) Engineer. These agents cooperate through planning, code localization, iterative development, and automated review, closely mimicking human workflows like GitHub Flow.

\textbf{Access Control and Smart Contract Security.} \cite{zhang2024acfixguidingllmsmined} addresses the challenge of automatically and appropriately repairing access control (AC) vulnerabilities in smart contracts. The innovation of this paper lies in combining mined RBAC practices with LLMs to create a context-aware repair framework for AC vulnerabilities. The model primarily uses GPT-4, enhanced by a new method called ACFIX, which mines common RBAC practices from existing smart contracts and employs a Multi-Agent Debate (MAD) mechanism to verify the generated patches through debates between generator and verifier agents to ensure correctness. Experimental results show that ACFIX successfully repaired 94.92\% of access control vulnerabilities, significantly outperforming the baseline GPT-4's 52.54\%. Another application in smart contracts ~\cite{hu2023largelanguagemodelpoweredsmart}, this paper introduces a two-stage adversarial framework, GPTLENS, which improves vulnerability detection accuracy through generation and discrimination phases. GPTLENS achieved a 76.9\% success rate in detecting smart contract vulnerabilities, better than the 38.5\% success rate of traditional methods. Another study,~\cite{ma2024combining} introduces the TrustLLM framework which increase the accuracy and interpretability of smart contract auditing by customizing LLM capabilities to the specific requirements of smart contract code. This paper conducts experiments on a balanced dataset comprising 1,734 positive samples and 1,810 negative samples, comparing TrustLLM with other models such as CodeBERT, GraphCodeBERT, and several versions of GPT and CodeLlama. TrustLLM achieves an F1 score of 91.21\% and an accuracy of 91.11\% which outperforming other models.

\textbf{LLMs for Offensive Security and Vulnerability Exploitation.} \cite{fang2024llm} investigates the use of GPT-4 to automatically exploit disclosed but unpatched vulnerabilities, the experiments showed that the LLM-based agent achieved an 87\% success rate in exploiting vulnerabilities when provided with CVE descriptions. Finally another LLM-based agent application in the penetration test, ~\cite{happe2023getting} employs GPT-3.5 to assist penetration testers by automating high-level task planning and low-level vulnerability discovery, thereby enhancing penetration testing capabilities. The experiments demonstrated successful automation of multiple stages of penetration testing, including high-level strategy formulation and low-level vulnerability discovery, showcasing the effectiveness of LLMs in penetration testing.

\textbf{LLMs for Security Engineering and System-level Assurance.} In the realm of software security, researchers have combined LLM and security engineering models to improve security analysis and design processes.~\cite{geissler2024concept} aims to propose a complex hybrid strategy to ensure the reliability and security of software systems, this involves a concept-guided approach where LLM-based agents interact with system model diagrams to perform tasks related to safety analysis. Beyond software-level security design, LLMs can also be integrated into autonomous driving systems.~\cite{nouri2024engineeringsafetyrequirementsautonomous} which has already been discussed in the \ref{cha:requirement}.

\subsection{Analysis}
Overall, the direction of LLM-based Agents represents significant innovative advancements in software security and maintenance, demonstrating improvements across all areas. LLM-based Agents, through multi-agent collaboration and runtime information tracking to help with debugging tasks, compared to traditional LLMs approaches are often rely on fixed prompts or feedback loops to debug a given code snippet or program. In vulnerability detection, LLM-based Agents combine RBAC practices and in-depth learning of complex code structures to improve the accuracy and efficiency of detecting vulnerabilities, traditional LLMs methods normally depend on extensive manual intervention and detailed guidance when handling tasks. LLM-based Agents also demonstrate effectiveness in penetration testing by automating high-level task planning and low-level vulnerability exploration, thereby enhancing penetration testing capabilities. In contrast, traditional LLM methods are more suited for passive detection and analysis, lacking proactive testing and defense capabilities. 

\begin{figure}[htbp]
    \centering
    \includegraphics[width=1\linewidth]{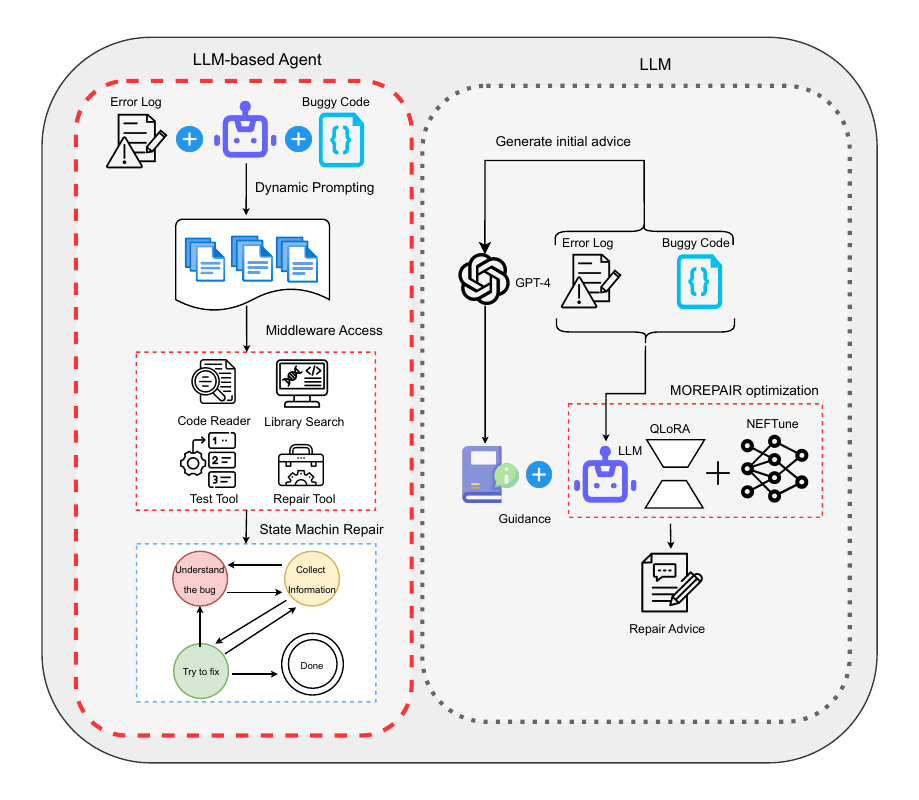}
    \caption{Illustration of Comparison Framework Between LLM-based Agent \cite{bouzenia2024repairagent} and LLM \cite{yang2024multi} in Software Security and Maintenance.}
    \label{fig:maint}
\end{figure}
From the perspective of automation, LLM-based agents automate the detection and repair of software errors through multi-agent frameworks and dynamic analysis tools, improving the automation and accuracy of the repair process. Traditional LLMs methods also have a good performance on various maintenance or debug tasks, but often lack autonomous decision-making and dynamic adjustment capabilities during the repair process. In terms of software security, intelligent agent become more flexibly by combining LLM and security engineering models to improve security analysis and design processes, thereby enhancing the reliability and security of software systems. when dealing with security tasks by LLMs only, often rely on static analysis lacking adaptability and optimization capabilities. As shown in the Figure.\ref{fig:maint}, the comparison using the MOREPAIR~\cite{yang2024multi} for LLMs and RepairAgent~\cite{bouzenia2024repairagent} for LLM-based agents. The LLM framework utilize the optimization techniques (QLoRA, NEFTune) to generate repair advices, the RepairAgent utilize multiple tools during the inspection which facilitate the precision and accuracy of the analysis before the repair process, the idea is quite similar with "reasoning before action". Then the agent framework utilize state machine and LLM to refine continuously, and if failed during the repair process, the RepairAgent will enter the self-reflection phase to understand the reason autonomously.

Thus, from the review, we can say that LLM-based agents brings more autonomy and flexibility in the field of software security and maintenance. These improvements can enhance task execution efficiency and accuracy, also extend the application scope of LLMs in complex software engineering tasks, demonstrating their strong potential in proactive defense, complex task handling, and meeting high reliability requirements.

\subsection{Benchmarks}
When analyzing the benchmarks used in LLM literature, several public datasets stand out due to their frequent use and presence across different application scenarios. Datasets such as Defects4J, Codeflaws, QuixBugs, and the Common Vulnerability and Exposure (CVE) database are commonly employed in the domains of vulnerability detection and software security. For instance, Defects4J is widely used in papers like~\cite{bouzenia2024repairagent} and~\cite{zhong2024ldb} to evaluate automated program repair tools. Similarly, Codeflaws and QuixBugs are used in papers like~\cite{lee2024unified} to test debugging capabilities, focusing on smaller algorithmic problems typically found in competitive programming and educational settings. These datasets effectively measure the ability of LLMs to detect vulnerabilities and modify code in specific code blocks. 

CVE is a critical benchmark for evaluating the security capabilities of LLMs, offering a repository of known vulnerabilities that allow LLMs to assess their ability to autonomously detect and exploit security flaws, bridging the gap between theoretical research and practical cybersecurity applications. Another notable dataset is ARepair, used in~\cite{alhanahnah2024empirical}. This dataset consists of defective specifications and tests the ability of LLMs to understand and repair formal specifications. More common datasets like HumanEval and MBPP are also frequently used to evaluate the functional correctness of code generated by LLMs. Similarly, Alloy4Fun is used to test the repair of declarative specifications in Alloy framework~\cite{alhanahnah2024empirical}, reflecting the LLM's performance in understanding and fixing logical errors in formal languages. 

Specialized datasets such as VulnHub and HackTheBox are used to evaluate the penetration testing capabilities of LLMs. Papers like~\cite{happe2023getting} utilize these environments to simulate real-world hacking scenarios, thereby assessing the practical applications of LLMs in cybersecurity. These benchmarks are crucial for evaluating the real-world efficacy of LLM-based agents in cyber-security environments, bridging the gap between theoretical capabilities and practical applications. In the context of smart contract security, datasets extracted from Etherscan and those compiled for tools like SmartFix provide benchmarks for evaluating LLMs' ability to identify and fix vulnerabilities in blockchain-based applications, emphasizing the reliability and security of decentralized applications.

When comparing the benchmarks used in LLM and LLM-based agent research, several key similarities and differences emerge. Both approaches frequently use datasets like Defects4J, CVE, and HumanEval, highlighting their foundational role in evaluating software engineering tasks. However, LLM-based agent research often combines these datasets with specialized benchmarks like VulnHub and HackTheBox to test more dynamic and interactive capabilities, especially in the context of cybersecurity. LLM-based agent research typically focuses more on real-time autonomous decision-making and action, reflected in their choice of benchmarks. These datasets test not only the agents' knowledge but also their ability to autonomously apply this knowledge in real-world scenarios. This contrasts with traditional LLMs research, which typically focuses on static tasks like vulnerability repair and code generation without requiring real-time interaction and further changes or decision-making. Moreover, the use of specialized benchmarks like the smart contract datasets from Etherscan in LLM-based agent research underscores the importance of blockchain technology and the need for robust security measures in decentralized applications, this trend highlights the adaptability and diversity of LLM-based agents in addressing emerging challenges in software security and maintenance. This distinction reflects the broader and more interactive application scenarios of LLM-based agents, also the public dataset may not suitable for LLM-based agent in particular structure designed, so there are a lot of self-collected benchmark emerged which provide more flexibility.

\subsection{Evaluation Metrics}
The evaluation metrics for the llm in the software security and maintenance are quite diverse. Researchers need to consider various factors such as coverage, efficiency, and reliability of the model or framework. Evaluation Metrics like success rate and pass rate are directly related to the performance of LLMs in different scenarios. In Table~\ref{tab:ss_metrics}, common standards such as success rate and change rate are frequently used to evaluate the robustness of models when faced with diverse inputs. Time overhead and query number are used to assess the efficiency and resource consumption of models when performing specific tasks. Additionally, ROC AUC, F1 score, and accuracy are important for evaluating the model's ability to identify vulnerabilities, especially in binary classification tasks. In code repair tasks, metrics such as compilability and plausibility are very common, these metrics ensure that the generated solutions are correct and deployable. Common standards like BLEU and CodeBLEU are used to evaluate the quality and human-likeness of generated code, which helps determine if the model's capabilities and performance are comparable to human performance. Furthermore, domain-specific metrics like tree edit distance and test pass rate are used to evaluate the effectiveness of LLM applications in specialized fields of software engineering, these metrics are used to address the limitations posed by software security and maintenance. In contrast, while LLM-based agents use evaluation metrics similar to those used for LLMs, such as success rate, they also incorporate more subjective metrics for evaluation. These include appropriateness, relevance, and adequacy, which are human-judged standards. Overall, the evaluation metrics used by agents tend to be simpler and easier to understand than those used for LLMs. This is likely because agents handle high-level tasks, such as the success rate of generating potential vulnerabilities and the frequency of agents calling external tools, so they also need to consider computational and time overheads of the overall architecture.

By comparing these metrics, we can see that LLMs emphasising the success rate of individual testing methods, LLM-based agents focus more on the overall task completion time/cost/effectiveness. LLMs typically use binary classification metrics like ROC, AUC, and F1 score, while agents tend to emphasize the success rate and accuracy during both the generation and validation phases, providing a comprehensive evaluation. For the time cost and performance, LLMs mainly focus on the execution time of testing methods and the number of queries to assess their efficiency. In contrast, LLM-based agents focus more on the completion time of repair tasks and the number of API calls, it will make sure the efficiency and practicality of overall architecture.
\begin{table*}
\centering
\caption{Evaluation Metrics in Software Security and Maintenance}
\scriptsize 
\label{tab:ss_metrics}
\begin{tabular}{|l|l|l|c|} 
\hline
\multicolumn{1}{|c|}{Reference Paper} & \multicolumn{1}{c|}{Benchmarks} 
& \multicolumn{1}{c|}{Evaluation Metrics} & \multicolumn{1}{l|}{Agent}  \\ 
\hline
\begin{tabular}[c]{@{}l@{}}{~\cite{xiao2024ritfis}}\end{tabular} 
& \begin{tabular}[c]{@{}l@{}}Financial Sentiment Analysis\\Movie Review Analysis~\\News Classification~\end{tabular} & \begin{tabular}[c]{@{}l@{}}Success Rate, Change Rate, Perplexity, Time Overhead, Query Number\end{tabular} & No \\ 

\hline
\begin{tabular}[c]{@{}l@{}}{~\cite{shestov2024finetuning}}\end{tabular}     & \begin{tabular}[c]{@{}l@{}}CVEfixes\\Manually-Curated Dataset \\(624 vulnerabilities across \\205 Java projects)\\VCMatch \\(10 popular repositories)\end{tabular} & \begin{tabular}[c]{@{}l@{}}ROC AUC, F1 Score, Accuracy, Optimal Classification, Threshold\end{tabular}          & No \\ 

\hline
\begin{tabular}[c]{@{}l@{}}{~\cite{li2023hitchhiker}}\end{tabular}          & \begin{tabular}[c]{@{}l@{}}CVEfixes\\Manually-Curated Dataset\\VCMatch\end{tabular} & \begin{tabular}[c]{@{}l@{}}Precision, Recall\end{tabular} & No \\ 

\hline
{~\cite{deng2023pentestgpt}} & \begin{tabular}[c]{@{}l@{}}HackTheBox\\VulnHub\end{tabular} & \begin{tabular}[c]{@{}l@{}}Overall Task Completion, Sub-task Completion, Task Variety, Challenge \\Levels, Progress Tracking\end{tabular} & No \\ 

\hline
\begin{tabular}[c]{@{}l@{}}{\cite{ruiz2024novel}}\end{tabular}              & \begin{tabular}[c]{@{}l@{}}Defects4J v1.2\\Defects4J v2.0\\QuixBugs\\HumanEval-Java\end{tabular} & \begin{tabular}[c]{@{}l@{}}Compilability, Plausibility, Test pass rate, Exact Match, BLEU\end{tabular} & No \\ 

\hline
{~\cite{lu2024grace}} & \begin{tabular}[c]{@{}l@{}}Devign\\Reveal\\Big-Vul\end{tabular} & \begin{tabular}[c]{@{}l@{}}F1 score, Accuracy, Precision,Recall.\end{tabular} & No \\ 

\hline
\begin{tabular}[c]{@{}l@{}}{~\cite{ding2024vulnerability}}\end{tabular}     & \begin{tabular}[c]{@{}l@{}}PRIMEVUL\\BigVul\end{tabular} & \begin{tabular}[c]{@{}l@{}}F1 score, Accuracy, Precision, Recall, VD-S, Pair-wise, evaluation metrics\end{tabular} & No \\ 

\hline
\begin{tabular}[c]{@{}l@{}}{~\cite{cheshkov2023evaluation}}\end{tabular}    & \begin{tabular}[c]{@{}l@{}}Customized GitHub dataset\\(308 binary classification and\\~120 multi-label classification)\end{tabular} & \begin{tabular}[c]{@{}l@{}}Precision, Recall, F1-Score, AUC, Accuracy\end{tabular} & No \\ 

\hline
\begin{tabular}[c]{@{}l@{}}{~\cite{hilario2024generative}}\end{tabular}  
& \begin{tabular}[c]{@{}l@{}}VulnHub\end{tabular}&
\begin{tabular}[c]{@{}l@{}}Output's Description\end{tabular} & No \\ 

\hline
\begin{tabular}[c]{@{}l@{}}{~\cite{thapa2022transformer}}\end{tabular}      & \begin{tabular}[c]{@{}l@{}}VulDeePecker\\SeVC\end{tabular} & \begin{tabular}[c]{@{}l@{}}False Positive Rate, False Negative Rate, Precision, Recall, F1-score\end{tabular} & No \\ 

\hline
{~\cite{wang2024navrepair}} & CVEFixes & \begin{tabular}[c]{@{}l@{}}CodeBLEU, Tree Edit Distance, Pass@k\end{tabular} & No \\ 

\hline
\begin{tabular}[c]{@{}l@{}}{~\cite{yang2024multi}}\end{tabular} & \begin{tabular}[c]{@{}l@{}}EvalRepair-C++\\EvalRepair-Java\end{tabular}     & \begin{tabular}[c]{@{}l@{}}TOP-5 and TOP-10, Repair\end{tabular} & No \\ 

\hline
\begin{tabular}[c]{@{}l@{}}{~\cite{zhang2024criticalreviewlargelanguage}}\end{tabular} 
& EvalGPTFix
& \begin{tabular}[c]{@{}l@{}}Number of Correctly Fixed Bugs,\\Fix Rate per Prompt Type,\\Success Rate over All Test Cases\end{tabular} 
& No \\

\hline
\begin{tabular}[c]{@{}l@{}}{~\cite{toth2024llms}}\end{tabular} 
& ChatPHP-DB (2,500 PHP websites) 
& \begin{tabular}[c]{@{}l@{}} Vulnerable Parameters (SQLi, XSS, File Upload).\\Dynamic + Static + Manual Analysis.\\Exploitability Rate, Prepared Statement Ratio.\\CWE-based Classification.\end{tabular} 
& No \\

\hline
{~\cite{zhang2024pydex}} & \begin{tabular}[c]{@{}l@{}}Introductory Python \\Assignments Dataset\end{tabular} & \begin{tabular}[c]{@{}l@{}}Repair Rate, Token Edit Distance\end{tabular} & No \\ 

\hline
{~\cite{joshi2023repair}} & \begin{tabular}[c]{@{}l@{}}Multi-languages dataset\\(Excel,Power Fx,Python,\\JavaScript,C andPowerShell)\end{tabular}  & Exact Matches~ & No \\ 

\hline
\begin{tabular}[c]{@{}l@{}}{~\cite{xiang2024far}}\end{tabular} & Defects4J 1.2 and 2.0 & \begin{tabular}[c]{@{}l@{}}Plausible Patches, Correct Fix\end{tabular} & No \\ 

\hline
\begin{tabular}[c]{@{}l@{}}{~\cite{happe2023getting}}\end{tabular}          & \begin{tabular}[c]{@{}l@{}}MITRE ATTCK framework \\(lin.security Linux VM)\end{tabular} & Ability Identify Vulnerabilities & Yes \\ 

\hline
\begin{tabular}[c]{@{}l@{}}{~\cite{zhang2024acfixguidingllmsmined}}\end{tabular} & 118 AC Vulnerabilities & \begin{tabular}[c]{@{}l@{}}Success Rate, Exploitation based evaluation, Manual inspection of the\\patches.\end{tabular} & Yes \\ 

\hline
\begin{tabular}[c]{@{}l@{}}{~\cite{hu2023largelanguagemodelpoweredsmart}}\end{tabular} & 13 Smart Contract Bugs & \begin{tabular}[c]{@{}l@{}}Success Rate, Contract level, Trial level\end{tabular} & Yes \\ 

\hline
\begin{tabular}[c]{@{}l@{}}{~\cite{lee2024unified}}\end{tabular} & \begin{tabular}[c]{@{}l@{}}Codeflaws,QuixBugs,\\ConDefects\end{tabular}     & \begin{tabular}[c]{@{}l@{}}Number of correctly fixed bugs, Number of plausibly patched bugs,\\Correctness rate of generated patches\end{tabular} & Yes \\ 

\hline
\begin{tabular}[c]{@{}l@{}}{~\cite{alhanahnah2024empirical}}\end{tabular}   & ARepair,Alloy4Fun & \begin{tabular}[c]{@{}l@{}}Correct@6, Runtime and Token Usage\end{tabular} & Yes \\

\hline
\begin{tabular}[c]{@{}l@{}}{~\cite{geissler2024concept}}\end{tabular} & System Model Graph & \begin{tabular}[c]{@{}l@{}}Accuracy, Effectiveness, Appropriateness\end{tabular} & Yes \\ 

\hline
\begin{tabular}[c]{@{}l@{}}{~\cite{ma2024combining}}\end{tabular} & \begin{tabular}[c]{@{}l@{}}263 real smart contract vulnerabilities\end{tabular} & \begin{tabular}[c]{@{}l@{}}F1 score, Accuracy, Consistency,Precision, Recall\end{tabular} & Yes \\ 

\hline
\begin{tabular}[c]{@{}l@{}}{~\cite{zhong2024debuglikehumanlarge}}\end{tabular} & \begin{tabular}[c]{@{}l@{}}HumanEval,MBPP,\\TransCoder\end{tabular} & \begin{tabular}[c]{@{}l@{}}Accuracy, Pass@1\end{tabular} & Yes \\ 

\hline

\hline
\begin{tabular}[c]{@{}l@{}}{~\cite{nouri2024engineeringsafetyrequirementsautonomous}}\end{tabular} & System Model Graph & \begin{tabular}[c]{@{}l@{}}Relevance, Adequacy\end{tabular} & Yes \\ 

\hline
\begin{tabular}[c]{@{}l@{}}{~\cite{fang2024llm}}\end{tabular} & 15 Vulnerabilities from CVE Lib & Success Rate, Cost & Yes \\ 

\hline
\begin{tabular}[c]{@{}l@{}}{~\cite{tao2024magisllmbasedmultiagentframework}}\end{tabular} & SWE-bench & \begin{tabular}[c]{@{}l@{}}Resolved Ratio, Planning Accuracy, \\ Line-level Localization Accuracy\end{tabular} & Yes \\

\hline
\begin{tabular}[c]{@{}l@{}}{~\cite{zhang2024autocoderover}}\end{tabular} & SWE-bench Lite & \begin{tabular}[c]{@{}l@{}}Resolved Rate, Pass@1, Pass@3,\\Correctness Rate, Average Time per Task,\\Token Cost\end{tabular} & Yes \\

\hline
\begin{tabular}[c]{@{}l@{}}{~\cite{bouzenia2024repairagent}}\end{tabular} & Defects4J & \begin{tabular}[c]{@{}l@{}}Plausible Fixes, Correct Fixes\end{tabular} & Yes \\

\hline
\end{tabular}
\end{table*}

%% file: pages/discussion.tex
\section{Discussion\label{cha:results}}
\subsection{Experiment Models}
In Sections III-IX, we reviewed and introduced the research on LLMs and LLM-based agent applications in software engineering in recent years. These studies have different research directions and we divided them into six subtopics for classification and discussion. With the advancement of large language models, to better understand their application in various fields, we summarized a total of 110 papers, focusing primarily on discussing the frequency of LLM usage in the field of software engineering

\begin{figure}[h]
\scriptsize
    \centering
    \includegraphics[width=1\linewidth]{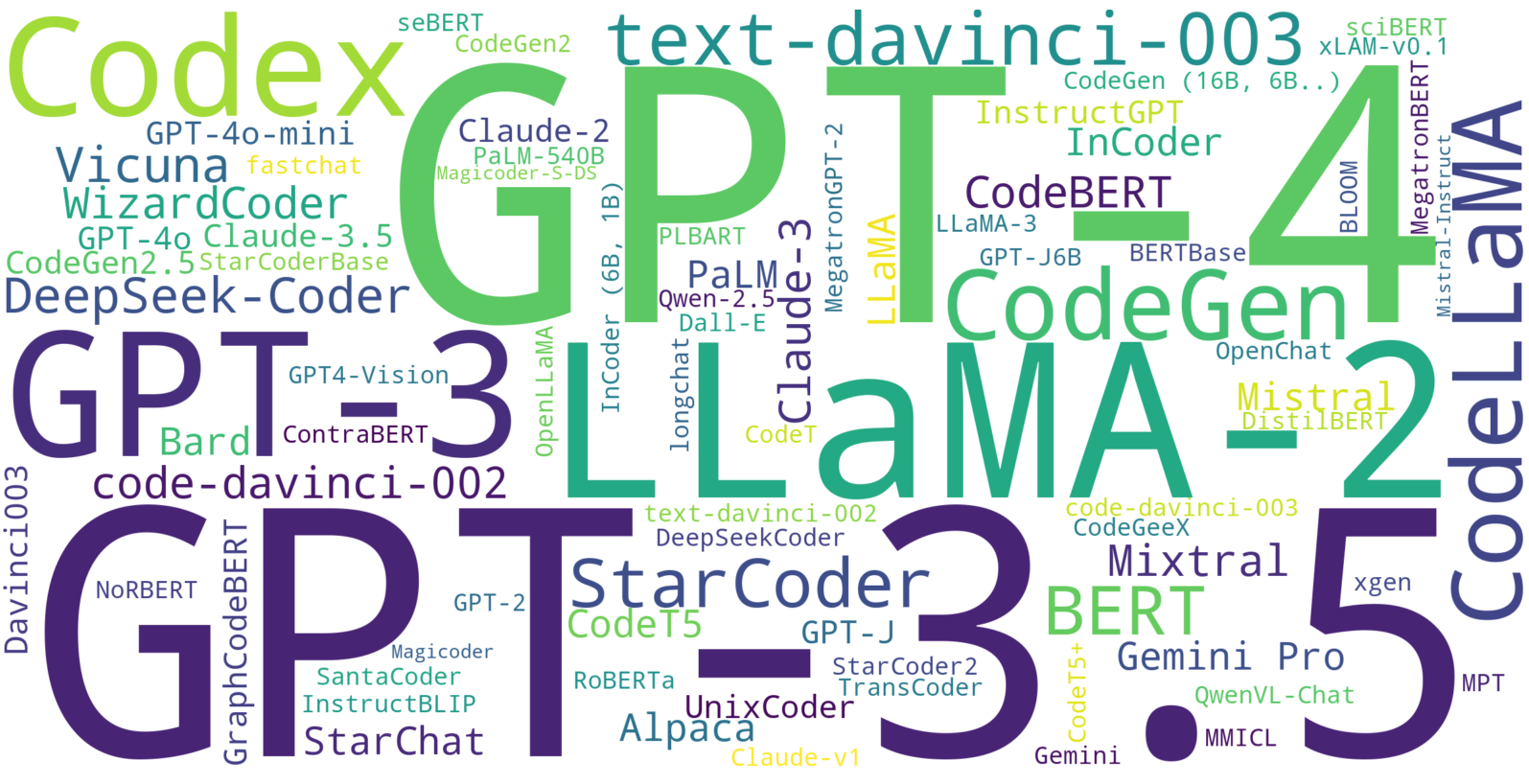} 
    \caption{Experiment Models Usage Word Cloud.} 
    \label{fig:wordcloud}
\end{figure}

Based on the review of 139 papers, our primary focus is on the models or frameworks utilized by the authors in their experiments. This is due to the fact that these papers often include tests of model performance in specific domains, such as evaluating LLaMA's performance in code generation. Therefore, during our data collection process, we also included models used for comparison purposes, as these models often represent the state-of-the-art capabilities in their respective fields at the time of the study. In summary, across the 139 papers, we identified a total of 82 unique large language models. We visualized the frequency of these model names in a word cloud for a more intuitive representation, as shown in Figure.\ref{fig:wordcloud}. From the figure, we can observe that models such as GPT-3.5, GPT-4, LLaMA2, and Codex are frequently used. Although close source LLMs cannot be locally deployed or further trained, their exceptional capabilities make them a popular choice for comparison in experiments or for data augmentation, where GPT-4 is used to generate additional data to support the research model frameworks. For instance, researchers might use OpenAI's API to generate initial text and then employ locally deployed models for further processing and optimization~\cite{dong2024selfcollaborationcodegenerationchatgpt,shen2024hugginggpt,jalil2023chatgpt,sridhara2023chatgpt}.

Therefore, it is not difficult to see that the use of general large models with superior performance to assist development or as a measurement standard has been increasingly used in the vertical field of software engineering in the past two years. In addition, for some fields that have never been touched by LLMs before, many researchers first refer to the model ChatGPT and conduct various performance experiments on the newer GPT-4~\cite{White2024,10.1007/978-3-031-48550-3_17,10371698}. Those models can be integrated into larger systems and combining with other machine learning models and tools, these models can be used to generate natural language responses, while another model handles intent recognition and dialogue management.

\begin{figure*}[htbp]
\scriptsize
    \centering
    \includegraphics[width=\textwidth]{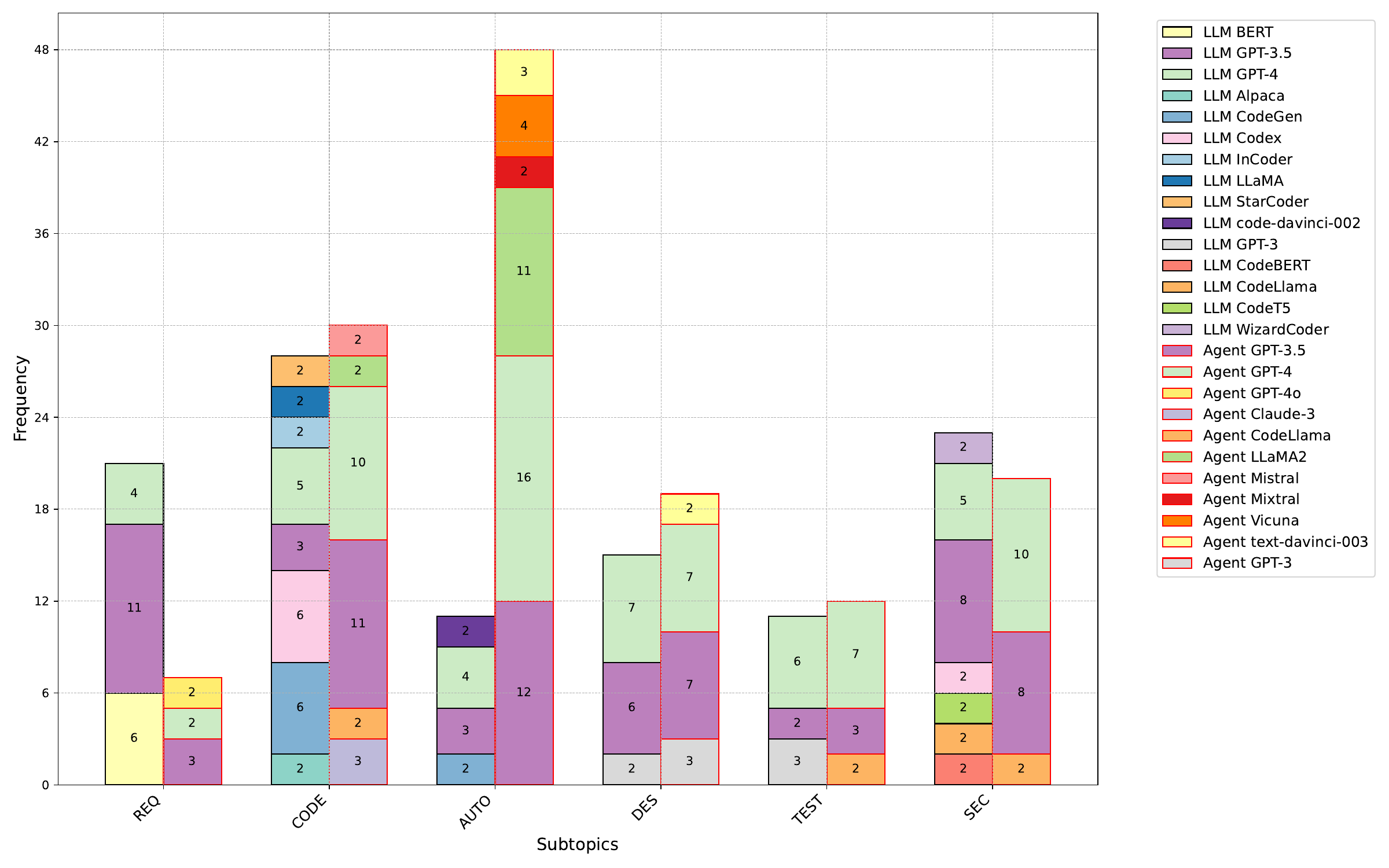}
    \caption{Experiment Models Usage in Different Subtopics.
    (REQ DENOTES "Requirement Engineering and Documentation", CODE DENOTES "Code Generation and Software Development", AUTO DENOTES "Autonomous Learning and Decision Making", DES DENOTES "Software Design and Decision Making", SEC DENOTES "Software Security and Maintenance")}
    \label{fig:models_usage}
\end{figure*}

Although the word cloud provides a rough overview of model usage frequency, it lacks detailed information. To gain deeper insights, we combined a grouped bar chart and a stacked bar chart to further analyze the usage of models in studies across different subtopics. The corresponding bar charts are presented in Figure.\ref{fig:models_usage}. During the analysis, we found that a large number of models appeared only once. Including these in the bar chart would have made the overall representation cluttered. Therefore, we \textbf{excluded models that appeared only once} and focused on the versatility of the remaining models. On the left side of each subtopic, we depict the models used in LLM-related studies, with the models used in LLM-based agent-related studies highlighted in red-bordered bars. From the figure, it is evident that in the Autonomous Learning and Decision Making subtopic, the number of models used in LLM-based agent-related studies is quite high. Specifically, GPT-4 and GPT-3.5 were used in 12 out of 23 papers and 16 out of 23 papers, respectively. In this subtopic, studies commonly utilized GPT-3.5/4 and LLaMA-2 for research and evaluation. During our analysis, we found that many studies on LLM-based agents evaluated the agents' ability to mimic human behavior and decision-making or perform some reasoning tasks~\cite{xie2024can,rasheed2024can,ma2024combining}. Since these studies do not require local deployment, they mainly assess the performance of state-of-the-art models in specific directions, leading to the frequent use of the GPT-family models. Frameworks like~\cite{shinn2024reflexion,yao2022react} constructed LLM-based agents by calling the GPT-4 API, using verbal reinforcement to help language agents learn from their mistakes. Due to the limitations of GPT models, many studies also used LLaMA as the LLM for agents, fine-tuning it on the generated datasets to evaluate the emergence of knowledge and capabilities. Overall, we found that in the Autonomous Learning and Decision Making subtopic, LLM-based agents often use multiple models for testing and performance evaluation in a single task, this results in a significantly higher model usage frequency in this topic compared to others.

Not only in the Autonomous Learning and Decision Making subtopic, but also across other themes, we observe that the variety of models (represented by the number of colors) used by LLM-based agents is relatively limited. For instance, in the requirement engineering and documentation subtopic, only GPT-3.5 and GPT-4 models were involved in the experiments. To analyze the reasons behind this phenomenon, we need to exclude the factors that models appearing only once were not considered and that there are inherently fewer studies on intelligent agents. We believe this primarily reflects the integration relationship between the agents and the large language models. The combination of these two technologies aims to address the limitations of large language models in specific tasks or aspects. Intelligent agents allow researchers to design a more flexible framework and incorporate the large language model into it. These models, having been trained on vast amounts of data, possess strong generalizability, making them suitable for a wide range of tasks and domains. 

Therefore, researchers and developers can use the same model to address multiple issues, reducing the need for various models. In code generation~\cite{zheng2024opencodeinterpreter,hong2023metagptmetaprogrammingmultiagent}, test case generation~\cite{li2024large,lee2024unified}, and software security~\cite{alshahwan2024assuredllmbasedsoftwareengineering,zhong2024ldb}, there are instances of using CodeLlama. This model is fine-tuned and optimized based on the LLaMA architecture. At its release, it was considered one of the state-of-the-art models for code generation and understanding tasks, showing strong performance and potential compared to other models like Codex. Another potential reason is the previous successful applications and research outcomes that have proven these models' effectiveness, further enhancing researchers' trust and reliance on them. Compared to models that perform well in specific domains, in intelligent agent development, there is a preference for using general-purpose large models to ensure that the core of the agent possesses excellent text comprehension abilities, allowing for further reasoning, planning, and task execution. From the Figure.\ref{fig:models_usage}, we can also observe that research in the code generation and software development fields adopts a wide variety of models, further indicating the extensive attention this area receives and the excellent performance of models in code generation task.
\begin{figure}[h]
    \centering
    \includegraphics[width=.5\textwidth]{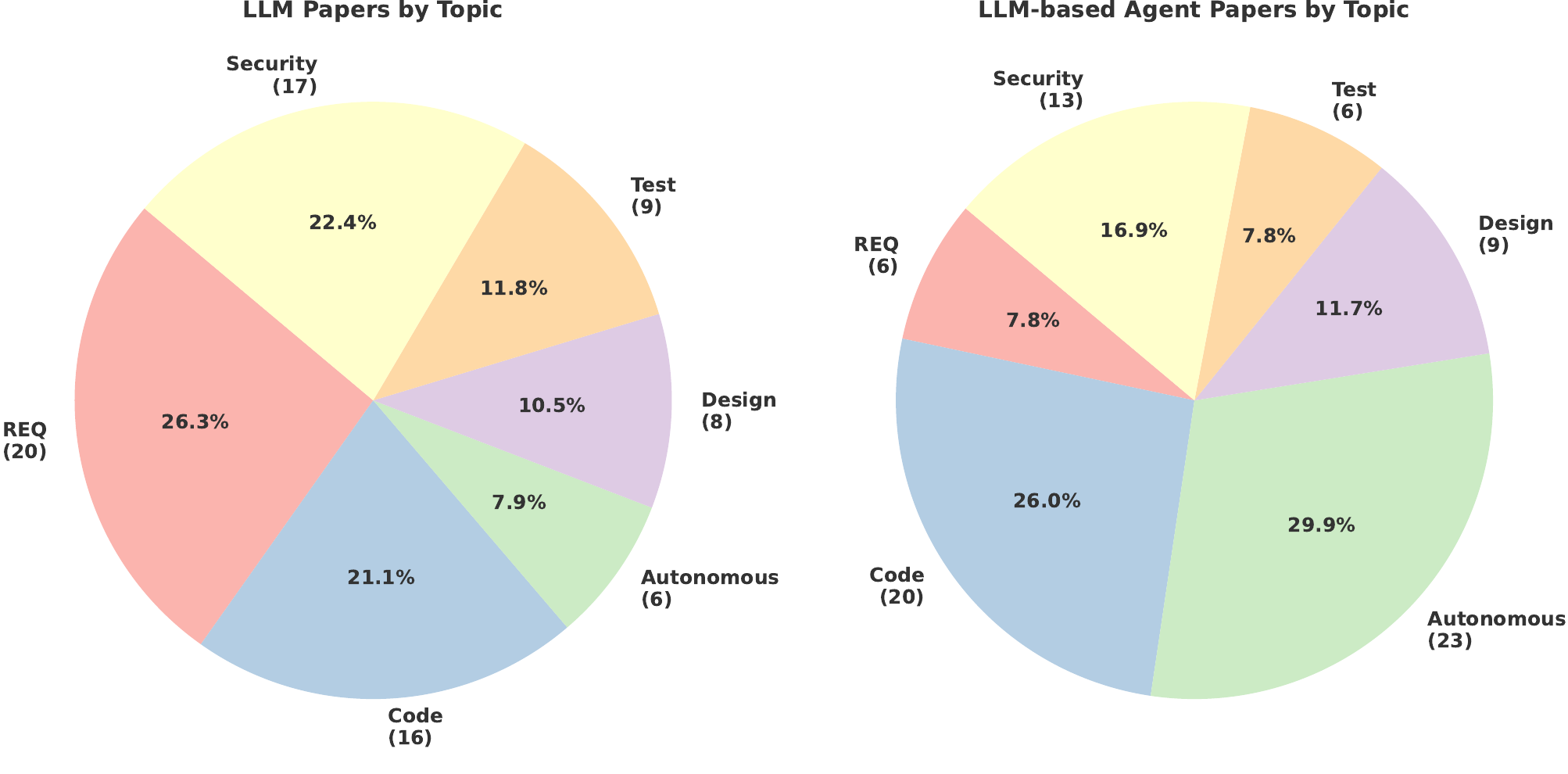} 
    \caption{Distribution of LLMs and Agents across six topics.} 
    \label{fig:piechart}
\end{figure}
\begin{table}[h!]
\scriptsize
    \centering
    \begin{tabular}{|c|c|c|}
        \hline
        Topic A & Topic B & Overlap Count \\
        \hline
        REQ & CODE & 1 \\
        REQ & SEC & 1 \\
        CODE & TEST & 3 \\
        CODE & AUTO & 3 \\
        AUTO & SEC & 3 \\
        TEST & SEC & 1 \\
        \hline
    \end{tabular}
    \caption{Selected Topic Overlaps: Number of Papers Addressing Two Topics Simultaneously 
    (REQ = Requirement Engineering and Documentation, CODE = Code Generation and Software Development, AUTO = Autonomous Learning and Decision Making, SEC = Software Security and Maintenance, TEST = Software Testing and Evaluation)}
    \label{tab:overlap}
\end{table}
\subsection{Topics Overlapping}


Figure~\ref{fig:piechart} shows the distribution of all collected literature across six themes. For LLM-type literature, the most represented theme is requirement engineering and documentation, followed closely by software security and maintenance, which together account for nearly half of all LLM papers. In contrast, themes such as test case generation and autonomous decision making are less emphasized. This trend shifts significantly in the LLM-based agent literature. The most prominent theme is autonomous learning and decision making, comprising nearly 30\% of all agent-related papers, followed by code generation and development. In comparison, research on using LLM-based agents for requirement engineering and test case generation remains relatively limited, this reflects a tendency to leverage agent frameworks primarily for tasks involving autonomy, adaptability, and iterative decision making, rather than for static tasks such as test generation. As such, more research efforts focus on understanding how agentified LLMs can enhance capabilities in dynamic domains like autonomous decision making and software maintenance.

Table~\ref{tab:overlap} presents the selected overlap of papers spanning multiple key themes. Notably, the most frequent overlaps are observed between code generation and software testing \cite{fakhoury2024llm,nguyen2024agilecoderdynamiccollaborativeagents,huang2023agentcoder}, as well as between code generation and autonomous learning \cite{manish2024autonomous,wang2024executable,zhang2024codeagentenhancingcodegeneration}. These overlaps suggest that tasks in code generation often co-occur with evaluation and learning capabilities, highlighting the integrative nature of LLM and LLM-based agent applications. Additionally, autonomous learning also shares a substantial intersection with software security and maintenance \cite{happe2023getting,ma2024combining,fang2024llm}, reflecting the importance of adaptive and learning-driven approaches in security-related contexts.

Other overlaps, such as those between requirements engineering and code generation (1 paper), and between testing and security (1 paper), show that even traditionally distinct areas are increasingly interwoven in LLM-related research. These interconnections indicate that achieving high performance in complex software engineering tasks often involves combining methodologies from multiple domains. For example, autonomous learning techniques may be utilized not only in optimizing decision-making processes but also in dynamically generating code or performing real-time security assessments. Similarly, enhancing software security often relies on leveraging capabilities from code generation and testing frameworks~\cite{hu2024leveragingprintdebuggingimprove,rasheed2023autonomous,zheng2024opencodeinterpreter,lu2024selfselfevolutionlanguagefeedback}. This interconnectedness reinforces the trend that LLM-based agents act as bridges across domains, promoting knowledge transfer and technique reuse within the broader landscape of software engineering.

\subsection{Benchmarks and Metrics}
\subsubsection{Benchmarks}
\begin{figure}[H]
\scriptsize
    \centering
    \includegraphics[width=1\linewidth]{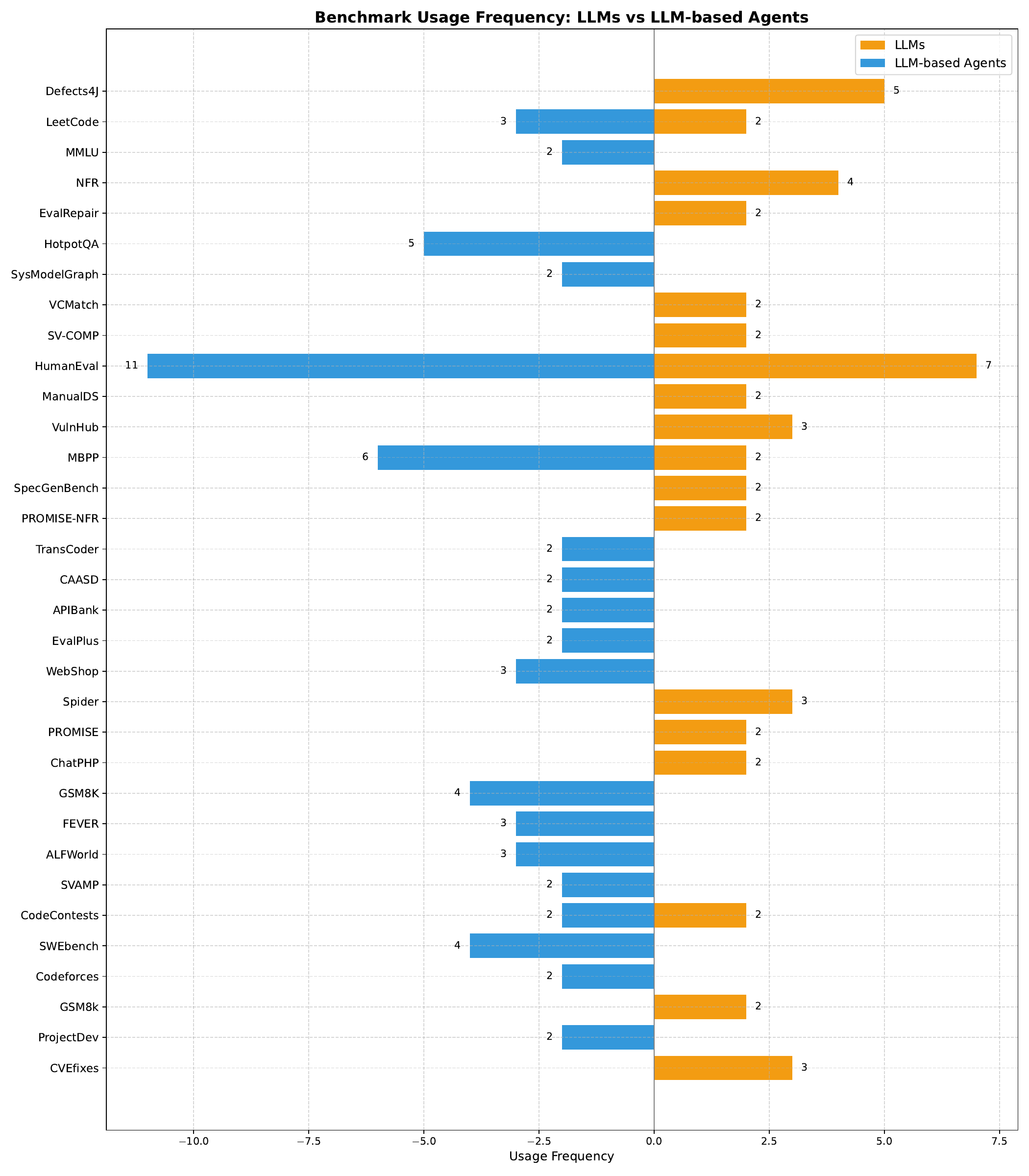}
    \caption{Distribution of Benchmarks.} 
    \label{fig:benchmark}
\end{figure}

As shown in Figure~\ref{fig:benchmark}, we illustrate the distribution of representative benchmark datasets across six major software engineering topics. It is important to note that the number of benchmark datasets used in practice is substantially greater than what is visualized in the figure. This is because many research studies rely on custom or domain-specific datasets that are not widely disseminated or publicly available. For instance, in requirements engineering, researchers often construct datasets from real-world requirement specifications or user stories~\cite{White2024,10.1007/978-3-031-61154-4_8}, which are typically not standardized nor openly accessible. Similarly, some studies refer vaguely to their datasets as “Customized GitHub Datasets”~\cite{cheshkov2023evaluationchatgptmodelvulnerability}, thereby excluding them from cross-paper quantitative statistics.

Therefore, the benchmark datasets shown in the figure primarily represent commonly-used public datasets across LLM and LLM-based agent tasks. Among these, \textbf{HumanEval} and \textbf{MBPP} are the most widely adopted across both LLMs and agents for code generation evaluation, given their structured format and popularity in prior benchmark studies. Interestingly, although some overlap exists, the benchmark usage patterns of LLMs and LLM-based agents diverge significantly in focus and diversit. For example, agent-related studies frequently adopt benchmarks that support complex, interactive, or knowledge-intensive reasoning tasks. The \textbf{FEVER}\footnote{\url{https://fever.ai/dataset/fever.html}} and \textbf{HotpotQA}\footnote{\url{https://hotpotqa.github.io/}} datasets are widely used in agent-based research for fact verification and multi-hop reasoning tasks, respectively. As seen in~\cite{zhao2024expel}, the FEVER dataset is employed to evaluate the factual consistency of responses generated by the ExpeL agent, while HotpotQA serves as a foundation for question-answering agents requiring long-range dependency tracking and contextual inference.

In the context of software security and maintenance, \textbf{Defects4J}\footnote{\url{https://github.com/rjust/defects4j}} remains a staple benchmark for evaluating code repair capabilities of LLMs. It consists of 835 real-world bugs drawn from multiple open-source Java projects and remains a standard for evaluating the precision and effectiveness of automated program repair. However, it is observed to be rarely used by LLM-based agent studies, potentially due to its focus on single-step repair evaluation, which is less aligned with the multi-turn, tool-augmented, and real-time characteristics of agentic systems. Instead, agent-based studies increasingly turn to more comprehensive benchmarks such as \textbf{SWE-bench}, \textbf{ProjectDev}, or \textbf{API-Bank}, which emphasize multi-stage workflows and tool invocation.

\textbf{Emerging benchmarks} like \textbf{ConDefects}~\cite{lee2024unified} and \textbf{EvalPlus} are designed to mitigate data leakage and improve real-world representativeness. They also provide fine-grained evaluation for sub-tasks like defect localization, which could be more suitable for future multi-agent testing scenarios.

\vspace{0.5em}
\noindent
\subsubsection{Evaluation Metrics.}  
\begin{figure}[H]
\scriptsize
    \centering
    \includegraphics[width=1\linewidth]{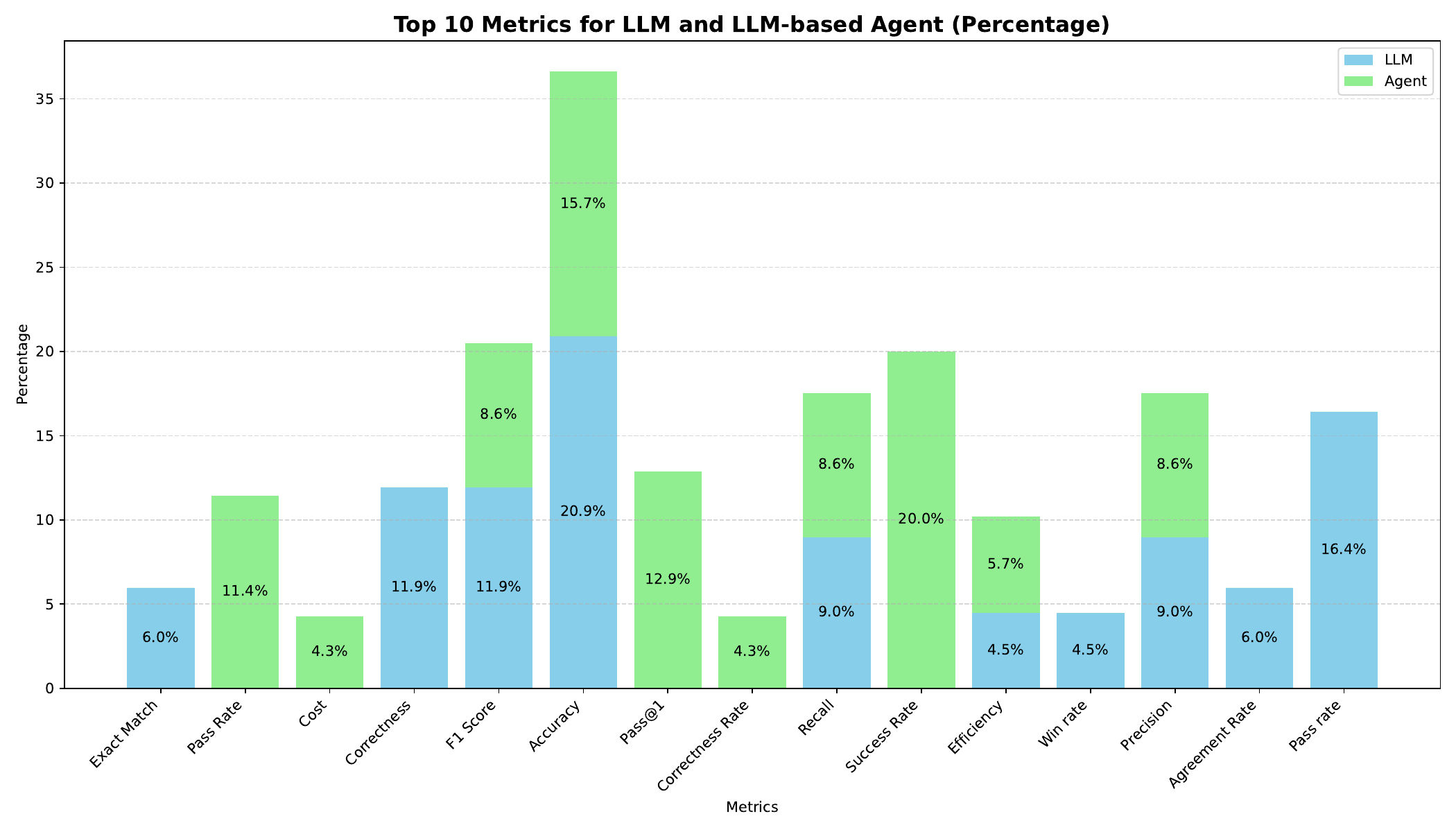}
    \caption{Top 10 Evaluation Metrics.} 
    \label{fig:top10Metrics}
\end{figure}

As shown in Figure~\ref{fig:top10Metrics}, we summarize the top 10 evaluation metrics most frequently used in LLM and LLM-based agent studies. The analysis reveals both convergence and divergence in metric preferences between the two paradigms. Traditional LLM research continues to emphasize structural correctness and task-specific quality, with top metrics including \textbf{Accuracy} (20.9\%), \textbf{F1 Score} (11.9\%), \textbf{Pass Rate} (16.4\%), \textbf{Correctness} (11.9\%), and \textbf{Exact Match} (6.0\%). These metrics are particularly common in structured tasks such as classification~\cite{fried2023incodergenerativemodelcode}, code generation~\cite{sun2023sql}, and bug repair~\cite{wang2023codet5+}.

In contrast, LLM-based agent research introduces broader, more dynamic evaluation dimensions. The most frequent metric used by agents is \textbf{Success Rate} (20.0\%), followed by \textbf{Pass@1} (12.9\%), \textbf{Accuracy} (15.7\%), and \textbf{Recall}/\textbf{Precision}/\textbf{F1 Score} (each 8.6\%). Furthermore, agent-based studies introduce agent-specific metrics such as \textbf{Efficiency} (5.7\%), \textbf{Cost} (4.3\%), \textbf{Correctness Rate} (4.3\%), and \textbf{Win Rate} (4.5\%), which reflect not only task completion, but also resource consumption and comparative agent performance.

It is noteworthy that although both paradigms rely on overlapping core metrics, their usage context differs. For instance, while \textbf{Accuracy} remains the dominant metric for both LLMs and agents too access many software task, it is often interpreted differently static classification accuracy for LLMs versus dynamic subtask completion accuracy for agents~\cite{ding2025toolcoder}. The use of \textbf{Success Rate} as the most frequent agent metric illustrates the community’s emphasis on end-to-end, multi-step success in goal-oriented workflows, such as tool usage or environment interaction~\cite{zhao2024expel, AutoCodeRover}.

Moreover, qualitative and user-centric evaluation metrics like \textbf{Alignment}, \textbf{Understandability}, and \textbf{Specifiability} have emerged in both LLM and agent studies. These metrics are crucial for measuring the model’s interpretability, instructional clarity, and alignment with user goals, especially in interactive multi-turn or development-assistive settings. In summary, while both LLMs and agents share a substantial subset of evaluation metrics, their distinct execution modes, static inference vs. autonomous orchestration, result in different emphases in metric design and interpretation. This trend underscores the need for context-aware evaluation frameworks to assess the next generation of LLM-based agents.




\subsection{Challenges and Opportunities}

As demonstrated across the six key software engineering domains covered in this survey, the transition from LLMs to LLM-based Agents not only introduces new capabilities but also brings about novel technical and theoretical challenges. This part consolidates insights from the reviewed literature, summarizing the core challenges faced by LLM-based Agents in software engineering, and explores the new opportunities enabled by this paradigm shift.

\textbf{1) Lack of standardized agent definitions and evaluation protocols.} Although a working definition is provided in Section~\ref{cha:Preliminaries}, there is currently no consensus in the academic community regarding the level of autonomy, planning capability, and tool usage required for an LLM to qualify as an agent. Existing studies (e.g., \cite{AutoCodeRover, zhao2024expel, patil2024goex}) have proposed varying degrees of agent behavior, such as the self-improvement in Reflexion \cite{shinn2024reflexion} and the reasoning mechanism in Graph-RAG \cite{edge2024local}, yet a unified evaluation benchmark remains absent. This fragmentation hinders reproducibility, horizontal comparisons, and the development of a cohesive theoretical foundation.

\textbf{2) Workflow complexity and error propagation in multi-agent systems.} While multi-agent collaboration enhances modularity and scalability~\cite{chen2023agentverse, wang2023voyager}, it also introduces challenges in synchronization, error recovery, and context sharing. As shown in \cite{sami2024aibasedmultiagentapproach, 10.1007/978-3-031-61154-4_8}, role-based collaboration can improve system performance, yet ensuring state consistency and semantic alignment among agents remains an open problem. Unlike traditional single LLM pipelines, asynchronous interactions between agents are more prone to cascading misunderstandings, especially in long-tail tasks such as requirement elicitation or security analysis.

\textbf{3) Tool integration bottlenecks and external dependency management.} Many LLM-based Agents rely on external tools to complete tasks (e.g., API-Bank in ToolCoder~\cite{ding2025toolcoder}, browser environments in AutoCodeRover~\cite{AutoCodeRover}), but integrating tools into different software engineering workflows (compilers, test frameworks, version control systems) often leads to fragile interfaces and poor adaptability. Moreover, there is a lack of standardized evaluation strategies for measuring agent performance under tool latency, failure, or partial observability.

\textbf{4) Lack of cross-task generalization and knowledge transfer capabilities.} Although some agents exhibit expertise in specific tasks, such as program repair or requirement compliance checking, agents capable of transferring knowledge across tasks remain rare. Frameworks like MAGIS~\cite{tao2024magis} and SWE-agent~\cite{yang2024sweagentagentcomputerinterfacesenable} attempt to accumulate experience through multiple interactions, but broader meta-learning or task-agnostic transfer mechanisms have yet to be established.

\textbf{5) Data scarcity and insufficient simulation environments constrain agent behavior evaluation.} Traditional LLM benchmarks (e.g.HumanEval, MBPP) focus on static input-output tasks, whereas agents in domains such as requirement engineering and software ssecurity, shown in works like \cite{sami2024aibasedmultiagentapproach, edge2024local, whitehouse2023llmpowereddataaugmentationenhanced} require multi-turn, multi-role simulations. The scarcity of such interactive datasets limits our ability to evaluate advanced agent behaviors such as negotiation, role conflict resolution, and tool-driven reasoning.

\textbf{6) Cognitive transparency and trustworthiness issues.} As agent autonomy increases, their reasoning and planning processes become increasingly opaque. Although studies like \cite{PathOCL, Genus} propose methods aligned with user expectations, the interpretability of decision traces, particularly in critical scenarios such as CoT + ToT planning for security sensitive tasks, it remains a key obstacle to deployment.

\subsection{Opportunities for Future Research}

\textbf{1) Towards a unified agent taxonomy and evaluation framework to enable end-to-end agent-driven software development pipelines.} The research community may establish shared definitions and multidimensional evaluation scales encompassing autonomy, tool usage, context management, and collaborative reasoning. Benchmarks such as CAASD~\cite{zhang2024experimentingnewprogrammingpractice} and API-Bank~\cite{wang2024executable} provide a solid foundation for agent-protocol-driven task construction. From requirement elicitation~\cite{Genus, 10.1007/978-3-031-61154-4_8}, system design~\cite{PathOCL}, testing~\cite{huang2023agentcoder}, to secure deployment~\cite{xiao2024ritfis}, agent modules can gradually be connected into an autonomous software engineering pipeline. This includes linking pre-development planning agents with post-deployment monitoring agents, forming a learning-enabled, decision-autonomous ecosystem.

\textbf{2) Enhancing agent memory and reflection mechanisms.} As illustrated by LLM-based Agents in this survey~\cite{islam2024mapcoder, shinn2024reflexion, zhao2024expel, masoudifard2024leveraging, zhang2023repocoderrag}, agents with explicit memory pools, experience tracking, and self-evaluation outperform stateless systems in complex debugging and requirement construction tasks. Future research may incorporate structured memory, error trajectory analysis, and goal-driven retrospection to enable stronger long-term planning capabilities.

\textbf{3) Technical collaboration between agents and humans, and embodied agents with multimodal capabilities in software engineering.} Several studies~\cite{josifoski2023flows, shen2024hugginggpt} emphasize the importance of designing workflows and evaluation processes with human-in-the-loop. Future agents should be able to support role-specific users (e.g., test engineers, product managers), achieving alignment in output granularity, decision traceability, and task handover mechanisms. Inspired by VOYAGER~\cite{qian2023communicative, wang2023voyager, yang2024sweagentagentcomputerinterfacesenable}, SE agents could be extended to handle graphical user interface operations, visual debugging, and even embedded system control. Such multimodal capabilities will allow agents to break free from purely text-based environments and operate in richer SE practice settings.

\textbf{4) Practical deployment of LLM-based Agents in software development.} Most current LLM-based Agents remain focused on function-level or module-level code completion, falling short in real-world project development tasks that require complex dependency management, version control, and multi-file structure organization~\cite{jin2025advancingcodegenerationlarge}. For LLM-based Agents to transition into practical engineering, they must possess project-level context modeling and dynamic tracking capabilities. For example, the Dynamic Code Graph Generator (DCGG) module~\cite{nguyen2024agilecoderdynamiccollaborativeagents} continuously models cross-file dependencies, class structures, and component versions to enhance adaptability to large-scale system construction tasks. Moreover, agents should evolve from being single-execution actuators to “development workflow collaborators” with role awareness. Frameworks like MapCoder~\cite{islam2024mapcoder}, MetaGPT~\cite{li2023metaagents}, and ClarifyGPT~\cite{mu2024clarifygpt} demonstrate diversified, role-specific agent task division mechanisms that significantly improve performance on complex tasks. Furthermore, frameworks such as AgentCoder and L2MAC emphasize the importance of task-decomposed and role-distributed agent collaboration as a key enabler for transitioning LLM-based Agents from static completion to real-world software engineering practice. Compared to traditional single-model or static scripting pipelines, such multi-agent architectures offer greater robustness and task adaptability, especially demonstrating superior scalability and debugging traceability in project-level engineering.

%% file: pages/conclusion.tex
\section{Conclusion} \label{cha:conclusion}
In this paper, we conducted a comprehensive literature review on the application of LLM and LLM-based agents in software engineering. We categorized software engineering into six topics: requirement engineering and documentation, code generation and software development, autonomous learning and decision making, software design and evaluation, software test generation, and software security and maintenance. For each topic, we analyzed the tasks, benchmarks, and evaluation metrics, distinguishing between LLM and LLM-based agents and discussing the differences and impacts they bring. We further analyzed and discussed the models used in the experiments of the 139 collected papers. Additionally, we provided statistics and distinctions between LLM and LLM-based agents regarding datasets and evaluation metrics. The analysis revealed that the emergence of LLM-based agents has led to extensive research and applications across various software engineering topics, demonstrating different emphases compared to traditional LLMs in terms of tasks, benchmarks, and evaluation metrics.

%% file: New_IEEEtran_how-to.bbl
\begin{thebibliography}{100}

\bibitem{wang2016bugram}
S.~Wang, D.~Chollak, D.~Movshovitz-Attias, and L.~Tan, ``Bugram: bug detection with n-gram language models,'' in {\em Proceedings of the 31st IEEE/ACM International Conference on Automated Software Engineering}, pp.~724--735, 2016.

\bibitem{vogelsang2019requirements}
A.~Vogelsang and M.~Borg, ``Requirements engineering for machine learning: Perspectives from data scientists,'' in {\em 2019 IEEE 27th International Requirements Engineering Conference Workshops (REW)}, (Jeju, Korea (South)), pp.~245--251, 2019.

\bibitem{chatgpt2022}
``Chatgpt: Optimizing language models for dialogue,'' 11 2022.
\newblock [Online; accessed 17-July-2024].

\bibitem{chen2021evaluating}
M.~Chen, J.~Tworek, H.~Jun, Q.~Yuan, H.~P. de~Oliveira~Pinto, J.~Kaplan, H.~Edwards, Y.~Burda, N.~Joseph, G.~Brockman, A.~Ray, R.~Puri, G.~Krueger, M.~Petrov, H.~Khlaaf, G.~Sastry, P.~Mishkin, B.~Chan, S.~Gray, N.~Ryder, M.~Pavlov, A.~Power, L.~Kaiser, M.~Bavarian, C.~Winter, P.~Tillet, F.~P. Such, D.~Cummings, M.~Plappert, F.~Chantzis, E.~Barnes, A.~Herbert-Voss, W.~H. Guss, A.~Nichol, A.~Paino, N.~Tezak, J.~Tang, I.~Babuschkin, S.~Balaji, S.~Jain, W.~Saunders, C.~Hesse, A.~N. Carr, J.~Leike, J.~Achiam, V.~Misra, E.~Morikawa, A.~Radford, M.~Knight, M.~Brundage, M.~Murati, K.~Mayer, P.~Welinder, B.~McGrew, D.~Amodei, S.~McCandlish, I.~Sutskever, and W.~Zaremba, ``Evaluating large language models trained on code,'' {\em arXiv preprint arXiv:2107.03374}, 2021.
\newblock arXiv:2107.03374 [cs.LG].

\bibitem{10.1145/3510003.3510203}
N.~Jain, S.~Vaidyanath, A.~Iyer, N.~Natarajan, S.~Parthasarathy, S.~Rajamani, and R.~Sharma, ``Jigsaw: large language models meet program synthesis,'' in {\em Proceedings of the 44th International Conference on Software Engineering}, ICSE '22, (New York, NY, USA), p.~1219–1231, Association for Computing Machinery, 2022.

\bibitem{li2024longcontextllmsstrugglelong}
T.~Li, G.~Zhang, Q.~D. Do, X.~Yue, and W.~Chen, ``Long-context llms struggle with long in-context learning,'' 2024.

\bibitem{10.1145/3649506}
J.~Yang, H.~Jin, R.~Tang, X.~Han, Q.~Feng, H.~Jiang, S.~Zhong, B.~Yin, and X.~Hu, ``Harnessing the power of llms in practice: A survey on chatgpt and beyond,'' {\em ACM Trans. Knowl. Discov. Data}, vol.~18, apr 2024.

\bibitem{10449667}
A.~Fan, B.~Gokkaya, M.~Harman, M.~Lyubarskiy, S.~Sengupta, S.~Yoo, and J.~M. Zhang, ``Large language models for software engineering: Survey and open problems,'' in {\em 2023 IEEE/ACM International Conference on Software Engineering: Future of Software Engineering (ICSE-FoSE)}, pp.~31--53, 2023.

\bibitem{10.1007/s11704}
L.~Wang, C.~Ma, X.~Feng, Z.~Zhang, H.~Yang, J.~Zhang, Z.~Chen, J.~Tang, X.~Chen, Y.~Lin, W.~X. Zhao, Z.~Wei, and J.~Wen, ``A survey on large language model based autonomous agents,'' {\em Frontiers of Computer Science}, vol.~18, no.~6, pp.~186345--, 2024.

\bibitem{xi2023risepotentiallargelanguage}
Z.~Xi, W.~Chen, X.~Guo, W.~He, Y.~Ding, B.~Hong, M.~Zhang, J.~Wang, S.~Jin, E.~Zhou, R.~Zheng, X.~Fan, X.~Wang, L.~Xiong, Y.~Zhou, W.~Wang, C.~Jiang, Y.~Zou, X.~Liu, Z.~Yin, S.~Dou, R.~Weng, W.~Cheng, Q.~Zhang, W.~Qin, Y.~Zheng, X.~Qiu, X.~Huang, and T.~Gui, ``The rise and potential of large language model based agents: A survey,'' 2023.

\bibitem{NEURIPS2020_6b493230}
P.~Lewis, E.~Perez, A.~Piktus, F.~Petroni, V.~Karpukhin, N.~Goyal, H.~K\"{u}ttler, M.~Lewis, W.-t. Yih, T.~Rockt\"{a}schel, S.~Riedel, and D.~Kiela, ``Retrieval-augmented generation for knowledge-intensive nlp tasks,'' in {\em Advances in Neural Information Processing Systems} (H.~Larochelle, M.~Ranzato, R.~Hadsell, M.~Balcan, and H.~Lin, eds.), vol.~33, pp.~9459--9474, Curran Associates, Inc., 2020.

\bibitem{copilot}
{GitHub, Inc.}, ``Github copilot: Your ai pair programmer.'' \url{https://github.com/features/copilot}, 2024.
\newblock [Online; accessed 17-July-2024].

\bibitem{russell2016artificial}
S.~Russell and P.~Norvig, {\em Artificial Intelligence: A Modern Approach}.
\newblock Pearson Education Limited, 2016.

\bibitem{jennings2000survey}
N.~R. Jennings, ``A survey of agent-oriented software engineering,'' {\em Knowledge Engineering Review}, vol.~15, no.~4, pp.~215--249, 2000.

\bibitem{bisk2020experiencegroundslanguage}
Y.~Bisk, A.~Holtzman, J.~Thomason, J.~Andreas, Y.~Bengio, J.~Chai, M.~Lapata, A.~Lazaridou, J.~May, A.~Nisnevich, N.~Pinto, and J.~Turian, ``Experience grounds language,'' 2020.

\bibitem{wei2022chain}
J.~Wei, X.~Wang, D.~Schuurmans, M.~Bosma, F.~Xia, E.~Chi, Q.~V. Le, D.~Zhou, {\em et~al.}, ``Chain-of-thought prompting elicits reasoning in large language models,'' {\em Advances in neural information processing systems}, vol.~35, pp.~24824--24837, 2022.

\bibitem{hou2024largelanguagemodelssoftware}
X.~Hou, Y.~Zhao, Y.~Liu, Z.~Yang, K.~Wang, L.~Li, X.~Luo, D.~Lo, J.~Grundy, and H.~Wang, ``Large language models for software engineering: A systematic literature review,'' 2024.

\bibitem{zheng2023understandinglargelanguagemodels}
Z.~Zheng, K.~Ning, J.~Chen, Y.~Wang, W.~Chen, L.~Guo, and W.~Wang, ``Towards an understanding of large language models in software engineering tasks,'' 2023.

\bibitem{nguyenduc2023generativeartificialintelligencesoftware}
A.~Nguyen-Duc, B.~Cabrero-Daniel, A.~Przybylek, C.~Arora, D.~Khanna, T.~Herda, U.~Rafiq, J.~Melegati, E.~Guerra, K.-K. Kemell, M.~Saari, Z.~Zhang, H.~Le, T.~Quan, and P.~Abrahamsson, ``Generative artificial intelligence for software engineering -- a research agenda,'' 2023.

\bibitem{ma2024lmsunderstandingcodesyntax}
W.~Ma, S.~Liu, Z.~Lin, W.~Wang, Q.~Hu, Y.~Liu, C.~Zhang, L.~Nie, L.~Li, and Y.~Liu, ``Lms: Understanding code syntax and semantics for code analysis,'' 2024.

\bibitem{yang2024robustnesssecurityprivacyexplainability}
Z.~Yang, Z.~Sun, T.~Z. Yue, P.~Devanbu, and D.~Lo, ``Robustness, security, privacy, explainability, efficiency, and usability of large language models for code,'' 2024.

\bibitem{huang2024generativesoftwareengineering}
Y.~Huang, Y.~Chen, X.~Chen, J.~Chen, R.~Peng, Z.~Tang, J.~Huang, F.~Xu, and Z.~Zheng, ``Generative software engineering,'' 2024.

\bibitem{DBLP2024}
``{DBLP}.'' \url{https://dblp.org}, 2024.

\bibitem{arXiv2024}
``{arXiv}.'' \url{https://arxiv.org/}, 2024.

\bibitem{thapa2022transformer}
C.~Thapa, S.~I. Jang, M.~E. Ahmed, S.~Camtepe, J.~Pieprzyk, and S.~Nepal, ``Transformer-based language models for software vulnerability detection,'' in {\em Proceedings of the 38th Annual Computer Security Applications Conference}, pp.~481--496, 2022.

\bibitem{desmond2024evalullm}
M.~Desmond, Z.~Ashktorab, Q.~Pan, C.~Dugan, and J.~M. Johnson, ``Evalullm: Llm assisted evaluation of generative outputs,'' in {\em Companion Proceedings of the 29th International Conference on Intelligent User Interfaces}, pp.~30--32, 2024.

\bibitem{rasheed2023autonomous}
Z.~Rasheed, M.~Waseem, K.-K. Kemell, W.~Xiaofeng, A.~N. Duc, K.~Syst{\"a}, and P.~Abrahamsson, ``Autonomous agents in software development: A vision paper,'' {\em arXiv preprint arXiv:2311.18440}, 2023.

\bibitem{manning1999foundations}
C.~Manning and H.~Schutze, {\em Foundations of statistical natural language processing}.
\newblock MIT press, 1999.

\bibitem{6795963}
S.~Hochreiter and J.~Schmidhuber, ``Long short-term memory,'' {\em Neural Computation}, vol.~9, no.~8, pp.~1735--1780, 1997.

\bibitem{hochreiter1997long}
S.~Hochreiter and J.~Schmidhuber, ``Long short-term memory,'' {\em Neural computation}, vol.~9, no.~8, pp.~1735--1780, 1997.

\bibitem{vaswani2017attention}
A.~Vaswani, N.~Shazeer, N.~Parmar, J.~Uszkoreit, L.~Jones, A.~N. Gomez, {\L}.~Kaiser, and I.~Polosukhin, ``Attention is all you need,'' {\em Advances in neural information processing systems}, vol.~30, 2017.

\bibitem{floridi2020gpt}
L.~Floridi and M.~Chiriatti, ``Gpt-3: Its nature, scope, limits, and consequences,'' {\em Minds and Machines}, vol.~30, pp.~681--694, 2020.

\bibitem{chowdhery2023palm}
A.~Chowdhery, S.~Narang, J.~Devlin, M.~Bosma, G.~Mishra, A.~Roberts, P.~Barham, H.~W. Chung, C.~Sutton, S.~Gehrmann, {\em et~al.}, ``Palm: Scaling language modeling with pathways,'' {\em Journal of Machine Learning Research}, vol.~24, no.~240, pp.~1--113, 2023.

\bibitem{zhang2022optopenpretrainedtransformer}
S.~Zhang, S.~Roller, N.~Goyal, M.~Artetxe, M.~Chen, S.~Chen, C.~Dewan, M.~Diab, X.~Li, X.~V. Lin, T.~Mihaylov, M.~Ott, S.~Shleifer, K.~Shuster, D.~Simig, P.~S. Koura, A.~Sridhar, T.~Wang, and L.~Zettlemoyer, ``Opt: Open pre-trained transformer language models,'' 2022.

\bibitem{wang2023codet5+}
Y.~Wang, H.~Le, A.~D. Gotmare, N.~D. Bui, J.~Li, and S.~C. Hoi, ``Codet5+: Open code large language models for code understanding and generation,'' {\em arXiv preprint arXiv:2305.07922}, 2023.

\bibitem{devlin2018bert}
J.~Devlin, M.-W. Chang, K.~Lee, and K.~Toutanova, ``Bert: Pre-training of deep bidirectional transformers for language understanding,'' {\em arXiv preprint arXiv:1810.04805}, 2018.

\bibitem{brown2020language}
T.~Brown, B.~Mann, N.~Ryder, M.~Subbiah, J.~D. Kaplan, P.~Dhariwal, A.~Neelakantan, P.~Shyam, G.~Sastry, A.~Askell, {\em et~al.}, ``Language models are few-shot learners,'' {\em Advances in neural information processing systems}, vol.~33, pp.~1877--1901, 2020.

\bibitem{touvron2023llama}
H.~Touvron, T.~Lavril, G.~Izacard, X.~Martinet, M.-A. Lachaux, T.~Lacroix, B.~Rozi{\`e}re, N.~Goyal, E.~Hambro, F.~Azhar, {\em et~al.}, ``Llama: Open and efficient foundation language models,'' {\em arXiv preprint arXiv:2302.13971}, 2023.

\bibitem{chen2016evolution}
J.~X. Chen, ``The evolution of computing: Alphago,'' {\em Computing in Science \& Engineering}, vol.~18, no.~4, pp.~4--7, 2016.

\bibitem{zhao2024expel}
A.~Zhao, D.~Huang, Q.~Xu, M.~Lin, Y.-J. Liu, and G.~Huang, ``Expel: Llm agents are experiential learners,'' in {\em Proceedings of the AAAI Conference on Artificial Intelligence}, vol.~38, pp.~19632--19642, 2024.

\bibitem{yao2022react}
S.~Yao, J.~Zhao, D.~Yu, N.~Du, I.~Shafran, K.~Narasimhan, and Y.~Cao, ``React: Synergizing reasoning and acting in language models,'' {\em arXiv preprint arXiv:2210.03629}, 2022.

\bibitem{huang2022language}
W.~Huang, P.~Abbeel, D.~Pathak, and I.~Mordatch, ``Language models as zero-shot planners: Extracting actionable knowledge for embodied agents,'' in {\em International conference on machine learning}, pp.~9118--9147, PMLR, 2022.

\bibitem{wang2023voyager}
G.~Wang, Y.~Xie, Y.~Jiang, A.~Mandlekar, C.~Xiao, Y.~Zhu, L.~Fan, and A.~Anandkumar, ``Voyager: An open-ended embodied agent with large language models,'' {\em arXiv preprint arXiv:2305.16291}, 2023.

\bibitem{whitehouse2023llmpowereddataaugmentationenhanced}
C.~Whitehouse, M.~Choudhury, and A.~F. Aji, ``Llm-powered data augmentation for enhanced cross-lingual performance,'' 2023.

\bibitem{white2023prompt}
J.~White, Q.~Fu, S.~Hays, M.~Sandborn, C.~Olea, H.~Gilbert, A.~Elnashar, J.~Spencer-Smith, and D.~C. Schmidt, ``A prompt pattern catalog to enhance prompt engineering with chatgpt,'' {\em arXiv preprint arXiv:2302.11382}, 2023.

\bibitem{reid2024gemini}
M.~Reid, N.~Savinov, D.~Teplyashin, D.~Lepikhin, T.~Lillicrap, J.-b. Alayrac, R.~Soricut, A.~Lazaridou, O.~Firat, J.~Schrittwieser, {\em et~al.}, ``Gemini 1.5: Unlocking multimodal understanding across millions of tokens of context,'' {\em arXiv preprint arXiv:2403.05530}, 2024.

\bibitem{ji2023survey}
Z.~Ji, N.~Lee, R.~Frieske, T.~Yu, D.~Su, Y.~Xu, E.~Ishii, Y.~J. Bang, A.~Madotto, and P.~Fung, ``Survey of hallucination in natural language generation,'' {\em ACM Computing Surveys}, vol.~55, no.~12, pp.~1--38, 2023.

\bibitem{an2024nissist}
K.~An, F.~Yang, L.~Li, Z.~Ren, H.~Huang, L.~Wang, P.~Zhao, Y.~Kang, H.~Ding, Q.~Lin, {\em et~al.}, ``Nissist: An incident mitigation copilot based on troubleshooting guides,'' {\em arXiv preprint arXiv:2402.17531}, 2024.

\bibitem{li2024agentsneed}
J.~Li, Q.~Zhang, Y.~Yu, Q.~Fu, and D.~Ye, ``More agents is all you need,'' 2024.

\bibitem{dubois2024alpacafarm}
Y.~Dubois, C.~X. Li, R.~Taori, T.~Zhang, I.~Gulrajani, J.~Ba, C.~Guestrin, P.~S. Liang, and T.~B. Hashimoto, ``Alpacafarm: A simulation framework for methods that learn from human feedback,'' {\em Advances in Neural Information Processing Systems}, vol.~36, 2024.

\bibitem{bouzenia2024repairagent}
I.~Bouzenia, P.~Devanbu, and M.~Pradel, ``Repairagent: An autonomous, llm-based agent for program repair,'' {\em arXiv preprint arXiv:2403.17134}, 2024.

\bibitem{musumeci2024llm}
E.~Musumeci, M.~Brienza, V.~Suriani, D.~Nardi, and D.~D. Bloisi, ``Llm based multi-agent generation of semi-structured documents from semantic templates in the public administration domain,'' in {\em International Conference on Human-Computer Interaction}, pp.~98--117, Springer, 2024.

\bibitem{yang2024sweagentagentcomputerinterfacesenable}
J.~Yang, C.~E. Jimenez, A.~Wettig, K.~Lieret, S.~Yao, K.~Narasimhan, and O.~Press, ``Swe-agent: Agent-computer interfaces enable automated software engineering,'' 2024.

\bibitem{zhang2024codeagentenhancingcodegeneration}
K.~Zhang, J.~Li, G.~Li, X.~Shi, and Z.~Jin, ``Codeagent: Enhancing code generation with tool-integrated agent systems for real-world repo-level coding challenges,'' 2024.

\bibitem{AutoCodeRover}
Y.~Zhang, H.~Ruan, Z.~Fan, and A.~Roychoudhury, ``Autocoderover: Autonomous program improvement,'' in {\em Proceedings of the 33rd ACM SIGSOFT International Symposium on Software Testing and Analysis}, ISSTA 2024, (New York, NY, USA), p.~1592–1604, Association for Computing Machinery, 2024.

\bibitem{patil2024goex}
S.~G. Patil, T.~Zhang, V.~Fang, R.~Huang, A.~Hao, M.~Casado, J.~E. Gonzalez, R.~A. Popa, I.~Stoica, {\em et~al.}, ``Goex: Perspectives and designs towards a runtime for autonomous llm applications,'' {\em arXiv preprint arXiv:2404.06921}, 2024.

\bibitem{hong2023metagptmetaprogrammingmultiagent}
S.~Hong, M.~Zhuge, J.~Chen, X.~Zheng, Y.~Cheng, C.~Zhang, J.~Wang, Z.~Wang, S.~K.~S. Yau, Z.~Lin, L.~Zhou, C.~Ran, L.~Xiao, C.~Wu, and J.~Schmidhuber, ``Metagpt: Meta programming for a multi-agent collaborative framework,'' 2023.

\bibitem{chen2023agentverse}
W.~Chen, Y.~Su, J.~Zuo, C.~Yang, C.~Yuan, C.~Qian, C.-M. Chan, Y.~Qin, Y.~Lu, R.~Xie, {\em et~al.}, ``Agentverse: Facilitating multi-agent collaboration and exploring emergent behaviors in agents,'' {\em arXiv preprint arXiv:2308.10848}, 2023.

\bibitem{shinn2024reflexion}
N.~Shinn, F.~Cassano, A.~Gopinath, K.~Narasimhan, and S.~Yao, ``Reflexion: Language agents with verbal reinforcement learning,'' {\em Advances in Neural Information Processing Systems}, vol.~36, 2024.

\bibitem{luo2022prcbert}
X.~Luo, Y.~Xue, Z.~Xing, and J.~Sun, ``Prcbert: Prompt learning for requirement classification using bert-based pretrained language models,'' in {\em Proceedings of the 37th IEEE/ACM International Conference on Automated Software Engineering}, pp.~1--13, 2022.

\bibitem{9218141}
T.~Hey, J.~Keim, A.~Koziolek, and W.~F. Tichy, ``Norbert: Transfer learning for requirements classification,'' in {\em 2020 IEEE 28th International Requirements Engineering Conference (RE)}, pp.~169--179, 2020.

\bibitem{zhang2023evaluation}
J.~Zhang, Y.~Chen, N.~Niu, and C.~Liu, ``Evaluation of chatgpt on requirements information retrieval under zero-shot setting,'' {\em Available at SSRN 4450322}, 2023.

\bibitem{ChatGPTvsSVMRE}
A.~El-Hajjami, N.~Fafin, and C.~Salinesi, ``Which ai technique is better to classify requirements? an experiment with svm, lstm, and chatgpt,'' 2024.

\bibitem{arora2024advancing}
C.~Arora, J.~Grundy, and M.~Abdelrazek, ``Advancing requirements engineering through generative ai: Assessing the role of llms,'' in {\em Generative AI for Effective Software Development}, pp.~129--148, Springer, 2024.

\bibitem{Genus}
T.~Rahman and Y.~Zhu, ``Automated user story generation with test case specification using large language model,'' 2024.

\bibitem{krishna2024usingllmssoftwarerequirements}
M.~Krishna, B.~Gaur, A.~Verma, and P.~Jalote, ``Using llms in software requirements specifications: An empirical evaluation,'' 2024.

\bibitem{ma2024specgenautomatedgenerationformal}
L.~Ma, S.~Liu, Y.~Li, X.~Xie, and L.~Bu, ``Specgen: Automated generation of formal program specifications via large language models,'' 2024.

\bibitem{10.1007/3-540-45251-6_29}
C.~Flanagan and K.~R.~M. Leino, ``Houdini, an annotation assistant for esc/java,'' in {\em FME 2001: Formal Methods for Increasing Software Productivity} (J.~N. Oliveira and P.~Zave, eds.), (Berlin, Heidelberg), pp.~500--517, Springer Berlin Heidelberg, 2001.

\bibitem{PathOCL}
S.~Abukhalaf, M.~Hamdaqa, and F.~Khomh, ``Pathocl: Path-based prompt augmentation for ocl generation with gpt-4,'' in {\em Proceedings of the 2024 IEEE/ACM First International Conference on AI Foundation Models and Software Engineering}, FORGE '24, (New York, NY, USA), p.~108–118, Association for Computing Machinery, 2024.

\bibitem{White2024}
J.~White, S.~Hays, Q.~Fu, J.~Spencer-Smith, and D.~C. Schmidt, {\em ChatGPT Prompt Patterns for Improving Code Quality, Refactoring, Requirements Elicitation, and Software Design}, pp.~71--108.
\newblock Cham: Springer Nature Switzerland, 2024.

\bibitem{luitel2024improving}
D.~Luitel, S.~Hassani, and M.~Sabetzadeh, ``Improving requirements completeness: Automated assistance through large language models,'' {\em Requirements Engineering}, vol.~29, no.~1, pp.~73--95, 2024.

\bibitem{10.1145/3528588.3528651}
A.~Moharil and A.~Sharma, ``Identification of intra-domain ambiguity using transformer-based machine learning,'' in {\em Proceedings of the 1st International Workshop on Natural Language-Based Software Engineering}, NLBSE '22, (New York, NY, USA), p.~51–58, Association for Computing Machinery, 2023.

\bibitem{10.1007/978-3-031-48550-3_17}
K.~Ronanki, B.~Cabrero-Daniel, and C.~Berger, ``Chatgpt as a tool for user story quality evaluation: Trustworthy out of the box?,'' in {\em Agile Processes in Software Engineering and Extreme Programming -- Workshops} (P.~Kruchten and P.~Gregory, eds.), (Cham), pp.~173--181, Springer Nature Switzerland, 2024.

\bibitem{poudel2023leveragingtransformerbasedlanguagemodels}
A.~Poudel, J.~Lin, and J.~Cleland-Huang, ``Leveraging transformer-based language models to automate requirements satisfaction assessment,'' 2023.

\bibitem{10371698}
K.~Ronanki, C.~Berger, and J.~Horkoff, ``Investigating chatgpt’s potential to assist in requirements elicitation processes,'' in {\em 2023 49th Euromicro Conference on Software Engineering and Advanced Applications (SEAA)}, pp.~354--361, 2023.

\bibitem{10459458}
J.~S. Yeow, M.~E. Rana, and N.~A. Abdul~Majid, ``An automated model of software requirement engineering using gpt-3.5,'' in {\em 2024 ASU International Conference in Emerging Technologies for Sustainability and Intelligent Systems (ICETSIS)}, pp.~1746--1755, 2024.

\bibitem{10628487}
B.~Wei, ``Requirements are all you need: From requirements to code with llms,'' in {\em 2024 IEEE 32nd International Requirements Engineering Conference (RE)}, pp.~416--422, 2024.

\bibitem{10.1007/978-3-031-60615-1_7}
E.~Musumeci, M.~Brienza, V.~Suriani, D.~Nardi, and D.~D. Bloisi, ``Llm based multi-agent generation of semi-structured documents from semantic templates in the public administration domain,'' in {\em Artificial Intelligence in HCI} (H.~Degen and S.~Ntoa, eds.), (Cham), pp.~98--117, Springer Nature Switzerland, 2024.

\bibitem{zhang2024experimentingnewprogrammingpractice}
S.~Zhang, J.~Wang, G.~Dong, J.~Sun, Y.~Zhang, and G.~Pu, ``Experimenting a new programming practice with llms,'' 2024.

\bibitem{nouri2024engineeringsafetyrequirementsautonomous}
A.~Nouri, B.~Cabrero-Daniel, F.~Törner, H.~Sivencrona, and C.~Berger, ``Engineering safety requirements for autonomous driving with large language models,'' 2024.

\bibitem{10.1007/978-3-031-61154-4_8}
Z.~Zhang, M.~Rayhan, T.~Herda, M.~Goisauf, and P.~Abrahamsson, ``Llm-based agents for automating the enhancement of user story quality: An early report,'' in {\em Agile Processes in Software Engineering and Extreme Programming} (D.~{\v{S}}mite, E.~Guerra, X.~Wang, M.~Marchesi, and P.~Gregory, eds.), (Cham), pp.~117--126, Springer Nature Switzerland, 2024.

\bibitem{masoudifard2024leveraging}
A.~Masoudifard, M.~M. Sorond, M.~Madadi, M.~Sabokrou, and E.~Habibi, ``Leveraging graph-rag and prompt engineering to enhance llm-based automated requirement traceability and compliance checks,'' {\em arXiv preprint arXiv:2412.08593}, 2024.

\bibitem{edge2024local}
D.~Edge, H.~Trinh, N.~Cheng, J.~Bradley, A.~Chao, A.~Mody, S.~Truitt, D.~Metropolitansky, R.~O. Ness, and J.~Larson, ``From local to global: A graph rag approach to query-focused summarization,'' {\em arXiv preprint arXiv:2404.16130}, 2024.

\bibitem{sami2024aibasedmultiagentapproach}
M.~A. Sami, M.~Waseem, Z.~Zhang, Z.~Rasheed, K.~Systä, and P.~Abrahamsson, ``Ai based multiagent approach for requirements elicitation and analysis,'' 2024.

\bibitem{tao2024magis}
W.~Tao, Y.~Zhou, Y.~Wang, W.~Zhang, H.~Zhang, and Y.~Cheng, ``Magis: Llm-based multi-agent framework for github issue resolution,'' {\em Advances in Neural Information Processing Systems}, vol.~37, pp.~51963--51993, 2024.

\bibitem{xie2023impactlargelanguagemodels}
D.~Xie, B.~Yoo, N.~Jiang, M.~Kim, L.~Tan, X.~Zhang, and J.~S. Lee, ``Impact of large language models on generating software specifications,'' 2023.

\bibitem{MOHARIL2023102994}
A.~Moharil and A.~Sharma, ``Tabasco: A transformer based contextualization toolkit,'' {\em Science of Computer Programming}, vol.~230, p.~102994, 2023.

\bibitem{chen2021evaluatinglargelanguagemodels}
M.~Chen, J.~Tworek, H.~Jun, Q.~Yuan, H.~P. de~Oliveira~Pinto, J.~Kaplan, H.~Edwards, Y.~Burda, N.~Joseph, G.~Brockman, A.~Ray, R.~Puri, G.~Krueger, M.~Petrov, H.~Khlaaf, G.~Sastry, P.~Mishkin, B.~Chan, S.~Gray, N.~Ryder, M.~Pavlov, A.~Power, L.~Kaiser, M.~Bavarian, C.~Winter, P.~Tillet, F.~P. Such, D.~Cummings, M.~Plappert, F.~Chantzis, E.~Barnes, A.~Herbert-Voss, W.~H. Guss, A.~Nichol, A.~Paino, N.~Tezak, J.~Tang, I.~Babuschkin, S.~Balaji, S.~Jain, W.~Saunders, C.~Hesse, A.~N. Carr, J.~Leike, J.~Achiam, V.~Misra, E.~Morikawa, A.~Radford, M.~Knight, M.~Brundage, M.~Murati, K.~Mayer, P.~Welinder, B.~McGrew, D.~Amodei, S.~McCandlish, I.~Sutskever, and W.~Zaremba, ``Evaluating large language models trained on code,'' 2021.

\bibitem{ni2023l2cevalevaluatinglanguagetocodegeneration}
A.~Ni, P.~Yin, Y.~Zhao, M.~Riddell, T.~Feng, R.~Shen, S.~Yin, Y.~Liu, S.~Yavuz, C.~Xiong, S.~Joty, Y.~Zhou, D.~Radev, and A.~Cohan, ``L2ceval: Evaluating language-to-code generation capabilities of large language models,'' 2023.

\bibitem{sun2023sql}
R.~Sun, S.~{\"O}. Arik, A.~Muzio, L.~Miculicich, S.~Gundabathula, P.~Yin, H.~Dai, H.~Nakhost, R.~Sinha, Z.~Wang, {\em et~al.}, ``Sql-palm: Improved large language model adaptation for text-to-sql (extended),'' {\em arXiv preprint arXiv:2306.00739}, 2023.

\bibitem{zheng2024codegeexpretrainedmodelcode}
Q.~Zheng, X.~Xia, X.~Zou, Y.~Dong, S.~Wang, Y.~Xue, Z.~Wang, L.~Shen, A.~Wang, Y.~Li, T.~Su, Z.~Yang, and J.~Tang, ``Codegeex: A pre-trained model for code generation with multilingual benchmarking on humaneval-x,'' 2024.

\bibitem{hu2024leveragingprintdebuggingimprove}
X.~Hu, K.~Kuang, J.~Sun, H.~Yang, and F.~Wu, ``Leveraging print debugging to improve code generation in large language models,'' 2024.

\bibitem{peng2023impactaideveloperproductivity}
S.~Peng, E.~Kalliamvakou, P.~Cihon, and M.~Demirer, ``The impact of ai on developer productivity: Evidence from github copilot,'' 2023.

\bibitem{fried2023incodergenerativemodelcode}
D.~Fried, A.~Aghajanyan, J.~Lin, S.~Wang, E.~Wallace, F.~Shi, R.~Zhong, W.~tau Yih, L.~Zettlemoyer, and M.~Lewis, ``Incoder: A generative model for code infilling and synthesis,'' 2023.

\bibitem{nijkamp2023codegenopenlargelanguage}
E.~Nijkamp, B.~Pang, H.~Hayashi, L.~Tu, H.~Wang, Y.~Zhou, S.~Savarese, and C.~Xiong, ``Codegen: An open large language model for code with multi-turn program synthesis,'' 2023.

\bibitem{10.1145/3649825}
Y.~Ding, M.~J. Min, G.~Kaiser, and B.~Ray, ``Cycle: Learning to self-refine the code generation,'' {\em Proc. ACM Program. Lang.}, vol.~8, apr 2024.

\bibitem{fakhoury2024llm}
S.~Fakhoury, A.~Naik, G.~Sakkas, S.~Chakraborty, and S.~K. Lahiri, ``Llm-based test-driven interactive code generation: User study and empirical evaluation,'' {\em IEEE Transactions on Software Engineering}, 2024.

\bibitem{ridnik2024codegenerationalphacodiumprompt}
T.~Ridnik, D.~Kredo, and I.~Friedman, ``Code generation with alphacodium: From prompt engineering to flow engineering,'' 2024.

\bibitem{zhang2023repocoderrag}
H.~Koziolek, S.~Gr\"{u}ner, R.~Hark, V.~Ashiwal, S.~Linsbauer, and N.~Eskandani, ``Llm-based and retrieval-augmented control code generation,'' in {\em Proceedings of the 1st International Workshop on Large Language Models for Code}, LLM4Code '24, (New York, NY, USA), p.~22–29, Association for Computing Machinery, 2024.

\bibitem{zhang2023algosynthesizingalgorithmicprograms}
K.~Zhang, D.~Wang, J.~Xia, W.~Y. Wang, and L.~Li, ``Algo: Synthesizing algorithmic programs with llm-generated oracle verifiers,'' 2023.

\bibitem{dong2024selfcollaborationcodegenerationchatgpt}
Y.~Dong, X.~Jiang, Z.~Jin, and G.~Li, ``Self-collaboration code generation via chatgpt,'' 2024.

\bibitem{lin2024llm}
F.~Lin, D.~J. Kim, {\em et~al.}, ``When llm-based code generation meets the software development process,'' {\em arXiv preprint arXiv:2403.15852}, 2024.

\bibitem{islam2024mapcoder}
M.~A. Islam, M.~E. Ali, and M.~R. Parvez, ``Mapcoder: Multi-agent code generation for competitive problem solving,'' {\em arXiv preprint arXiv:2405.11403}, 2024.

\bibitem{holt2024lmac}
S.~Holt, M.~R. Luyten, and M.~van~der Schaar, ``L2{MAC}: Large language model automatic computer for extensive code generation,'' in {\em The Twelfth International Conference on Learning Representations}, 2024.

\bibitem{ishibashi2024selforganizedagentsllmmultiagent}
Y.~Ishibashi and Y.~Nishimura, ``Self-organized agents: A llm multi-agent framework toward ultra large-scale code generation and optimization,'' 2024.

\bibitem{rasheed2024codepori}
Z.~Rasheed, M.~Waseem, M.~Saari, K.~Syst{\"a}, and P.~Abrahamsson, ``Codepori: Large scale model for autonomous software development by using multi-agents,'' {\em arXiv preprint arXiv:2402.01411}, 2024.

\bibitem{huang2023agentcoder}
D.~Huang, Q.~Bu, J.~M. Zhang, M.~Luck, and H.~Cui, ``Agentcoder: Multi-agent-based code generation with iterative testing and optimisation,'' {\em arXiv preprint arXiv:2312.13010}, 2023.

\bibitem{nguyen2024agilecoderdynamiccollaborativeagents}
M.~H. Nguyen, T.~P. Chau, P.~X. Nguyen, and N.~D.~Q. Bui, ``Agilecoder: Dynamic collaborative agents for software development based on agile methodology,'' 2024.

\bibitem{qian2023communicative}
C.~Qian, X.~Cong, C.~Yang, W.~Chen, Y.~Su, J.~Xu, Z.~Liu, and M.~Sun, ``Communicative agents for software development,'' {\em arXiv preprint arXiv:2307.07924}, 2023.

\bibitem{manish2024autonomous}
S.~Manish, ``An autonomous multi-agent llm framework for agile software development,'' {\em International Journal of Trend in Scientific Research and Development}, vol.~8, no.~5, pp.~892--898, 2024.

\bibitem{zheng2024opencodeinterpreter}
T.~Zheng, G.~Zhang, T.~Shen, X.~Liu, B.~Y. Lin, J.~Fu, W.~Chen, and X.~Yue, ``Opencodeinterpreter: Integrating code generation with execution and refinement,'' {\em arXiv preprint arXiv:2402.14658}, 2024.

\bibitem{schick2024toolformer}
T.~Schick, J.~Dwivedi-Yu, R.~Dess{\`\i}, R.~Raileanu, M.~Lomeli, E.~Hambro, L.~Zettlemoyer, N.~Cancedda, and T.~Scialom, ``Toolformer: Language models can teach themselves to use tools,'' {\em Advances in Neural Information Processing Systems}, vol.~36, 2024.

\bibitem{qin2023toolllm}
Y.~Qin, S.~Liang, Y.~Ye, K.~Zhu, L.~Yan, Y.~Lu, Y.~Lin, X.~Cong, X.~Tang, B.~Qian, {\em et~al.}, ``Toolllm: Facilitating large language models to master 16000+ real-world apis,'' {\em arXiv preprint arXiv:2307.16789}, 2023.

\bibitem{wang2024executable}
X.~Wang, Y.~Chen, L.~Yuan, Y.~Zhang, Y.~Li, H.~Peng, and H.~Ji, ``Executable code actions elicit better llm agents,'' in {\em Proceedings of the 41st International Conference on Machine Learning}, ICML'24, JMLR.org, 2024.

\bibitem{mu2024clarifygpt}
F.~Mu, L.~Shi, S.~Wang, Z.~Yu, B.~Zhang, C.~Wang, S.~Liu, and Q.~Wang, ``Clarifygpt: A framework for enhancing llm-based code generation via requirements clarification,'' {\em Proceedings of the ACM on Software Engineering}, vol.~1, no.~FSE, pp.~2332--2354, 2024.

\bibitem{10.1145/3672456}
X.~Jiang, Y.~Dong, L.~Wang, F.~Zheng, Q.~Shang, G.~Li, Z.~Jin, and W.~Jiao, ``Self-planning code generation with large language models,'' {\em ACM Trans. Softw. Eng. Methodol.}, jun 2024.
\newblock Just Accepted.

\bibitem{li2023metaagents}
Y.~Li, Y.~Zhang, and L.~Sun, ``Metaagents: Simulating interactions of human behaviors for llm-based task-oriented coordination via collaborative generative agents,'' {\em arXiv preprint arXiv:2310.06500}, 2023.

\bibitem{murali2023codecompose}
V.~Murali, C.~Maddila, I.~Ahmad, M.~Bolin, D.~Cheng, N.~Ghorbani, R.~Fernandez, and N.~Nagappan, ``Codecompose: A large-scale industrial deployment of ai-assisted code authoring,'' {\em arXiv preprint arXiv:2305.12050}, 2023.

\bibitem{huang2022large}
J.~Huang, S.~S. Gu, L.~Hou, Y.~Wu, X.~Wang, H.~Yu, and J.~Han, ``Large language models can self-improve,'' {\em arXiv preprint arXiv:2210.11610}, 2022.

\bibitem{chen2024more}
L.~Chen, J.~Q. Davis, B.~Hanin, P.~Bailis, I.~Stoica, M.~Zaharia, and J.~Zou, ``Are more llm calls all you need? towards scaling laws of compound inference systems,'' {\em arXiv preprint arXiv:2403.02419}, 2024.

\bibitem{chen2023teaching}
X.~Chen, M.~Lin, N.~Sch{\"a}rli, and D.~Zhou, ``Teaching large language models to self-debug,'' {\em arXiv preprint arXiv:2304.05128}, 2023.

\bibitem{kang2023explainable}
S.~Kang, B.~Chen, S.~Yoo, and J.-G. Lou, ``Explainable automated debugging via large language model-driven scientific debugging,'' {\em arXiv preprint arXiv:2304.02195}, 2023.

\bibitem{franceschelli2023creativity}
G.~Franceschelli and M.~Musolesi, ``On the creativity of large language models,'' {\em arXiv preprint arXiv:2304.00008}, 2023.

\bibitem{lai2023largelanguagemodelslaw}
J.~Lai, W.~Gan, J.~Wu, Z.~Qi, and P.~S. Yu, ``Large language models in law: A survey,'' 2023.

\bibitem{zheng2024judging}
L.~Zheng, W.-L. Chiang, Y.~Sheng, S.~Zhuang, Z.~Wu, Y.~Zhuang, Z.~Lin, Z.~Li, D.~Li, E.~Xing, {\em et~al.}, ``Judging llm-as-a-judge with mt-bench and chatbot arena,'' {\em Advances in Neural Information Processing Systems}, vol.~36, 2024.

\bibitem{wang2024rethinking}
Q.~Wang, Z.~Wang, Y.~Su, H.~Tong, and Y.~Song, ``Rethinking the bounds of llm reasoning: Are multi-agent discussions the key?,'' {\em arXiv preprint arXiv:2402.18272}, 2024.

\bibitem{chen2023towards}
L.~Chen, Y.~Zhang, S.~Ren, H.~Zhao, Z.~Cai, Y.~Wang, P.~Wang, T.~Liu, and B.~Chang, ``Towards end-to-end embodied decision making via multi-modal large language model: Explorations with gpt4-vision and beyond,'' {\em arXiv preprint arXiv:2310.02071}, 2023.

\bibitem{li2023camel}
G.~Li, H.~Hammoud, H.~Itani, D.~Khizbullin, and B.~Ghanem, ``Camel: Communicative agents for" mind" exploration of large language model society,'' {\em Advances in Neural Information Processing Systems}, vol.~36, pp.~51991--52008, 2023.

\bibitem{liu2023bolaa}
Z.~Liu, W.~Yao, J.~Zhang, L.~Xue, S.~Heinecke, R.~Murthy, Y.~Feng, Z.~Chen, J.~C. Niebles, D.~Arpit, {\em et~al.}, ``Bolaa: Benchmarking and orchestrating llm-augmented autonomous agents,'' {\em arXiv preprint arXiv:2308.05960}, 2023.

\bibitem{lu2024selfselfevolutionlanguagefeedback}
J.~Lu, W.~Zhong, W.~Huang, Y.~Wang, Q.~Zhu, F.~Mi, B.~Wang, W.~Wang, X.~Zeng, L.~Shang, X.~Jiang, and Q.~Liu, ``Self: Self-evolution with language feedback,'' 2024.

\bibitem{xie2024can}
C.~Xie, C.~Chen, F.~Jia, Z.~Ye, K.~Shu, A.~Bibi, Z.~Hu, P.~Torr, B.~Ghanem, and G.~Li, ``Can large language model agents simulate human trust behaviors?,'' {\em arXiv preprint arXiv:2402.04559}, 2024.

\bibitem{liu2024agentlite}
Z.~Liu, W.~Yao, J.~Zhang, L.~Yang, Z.~Liu, J.~Tan, P.~K. Choubey, T.~Lan, J.~Wu, H.~Wang, {\em et~al.}, ``Agentlite: A lightweight library for building and advancing task-oriented llm agent system,'' {\em arXiv preprint arXiv:2402.15538}, 2024.

\bibitem{zhuge2024language}
M.~Zhuge, W.~Wang, L.~Kirsch, F.~Faccio, D.~Khizbullin, and J.~Schmidhuber, ``Language agents as optimizable graphs,'' {\em arXiv preprint arXiv:2402.16823}, 2024.

\bibitem{bassner2024iris}
P.~Bassner, E.~Frankford, and S.~Krusche, ``Iris: An ai-driven virtual tutor for computer science education,'' in {\em Proceedings of the 2024 on Innovation and Technology in Computer Science Education V. 1}, pp.~394--400, 2024.

\bibitem{feldt2023towards}
R.~Feldt, S.~Kang, J.~Yoon, and S.~Yoo, ``Towards autonomous testing agents via conversational large language models,'' in {\em 2023 38th IEEE/ACM International Conference on Automated Software Engineering (ASE)}, pp.~1688--1693, IEEE, 2023.

\bibitem{happe2023getting}
A.~Happe and J.~Cito, ``Getting pwn'd by ai: Penetration testing with large language models,'' in {\em Proceedings of the 31st ACM Joint European Software Engineering Conference and Symposium on the Foundations of Software Engineering}, pp.~2082--2086, 2023.

\bibitem{ma2024combining}
W.~Ma, D.~Wu, Y.~Sun, T.~Wang, S.~Liu, J.~Zhang, Y.~Xue, and Y.~Liu, ``Combining fine-tuning and llm-based agents for intuitive smart contract auditing with justifications,'' {\em arXiv preprint arXiv:2403.16073}, 2024.

\bibitem{fang2024llm}
R.~Fang, R.~Bindu, A.~Gupta, and D.~Kang, ``Llm agents can autonomously exploit one-day vulnerabilities,'' {\em arXiv preprint arXiv:2404.08144}, 2024.

\bibitem{ding2025toolcoder}
H.~Ding, S.~Tao, L.~Pang, Z.~Wei, J.~Gao, B.~Ding, H.~Shen, and X.~Chen, ``Toolcoder: A systematic code-empowered tool learning framework for large language models,'' {\em arXiv preprint arXiv:2502.11404}, 2025.

\bibitem{yao2024tree}
S.~Yao, D.~Yu, J.~Zhao, I.~Shafran, T.~Griffiths, Y.~Cao, and K.~Narasimhan, ``Tree of thoughts: Deliberate problem solving with large language models,'' {\em Advances in Neural Information Processing Systems}, vol.~36, 2024.

\bibitem{rasheed2024can}
Z.~Rasheed, M.~Waseem, A.~Ahmad, K.-K. Kemell, W.~Xiaofeng, A.~N. Duc, and P.~Abrahamsson, ``Can large language models serve as data analysts? a multi-agent assisted approach for qualitative data analysis,'' {\em arXiv preprint arXiv:2402.01386}, 2024.

\bibitem{ataei2024elicitron}
M.~Ataei, H.~Cheong, D.~Grandi, Y.~Wang, N.~Morris, and A.~Tessier, ``Elicitron: An llm agent-based simulation framework for design requirements elicitation,'' {\em arXiv preprint arXiv:2404.16045}, 2024.

\bibitem{sridhara2023chatgpt}
G.~Sridhara, S.~Mazumdar, {\em et~al.}, ``Chatgpt: A study on its utility for ubiquitous software engineering tasks,'' {\em arXiv preprint arXiv:2305.16837}, 2023.

\bibitem{gao2024llm}
M.~Gao, X.~Hu, J.~Ruan, X.~Pu, and X.~Wan, ``Llm-based nlg evaluation: Current status and challenges,'' {\em arXiv preprint arXiv:2402.01383}, 2024.

\bibitem{wan2024software}
L.~J. Wan, Y.~Huang, Y.~Li, H.~Ye, J.~Wang, X.~Zhang, and D.~Chen, ``Software/hardware co-design for llm and its application for design verification,'' in {\em 2024 29th Asia and South Pacific Design Automation Conference (ASP-DAC)}, pp.~435--441, IEEE, 2024.

\bibitem{kolthoff2023data}
K.~Kolthoff, C.~Bartelt, and S.~P. Ponzetto, ``Data-driven prototyping via natural-language-based gui retrieval,'' {\em Automated software engineering}, vol.~30, no.~1, p.~13, 2023.

\bibitem{kirova2024software}
V.~D. Kirova, C.~S. Ku, J.~R. Laracy, and T.~J. Marlowe, ``Software engineering education must adapt and evolve for an llm environment,'' in {\em Proceedings of the 55th ACM Technical Symposium on Computer Science Education V. 1}, pp.~666--672, 2024.

\bibitem{jalil2023chatgpt}
S.~Jalil, S.~Rafi, T.~D. LaToza, K.~Moran, and W.~Lam, ``Chatgpt and software testing education: promises \& perils (2023),'' {\em arXiv preprint arXiv:2302.03287}, 2023.

\bibitem{suri2023software}
S.~Suri, S.~N. Das, K.~Singi, K.~Dey, V.~S. Sharma, and V.~Kaulgud, ``Software engineering using autonomous agents: Are we there yet?,'' in {\em 2023 38th IEEE/ACM International Conference on Automated Software Engineering (ASE)}, pp.~1855--1857, IEEE, 2023.

\bibitem{chen2024llmarena}
J.~Chen, X.~Hu, S.~Liu, S.~Huang, W.-W. Tu, Z.~He, and L.~Wen, ``Llmarena: Assessing capabilities of large language models in dynamic multi-agent environments,'' {\em arXiv preprint arXiv:2402.16499}, 2024.

\bibitem{vallecillos2024agent}
F.~Vallecillos~Ruiz, ``Agent-driven automatic software improvement,'' in {\em Proceedings of the 28th International Conference on Evaluation and Assessment in Software Engineering}, pp.~470--475, 2024.

\bibitem{josifoski2023flows}
M.~Josifoski, L.~Klein, M.~Peyrard, Y.~Li, S.~Geng, J.~P. Schnitzler, Y.~Yao, J.~Wei, D.~Paul, and R.~West, ``Flows: Building blocks of reasoning and collaborating ai,'' {\em arXiv preprint arXiv:2308.01285}, 2023.

\bibitem{shen2024hugginggpt}
Y.~Shen, K.~Song, X.~Tan, D.~Li, W.~Lu, and Y.~Zhuang, ``Hugginggpt: Solving ai tasks with chatgpt and its friends in hugging face,'' {\em Advances in Neural Information Processing Systems}, vol.~36, 2024.

\bibitem{weber2024large}
I.~Weber, ``Large language models as software components: A taxonomy for llm-integrated applications,'' {\em arXiv preprint arXiv:2406.10300}, 2024.

\bibitem{cheng2023batch}
Z.~Cheng, J.~Kasai, and T.~Yu, ``Batch prompting: Efficient inference with large language model apis,'' {\em arXiv preprint arXiv:2301.08721}, 2023.

\bibitem{shankar2024validates}
S.~Shankar, J.~Zamfirescu-Pereira, B.~Hartmann, A.~G. Parameswaran, and I.~Arawjo, ``Who validates the validators? aligning llm-assisted evaluation of llm outputs with human preferences,'' {\em arXiv preprint arXiv:2404.12272}, 2024.

\bibitem{roy2024exploring}
D.~Roy, X.~Zhang, R.~Bhave, C.~Bansal, P.~Las-Casas, R.~Fonseca, and S.~Rajmohan, ``Exploring llm-based agents for root cause analysis,'' in {\em Companion Proceedings of the 32nd ACM International Conference on the Foundations of Software Engineering}, pp.~208--219, 2024.

\bibitem{tufano2020unit}
M.~Tufano, D.~Drain, A.~Svyatkovskiy, S.~K. Deng, and N.~Sundaresan, ``Unit test case generation with transformers and focal context,'' {\em arXiv preprint arXiv:2009.05617}, 2020.

\bibitem{zhang2023well}
Y.~Zhang, W.~Song, Z.~Ji, N.~Meng, {\em et~al.}, ``How well does llm generate security tests?,'' {\em arXiv preprint arXiv:2310.00710}, 2023.

\bibitem{kang2022test}
H.~J. Kang, T.~G. Nguyen, B.~Le, C.~S. P{\u{a}}s{\u{a}}reanu, and D.~Lo, ``Test mimicry to assess the exploitability of library vulnerabilities,'' in {\em Proceedings of the 31st ACM SIGSOFT International Symposium on Software Testing and Analysis}, pp.~276--288, 2022.

\bibitem{feng2024prompting}
S.~Feng and C.~Chen, ``Prompting is all you need: Automated android bug replay with large language models,'' in {\em Proceedings of the 46th IEEE/ACM International Conference on Software Engineering}, pp.~1--13, 2024.

\bibitem{kang2023large}
S.~Kang, J.~Yoon, and S.~Yoo, ``Large language models are few-shot testers: Exploring llm-based general bug reproduction,'' in {\em 2023 IEEE/ACM 45th International Conference on Software Engineering (ICSE)}, pp.~2312--2323, IEEE, 2023.

\bibitem{xia2024fuzz4all}
C.~S. Xia, M.~Paltenghi, J.~Le~Tian, M.~Pradel, and L.~Zhang, ``Fuzz4all: Universal fuzzing with large language models,'' in {\em Proceedings of the IEEE/ACM 46th International Conference on Software Engineering}, pp.~1--13, 2024.

\bibitem{ryan2024code}
G.~Ryan, S.~Jain, M.~Shang, S.~Wang, X.~Ma, M.~K. Ramanathan, and B.~Ray, ``Code-aware prompting: A study of coverage guided test generation in regression setting using llm,'' {\em arXiv preprint arXiv:2402.00097}, 2024.

\bibitem{pizzorno2024coverup}
J.~A. Pizzorno and E.~D. Berger, ``Coverup: Coverage-guided llm-based test generation,'' {\em arXiv preprint arXiv:2403.16218}, 2024.

\bibitem{liu2024llm}
K.~Liu, Y.~Liu, Z.~Chen, J.~M. Zhang, Y.~Han, Y.~Ma, G.~Li, and G.~Huang, ``Llm-powered test case generation for detecting tricky bugs,'' {\em arXiv preprint arXiv:2404.10304}, 2024.

\bibitem{tang2024chatgpt}
Y.~Tang, Z.~Liu, Z.~Zhou, and X.~Luo, ``Chatgpt vs sbst: A comparative assessment of unit test suite generation,'' {\em IEEE Transactions on Software Engineering}, 2024.

\bibitem{li2024large}
K.~Li and Y.~Yuan, ``Large language models as test case generators: Performance evaluation and enhancement,'' {\em arXiv preprint arXiv:2404.13340}, 2024.

\bibitem{wang2024xuat}
Z.~Wang, W.~Wang, Z.~Li, L.~Wang, C.~Yi, X.~Xu, L.~Cao, H.~Su, S.~Chen, and J.~Zhou, ``Xuat-copilot: Multi-agent collaborative system for automated user acceptance testing with large language model,'' {\em arXiv preprint arXiv:2401.02705}, 2024.

\bibitem{lee2024unified}
C.~Lee, C.~S. Xia, J.-t. Huang, Z.~Zhu, L.~Zhang, and M.~R. Lyu, ``A unified debugging approach via llm-based multi-agent synergy,'' {\em arXiv preprint arXiv:2404.17153}, 2024.

\bibitem{deng2023pentestgpt}
G.~Deng, Y.~Liu, V.~Mayoral-Vilches, P.~Liu, Y.~Li, Y.~Xu, T.~Zhang, Y.~Liu, M.~Pinzger, and S.~Rass, ``Pentestgpt: An llm-empowered automatic penetration testing tool,'' {\em arXiv preprint arXiv:2308.06782}, 2023.

\bibitem{xiao2024ritfis}
M.~Xiao, Y.~Xiao, H.~Dong, S.~Ji, and P.~Zhang, ``Ritfis: Robust input testing framework for llms-based intelligent software,'' {\em arXiv preprint arXiv:2402.13518}, 2024.

\bibitem{wang2024navrepair}
R.~Wang, Z.~Li, C.~Wang, Y.~Xiao, and C.~Gao, ``Navrepair: Node-type aware c/c++ code vulnerability repair,'' {\em arXiv preprint arXiv:2405.04994}, 2024.

\bibitem{shestov2024finetuning}
A.~Shestov, A.~Cheshkov, R.~Levichev, R.~Mussabayev, P.~Zadorozhny, E.~Maslov, C.~Vadim, and E.~Bulychev, ``Finetuning large language models for vulnerability detection,'' {\em arXiv preprint arXiv:2401.17010}, 2024.

\bibitem{cheshkov2023evaluation}
A.~Cheshkov, P.~Zadorozhny, and R.~Levichev, ``Evaluation of chatgpt model for vulnerability detection,'' {\em arXiv preprint arXiv:2304.07232}, 2023.

\bibitem{zhang2024criticalreviewlargelanguage}
Q.~Zhang, T.~Zhang, J.~Zhai, C.~Fang, B.~Yu, W.~Sun, and Z.~Chen, ``A critical review of large language model on software engineering: An example from chatgpt and automated program repair,'' 2024.

\bibitem{wang2021codet5}
Y.~Wang, W.~Wang, S.~Joty, and S.~C. Hoi, ``Codet5: Identifier-aware unified pre-trained encoder-decoder models for code understanding and generation,'' {\em arXiv preprint arXiv:2109.00859}, 2021.

\bibitem{lu2024grace}
G.~Lu, X.~Ju, X.~Chen, W.~Pei, and Z.~Cai, ``Grace: Empowering llm-based software vulnerability detection with graph structure and in-context learning,'' {\em Journal of Systems and Software}, vol.~212, p.~112031, 2024.

\bibitem{li2023hitchhiker}
H.~Li, Y.~Hao, Y.~Zhai, and Z.~Qian, ``The hitchhiker's guide to program analysis: A journey with large language models,'' {\em arXiv preprint arXiv:2308.00245}, 2023.

\bibitem{ding2024vulnerability}
Y.~Ding, Y.~Fu, O.~Ibrahim, C.~Sitawarin, X.~Chen, B.~Alomair, D.~Wagner, B.~Ray, and Y.~Chen, ``Vulnerability detection with code language models: How far are we?,'' {\em arXiv preprint arXiv:2403.18624}, 2024.

\bibitem{toth2024llms}
R.~T{\'o}th, T.~Bisztray, and L.~Erd{\H{o}}di, ``Llms in web development: Evaluating llm-generated php code unveiling vulnerabilities and limitations,'' in {\em International Conference on Computer Safety, Reliability, and Security}, pp.~425--437, Springer, 2024.

\bibitem{ruiz2024novel}
F.~V. Ruiz, A.~Grishina, M.~Hort, and L.~Moonen, ``A novel approach for automatic program repair using round-trip translation with large language models,'' {\em arXiv preprint arXiv:2401.07994}, 2024.

\bibitem{yang2024multi}
B.~Yang, H.~Tian, J.~Ren, H.~Zhang, J.~Klein, T.~F. Bissyand{\'e}, C.~L. Goues, and S.~Jin, ``Multi-objective fine-tuning for enhanced program repair with llms,'' {\em arXiv preprint arXiv:2404.12636}, 2024.

\bibitem{dettmers2023qloraefficientfinetuningquantized}
T.~Dettmers, A.~Pagnoni, A.~Holtzman, and L.~Zettlemoyer, ``Qlora: Efficient finetuning of quantized llms,'' 2023.

\bibitem{jain2023neftunenoisyembeddingsimprove}
N.~Jain, P.~yeh Chiang, Y.~Wen, J.~Kirchenbauer, H.-M. Chu, G.~Somepalli, B.~R. Bartoldson, B.~Kailkhura, A.~Schwarzschild, A.~Saha, M.~Goldblum, J.~Geiping, and T.~Goldstein, ``Neftune: Noisy embeddings improve instruction finetuning,'' 2023.

\bibitem{zhang2024pydex}
J.~Zhang, J.~P. Cambronero, S.~Gulwani, V.~Le, R.~Piskac, G.~Soares, and G.~Verbruggen, ``Pydex: Repairing bugs in introductory python assignments using llms,'' {\em Proceedings of the ACM on Programming Languages}, vol.~8, no.~OOPSLA1, pp.~1100--1124, 2024.

\bibitem{joshi2023repair}
H.~Joshi, J.~C. Sanchez, S.~Gulwani, V.~Le, G.~Verbruggen, and I.~Radi{\v{c}}ek, ``Repair is nearly generation: Multilingual program repair with llms,'' in {\em Proceedings of the AAAI Conference on Artificial Intelligence}, vol.~37, pp.~5131--5140, 2023.

\bibitem{xiang2024far}
J.~Xiang, X.~Xu, F.~Kong, M.~Wu, H.~Zhang, and Y.~Zhang, ``How far can we go with practical function-level program repair?,'' {\em arXiv preprint arXiv:2404.12833}, 2024.

\bibitem{hilario2024generative}
E.~Hilario, S.~Azam, J.~Sundaram, K.~Imran~Mohammed, and B.~Shanmugam, ``Generative ai for pentesting: the good, the bad, the ugly,'' {\em International Journal of Information Security}, vol.~23, no.~3, pp.~2075--2097, 2024.

\bibitem{zhong2024ldb}
L.~Zhong, Z.~Wang, and J.~Shang, ``Ldb: A large language model debugger via verifying runtime execution step-by-step,'' {\em arXiv preprint arXiv:2402.16906}, 2024.

\bibitem{alhanahnah2024empirical}
M.~Alhanahnah, M.~R. Hasan, and H.~Bagheri, ``An empirical evaluation of pre-trained large language models for repairing declarative formal specifications,'' {\em arXiv preprint arXiv:2404.11050}, 2024.

\bibitem{zhang2024autocoderover}
Y.~Zhang, H.~Ruan, Z.~Fan, and A.~Roychoudhury, ``Autocoderover: Autonomous program improvement,'' in {\em Proceedings of the 33rd ACM SIGSOFT International Symposium on Software Testing and Analysis}, pp.~1592--1604, 2024.

\bibitem{tao2024magisllmbasedmultiagentframework}
W.~Tao, Y.~Zhou, Y.~Wang, W.~Zhang, H.~Zhang, and Y.~Cheng, ``Magis: Llm-based multi-agent framework for github issue resolution,'' 2024.

\bibitem{zhang2024acfixguidingllmsmined}
L.~Zhang, K.~Li, K.~Sun, D.~Wu, Y.~Liu, H.~Tian, and Y.~Liu, ``Acfix: Guiding llms with mined common rbac practices for context-aware repair of access control vulnerabilities in smart contracts,'' 2024.

\bibitem{hu2023largelanguagemodelpoweredsmart}
S.~Hu, T.~Huang, F.~İlhan, S.~F. Tekin, and L.~Liu, ``Large language model-powered smart contract vulnerability detection: New perspectives,'' 2023.

\bibitem{geissler2024concept}
F.~Geissler, K.~Roscher, and M.~Trapp, ``Concept-guided llm agents for human-ai safety codesign,'' in {\em Proceedings of the AAAI Symposium Series}, vol.~3, pp.~100--104, 2024.

\bibitem{zhong2024debuglikehumanlarge}
L.~Zhong, Z.~Wang, and J.~Shang, ``Debug like a human: A large language model debugger via verifying runtime execution step-by-step,'' 2024.

\bibitem{alshahwan2024assuredllmbasedsoftwareengineering}
N.~Alshahwan, M.~Harman, I.~Harper, A.~Marginean, S.~Sengupta, and E.~Wang, ``Assured llm-based software engineering,'' 2024.

\bibitem{cheshkov2023evaluationchatgptmodelvulnerability}
A.~Cheshkov, P.~Zadorozhny, and R.~Levichev, ``Evaluation of chatgpt model for vulnerability detection,'' 2023.

\bibitem{jin2025advancingcodegenerationlarge}
H.~Jin, H.~Chen, Q.~Lu, and L.~Zhu, ``Towards advancing code generation with large language models: A research roadmap,'' 2025.

\end{thebibliography}
